%% file: GLPR14Xiv.tex
\RequirePackage{fix-cm}
\documentclass[natbib,twocolumn]{svjour3}          % twocolumn
\smartqed  % flush right qed marks, e.g. at end of proof
\usepackage{graphicx}
\usepackage{amsfonts,amsmath,amssymb,latexsym,theorem,color,natbib}
\usepackage{aps-bibstyle}
\usepackage{algorithm}
\usepackage{algorithmic}
\usepackage{nicefrac}

\newcommand{\btheta}{\boldsymbol{\theta}}

\newcommand{\bpsi}{\boldsymbol{\psi}}
\newcommand{\bvphi}{\boldsymbol{\varphi}}

\newcommand\bu{\mathbf{u}}

\newcommand\by{\mathbf{y}}
\newcommand\bz{\mathbf{z}}

\newcommand\modl{{\mathfrak{M}}}

\newcommand{\argmax}{\operatornamewithlimits{argmax}}
\newcommand{\argmin}{\operatornamewithlimits{argmin}}
\newcommand{\prox}{\operatorname{prox}}
\newcommand\findex{\hfill$\blacktriangleleft$ \bigskip}

\journalname{Statistics and Computing}
\begin{document}

\title{Bayesian computation: a summary of the current state, and samples backwards and forwards\thanks{Supported in
part by ``SuSTaIn'', EPSRC grant EP/D063485/1, at the University of Bristol, and ``i-like", EPSRC grant EP/K014463/1, at
the University of Warwick. Krzysztof {\L}atuszy\'nski holds a Royal Society University Research Fellowship, and Marcelo
Pereyra a Marie Curie Intra-European Fellowship for Career Development. Peter Green also holds a Distinguished
Professorship at UTS, Sydney, and Christian Robert an Institut Universitaire de France chair at CEREMADE, Universit\'e Paris-Dauphine.}
}


\author{Peter J. Green \and
Krzysztof {\L}atuszy\'nski \and 
Marcelo Pereyra \and\\
Christian P.~Robert
}

\authorrunning{P. J. Green, K. {\L}atuszy\'nski, M. Pereyra, \&~C. P. Robert} % if too long for running head

\institute{Peter Green and Marcelo Pereyra\at
             School of Mathematics, University of Bristol \\
              %%Tel.: +44 117 928 7967\\
              %%Fax: +44 117 928 7999\\
              \email{P.J.Green, Marcelo.Pereyra@bristol.ac.uk}           %  \\
           \and
           Krzysztof {\L}atuszy\'nski and Christian~P.~Robert \at
              Dept. of Statistics, University of Warwick\\
	      \email{K.G.Latuszynski@warwick.ac.uk, robert@ensae.fr}
}

\date{Received: date / Accepted: date}

\maketitle

\begin{abstract}
Recent decades have seen enormous improvements in computational inference for statistical models; there have been competitive continual enhancements in a wide range of
computational tools. In Bayesian inference, first and foremost, MCMC techniques have continued to
evolve, moving from random walk proposals to Langevin drift, to Hamiltonian Monte Carlo, and so on, with both theoretical and
algorithmic innovations opening new opportunities to practitioners.  However, this impressive evolution
in capacity is confronted by an even steeper increase in the complexity of the datasets to be addressed. The difficulties of modelling and then handling ever more complex datasets most likely
call for a new type of tool for computational inference that dramatically reduces the dimension and size of the raw data while
capturing its essential aspects. Approximate models and algorithms may thus be at the core of the next computational
revolution.

\end{abstract}

\section{Introduction}
\label{intro}

One may reasonably balk at the terms ``computational statistics" and ``Bayesian computation'' since, from its very
start, statistics has always involved some computational step to extract information, something manageable like an
estimator or a prediction, from raw data. This necessarily incomplete and unavoidably biased review of the recent past,
current state, and immediate future of algorithms for Bayesian inference thus first requires us to explain what we mean by
computation in a statistical context, before turning to what we perceive as medium term solutions and possible deadends.

Computations are an issue in statistics whenever processing a dataset becomes a difficulty, a liability, or even an
impossibility. Obviously, the computational challenge varies according to the time when it is faced: what was an issue
in the 19th century is most likely not so any longer (take for instance the derivation of the moment estimates of a
mixture of two normal distributions so painstakenly set by \cite{pearson:1894} for estimating the ratio of ``forehead"
breadth to body length on a dataset of 1,000 crabs or the intense algebraic derivations found in the analysis of
variance of the 1950s and 1960s \citep{searle:casella:mcculloch:1992}). 

The introduction of simulation tools in the 1940s followed hard on the heels of the invention of the computer and certainly contributed an impetus towards faster and better computers, at least in the first decade of this revolution. This shows that these tools were both needed, and unavailable without electronic calculators. The introduction of Markov chain Monte Carlo is harder to pin down as some partial versions can be traced all the way back to 1944--45 and the Manhattan project at Los Alamos \citep{metropolis:1987}. 
It is surprisingly much later, i.e., only by the early 1990s, that such methods became part of the Bayesian toolbox, that is, some time after the devising of other computer-dependent tools like the bootstrap or the EM algorithm, and despite the availability of personal computers that considerably eased programming and experimenting \citep{robert:casella:2011}.
It is presumably pointless to try to attribute this delay to a definite cause but a certain lack of probabilistic culture within the statistics community is probably partly to blame.

What makes this time-lag in MCMC methods becoming assimilated into the statistics community even more surprising is that fact that Bayesian inference having a significant role in statistical practice was really on hold pending the discovery of flexible computational tools that (implictly or explicitly) delivered values for the medium- to high-dimensional integrals that underpin the calculation of posterior distributions, in all but toy problems where conjugacy provided explicit answers. In fact, until Bayesians discovered MCMC, the only computational methodology that seemed to offer much chance of making practical Bayesian statistics practical was the portfolio of quadrature methods developed under Adrian Smith's leadership at Nottingham \citep{naylor:smith:1982,smith:sken:1985,smithssn}.

The very first article in the first issue of \emph{Statistics and Computing}, whose quarter-century we celebrate in this special issue, was (to the journal's credit!) on Bayesian analysis, and was precisely in this direction of using clever quadrature methods to approach moderately high-dimensional posterior analysis \citep{dellaportas:wright:2001}. By the next (second) issue, sampling-based methods had started to appear, with three papers out of five in the issue on or related to Gibbs
sampling \citep{verdinelli:wasserman:1991,carlin:gelfand:1991b,wakefield:gelfand:smith:1991}.

Now, reflecting upon the evolution of MCMC methods over the 25 or so years they have been at the forefront of Bayesian inference, the focus has evolved a long way, from hierarchical models that extended the linear, mixed and generalised linear models \citep{albert:1988,carlin:gelfand:smith:1992,bennett:racine-poon:wakefield:1996} which were initially the focus, and graphical models that stemmed from image analysis 
\citep{geman:geman:1984} and artificial intelligence, to dynamical models driven by ODE's 
\citep{wilkinson:2011b} and diffusions 
\citep{roberts2001inference,roberts:papaspiliopoulos:dellaportas:2001,beskos:papaspiliopoulos:roberts:fearnhead:2006}, hidden trees
\citep{largetetsimon99,hueron2001a,chipman:george:mcculoch:2006,aldkripop2008a} and graphs, aside with decision making in highly complex graphical models. While research on MCMC theory and methodology is still active and continually branching 
\citep{papa:robe:skol:2007,andrieu:roberts:2009,latuszynski:kosmidis:papa:roberts:2010, douc:robert:2010}, progressively incorporating the capacities of parallel processors and GPUs 
\citep{lee2009a,jacob:robert:smith:2010,strid:2010,suchard2010,scott:etal:2013,calderhead:2014}, we wonder if we are not currently facing a new era where those methods are no longer appropriate to undertake the analysis of new models, and of new formulations where models are no longer completely defined. We indeed believe that imprecise models, incomplete information and summarised data will become, if not already, a central aspect of statistical analysis, due to the massive influx of data and the need to provide non-statisticians with efficient tools. This is why we also cover in this survey the notions of approximate Bayesian computation (ABC) and comment on the use of optimisation tools.

The plan of the paper is that in Sections 2 and 3 we discuss recent progress and current issues in Markov chain Monte Carlo and Approximate Bayesian Computation respectively. In Section 4, we highlight some araes of modern optimisation that, through lack of familiarity, are making less impact in the mainstream of Bayesian computation than we think justified. Our Discussion in Section 5 raises issues about data science and relevance to applications, and looks to the future.

\section{MCMC, targetting the posterior}\label{sec:mcmc}

When MCMC techniques were introduced to the mainstream statistical (Bayesian)
community in 1990, they were received with skepticism that they could one
day become the central tool of Bayesian inference. For instance, despite the
assurance provided by the ergodic theorem, many researchers thought at first
that the convergence of those algorithms was a mere theoretical anticipation rather than a practical reality, in
contrast to traditional Monte Carlo methods, and hence that they could not be trusted to provide
``exact" answers. This perspective is obviously obsolete by now, when MCMC
output is considered as ``exact" as regular Monte Carlo, if possibly less
efficient in some settings.  Nowadays, MCMC is again attracting more
attention (than in the past decade, say, where developments were more
about alternatives, some of which described in the following sections), both
because of methodological developments linked to better theoretical tools, for
instance in the handling of stochastic processes, and because of new advances in
accelerated computing via parallel and cloud computing.

\subsection{Basics of MCMC}\label{sub:basimc}

%% {\sc More traditional part on the revolution(s) represented by MCMC in terms of
%% diffusion of Bayesian and hierarchical modelling. MALA, RJMCMC, MCMC for diffusion}

The introduction of Markov chain based methods within Monte Carlo thus took a certain amount of argument to reach
the mainstream statistical community, when compared with other groups who were using MCMC methods 10 to 30 years
earlier. It may sound unlikely at the current stage of our knowledge, but using methods that (a) generated correlated
output, (b) required some burnin time to remove the impact of the initial distribution and (c) did not lead to a closed
form expression for asymptotic variances were indeed met with resistance at first. As often, the immense computing
advantages offered by this versatile tool soon overcame the reluctance to accept those methods as similarly ``exact" as
other Monte Carlo techniques, applications driving the move from the early 1990s. We reproduce below the generic
version of the ``all variables at once" Metropolis--Hastings algorithm \citep{metropolis:1953,hastings:1970,bghm:1995,robert:casella:2011} as it
(still) constitutes in our opinion a fundamental advance in computational statistics, namely that, given a computable
density $\pi$ (up to a normalising constant) on
$\Theta$, and a proposal Markov kernel $q(\cdot|\cdot)$, there exists a universal
machine that returns a Markov chain with the proper stationary distribution, hence an associated operational MCMC algorithm.

\begin{footnotesize}
\begin{algorithm}[H]
\caption{Metropolis--Hastings algorithm (generic version)}
\begin{algorithmic}
\STATE Choose a starting value $\theta^{(0)})$
\FOR {$n=1$ to $N$}
\STATE Generate $\theta^{*}$ from a proposal $q (\cdot|\theta^{(n-1)})$
\STATE Compute the acceptance probability $$\rho^{(n)}=1 \wedge \pi(\theta^{*})\,q (\theta^{(n-1)}|\theta^{*})) \big/
\pi(\theta^{(n-1)}q(\theta^{*}|\theta^{(n-1)})$$
\STATE Generate $u_n\sim\mathcal{U}(0,1)$ and take $\theta^{(n)}=\theta^{*}$ if $u_n\le\rho^{(n)}$, 
$\theta^{(n)}=\theta^{(n-1)}$ otherwise.
\ENDFOR
\end{algorithmic}
\label{algo:MH.0}
\end{algorithm}
\end{footnotesize}

The first observation about the Metropolis--Hastings is that the flexibility in
choosing $q$ is a blessing, but also a curse since the choice determines the
performance of the algorithm. Hence 
a large part of the research on MCMC along the past 30 years (if we arbitrarily set the starting date at
\cite{geman:geman:1984}) has been on choice of the proposal $q$ to improve the efficiency of the algorithm, and in characterising its convergence
properties. This typically requires gathering or
computing additional information about $\pi$ and we discuss some of the
fundamental strategies in subsequent
sections. Algorithm \ref{algo:MH.0}, and its variants in which variables are updated singly or in blocks according to some schedule, remains a keystone in standard use of MCMC methodology,
even though the newer Hamiltonian Monte Carlo approach (see Section \ref{sub:hamish}) may sooner or later come to replace it.
While there is nothing intrinsically unique to the nature of this algorithm, or optimal in its convergence properties (other than the result of \cite{peskun1973optimum} on the optimality of the acceptance ratio),
attempts to bypass Metropolis--Hastings are few and limited. For instance, the
birth-and-death process developed by \cite{stephens:2000} used a continuous time jump process to explore a set of
models, only to be later shown \citep{cappe:robert:ryden:2002} to be equivalent to the (Metropolis--Hastings) reversible jump approach of
\cite{green:1995}.

Another aspect of the generic Metropolis--Hastings that became central more recently is that while the accept--reject
step does overcome need to know the normalising constant, it still requires $\pi$, if unnormalised, and this may be too
expensive to compute or even intractable for complicated models and large datasets. Much recent research effort has been
devoted to the design and understanding of appropriate modifications that use estimators or approximations of $\pi$ instead
and we will take the opportunity to summarise some of the progress in this direction.

\subsection{MALA and Manifold MALA}\label{sub:MALA}

Stochastic differential equations (SDEs) have been and still are informing Monte
Carlo development in a number of seminal ways. A key insight is that
the Langevin diffusion on $\Theta$ solving
\begin{equation} \label{eqn:Langevin1}
d \theta_t = {1 \over 2 } \nabla \log \pi(\theta_t) d t + dB_t
\end{equation}
has $ \pi $ as its stationary and limiting distribution. Here
$B_t$ is the standard Brownian motion and $\nabla $ denotes
gradient. The crude approach of sampling an Euler discretisation
(\cite{MR1214374}) of
\eqref{eqn:Langevin1} and using it as an approximate sample from $\pi $ was introduced in the applied literature
(\cite{ermak1975computer, doll1976generalized}). The method results in a
Markov chain evolving according to the dynamics
\begin{eqnarray}
  \nonumber 
  \theta^{(n)} | \theta^{(n-1)} & \sim & Q(\theta^{(n-1)}, \cdot) \\ \label{eqn:crude} 
& := & \theta^{(n-1)} + {h \over 2 } \nabla \log \pi (\theta^{(n-1)})
  \\ \nonumber  & & \qquad \qquad + \; h^{1/2}N(0,I_{d\times d}),
\end{eqnarray}
for a chosen discretisation step $h$.
There is a delicate tradeoff between accuracy of the approximation
improving as $h \to 0$ and sampling efficiency (as measured e.g. by
the effective sample size) improving when $h$
increases. This solution was soon followed by
its Me\-tro\-po\-li\-sed version (\cite{Rossky1978}) that uses the Euler
approximation of \eqref{eqn:crude} to produce a proposal in the
Metro\-polis--Has\-tings algorithm  \ref{algo:MH.0}, by letting $q
(\cdot|\theta^{(n-1)}) := \theta^{(n-1)} + {h \over 2 } \nabla \log
\pi(\theta^{(n-1)} ) + h^{1/2}N(0,I_{d\times d}).$ While in the probability
community Langevin diffusions and their equilibrium distributions had also
been around for some time (\cite{Kent1978}), it was the \cite{roberts1996exponential}
paper (motivated by \cite{besag1994comments} comment on \cite{MR1293234}) that brought the approach to the centre of interest of the
computational statistics community and sparked systematic study,
development and applications of Metropolis adjusted Langevin
algorithms (hence MALA) and their cousins.

There is a large body of empirical evidence that at the extra price of computing the gradient, MALA algorithms typically
provide a substantial 
speed-up in convergence on certain types of problems. However for very
light-tailed 
distributions the
drift term may grow to infinity and cause additional
instability. More precisely, for distributions with sufficiently
smooth contours, MALA is geometrically ergodic
(c.f. \cite{roberts2004general}) if the tails of $\pi $ decay as
$\exp\{- |\theta|^{\beta}\}$ with $\beta \in [1,2]$, while the random
walk
Metropolis algorithm is geometrically ergodic for all $\beta \geq 1$
(\cite{roberts1996exponential, mengersen:tweedie:1996}). The lack of
geometrical ergodicity has been precisely quantified by \cite{bou2012nonasymptotic}. 

Various refinements and extensions have been proposed. These include optimal scaling
and choice of the discretisation step $h$, adaptive versions (both discussed in
Section \ref{sub:adapmc}), combinations with proximal operators
\citep{pereyra:2014,schreck2013shrinkage}, and applications and algorithm
development for the infinite-dimensional context \citep{pillai2012optimal,cotter2013mcmc}. One particular
direction of active research is considering a more general version of equation
\eqref{eqn:Langevin1} with state-dependent drift and diffusion coefficient

\small
\begin{eqnarray} \label{eqn:Langevin2}
d \theta_t & = &\Big( {\sigma(\theta_t) \over 2 }\nabla \log \pi(\theta_t)
+{\gamma(\theta_t) \over 2}\Big) d t +  \sqrt{\sigma}(\theta_t)dB_t \quad \\
\nonumber 
\gamma_i(\theta_t) &=& %\frac{1}{2} 
\sum_j \frac{\partial
                  \sigma_{ij}(\theta_t)}{\partial \theta_j},
\end{eqnarray}
\normalsize
which also has $\pi$ as invariant distribution
(\cite{xifara2014langevin}, c.f. \cite{Kent1978}). 
The resulting proposals are 
\begin{eqnarray*}
q(\cdot|\theta^{(n-1)}) & := & {h \over 2 } \Big(\sigma(\theta^{(n-1)})\nabla \log
\pi(\theta^{(n-1)} ) + \gamma(\theta^{(n-1)})\Big) \\ && \qquad \qquad
                                                         +
                                                         h^{1/2}N(0,\sigma(\theta^{(n-1)})) +  \theta^{(n-1)}.
\end{eqnarray*}
Choosing
appropriate $\sigma$ for improved ergodicity is however nontrivial. The idea
has been explored in
\cite{stramer1999langevinI,stramer1999langevinII,roberts2002langevin}
and more recently \cite{girolami:2011} introduced a mathematically-coherent
approach of relating $\sigma$ to a metric tensor on a Riemannian
manifold of probability distributions. The resulting algorithms are termed
Manifold MALA (MMALA), Simplified MMALA \citep{girolami:2011}, and position-dependent MALA
(PMALA) \citep{xifara2014langevin}, and differ in implementation cost, depending on how precise is the
use they make of versions of equation \eqref{eqn:Langevin2}.  The
approach still leaves the specification of the metric to be used in
the space of probability distributions to the user, however there are
some natural choices. 
One can, for example, take the Hessian of $\pi$ and replace its
eigenvalues by their absolute values $\lambda_i \to |\lambda_i|$. Building the metric tensor from this spectrally-positive
version of the Hessian of $\pi$ and randomising the discretisation
step size $h$ results in an algorithm that is as robust as random walk Metropolis,
in the sense that it is geometrically ergodic for
targets with tail decay of $\exp\{- |\theta|^{\beta}\}$ for $\beta
>1$ (see \cite{WolnyPHD}). A robustified version of such a metric
has been introduced in \cite{betancourt2013general} and termed
SoftAbs. Here one approximates the absolute value of the eigenspectrum
of the Hessian of $\pi$ with a smooth strictly positive function $\lambda_i \to \lambda_i \frac{\exp{\{\alpha \lambda_i\}} +
  \exp{\{-\alpha \lambda_i\}}}{\exp{\{\alpha \lambda_i\}} - \exp{\{-\alpha
    \lambda_i\}}},$ where $\alpha$ is a smoothing parameter. The metric
  stabilises the behaviour of both MMALA, and 
  Hamiltonian Monte Carlo algorithms (discussed in the sequel), in the neighbourhoods
  where the signature of the Hessian changes.

\subsection{Hamiltonian Monte Carlo}\label{sub:hamish}

As with many improvements in the literature, starting with the very notion of MCMC, Hamiltonian (or hybrid) Monte Carlo (HMC)
stems from Physics \citep{duane:etal:1987}. After a slow emergence into the statistical community \citep{neal:1996}, it is
now central in statistical software like STAN \citep{stan-software:2014}. For a complete account of this important flavour of MCMC, the reader is referred to \cite{neal:2013}, which inspired the description below; see also \citealp{betancourt:byrne:linvingstone:girolami:2014} for a highly mathematical differential-geometric
approach to HMC.

This method can be seen as a particular and efficient instance of auxiliary variables (see, e.g., \citealp{besag:green:1993} and
\citealp{rubinstein81}), in which we apply a deterministic-proposal Metropolis method to the augmented target. In physical terms, the idea behind
HMC is to add a ``kinetic energy" term to the ``potential energy" (negative log-target), leading to the Hamiltonian
$$
H(\theta,p) = - \log\pi(\theta)+p^\text{T} M^{-1} p / 2
$$
where $\theta$ denotes the object to be simulated (i.e., the parameter), $p$ its speed or momentum and $M$ the Hamiltonian
matrix of $\pi$. In more statistical language, HMC creates an auxiliary variable $p$ such that moving according to
Hamilton's equations
\begin{align*}
\dfrac{\theta}{dt} &= \dfrac{\partial H}{\partial p} =  \dfrac{\partial H}{\partial p} = M^{-1} p\\
\dfrac{dp}{dt} &= -\dfrac{\partial H}{\partial \theta} =  \dfrac{\partial \log\pi}{\partial \theta}
\end{align*}
preserves the joint distribution with density $\exp\{ -H(\theta,\allowbreak p)\}$, hence the marginal distribution of $\theta$, that is, $\pi(\theta)$. 
Hence, if we could simulate exactly this joint distribution of $(\theta,p)$, a sample from $\pi(\theta)$ would be a
by-product. However, in practice, the equation is solved approximately and hence requires a Metropolis correction. As
discussed in, e.g., \cite{neal:2013}, the dynamics induced by Hamilton's equations is reversible and volume-preserving
in the $(\theta,p)$ space, which means in practice that there is no need for a Jacobian in Metropolis updates. The practical
implementation relies on a $k-$th order symplectic integrator \citep{MR2221614}, most commonly on the $2$-nd order {\em leapfrog approximation} that relies on a small step level
$\epsilon$, updating $p$ and $\theta$ via a modified Euler's method called the leapfrog that is reversible and being symplectic, preserves
volume as well. This discretised update can be repeated for an arbitrary number of steps.

When considering the implementation via a Metropolis algorithm, a new value of the momentum $p$ is drawn from the
pseudo-prior $\propto \exp\{ -p^\text{T} M^{-1} p / 2 \}$ and it is followed by a Metropolis step, which proposal
is driven by the leapfrog approximation to the Hamiltonian dynamics on $(\theta,p)$ and which acceptance is governed by the
Metropolis acceptance probability. What makes the potential strength of this augmentation (or dis-integ\-ra\-tion) scheme is
that the value of $H(\theta,p)$ hardly changes during the Metropolis move, which means that it is most likely to be accepted
and that it may produce a very different value of $\pi(\theta)$ without modifying the overall acceptance probability. In
other words, moving along level sets is almost energy-free, but if the move proceeds for ``long enough", the chain can reach 
far-away regions of the parameter space, thus avoid the myopia of standard MCMC algorithms. As explained in
\cite{neal:2013}, this means that Hamiltonian Monte Carlo avoids the inefficient random walk behaviour of most
Metropolis--Hastings algorithms. What drives the exploration of the different values of $H(\theta,p)$ is therefore the
simulation of the momentum, which makes its calibration both quite influential and delicate (\cite{betancourt2014optimizing}) as it depends on the unknown
normalising constant of the target. (By calibration, we mean primarily the choice of the time discretisation step
$\varepsilon$ in the leapfrog approximation and of the number $L$ of leapfrog leaps, but also the choice of the precision matrix $M$.)

\subsection{Optimal scaling and Adaptive MCMC}\label{sub:adapmc}

The convergence of the Metropolis-Hastings algorithm~\ref{algo:MH.0}
depends crucially on the choice of the proposal distribution $q$, as does the performance of 
both more complex MCMC and SMC algorithms,
that often are hybrids using Metropolis--Hastings as simulation substeps.

Optimising over all implementable $q$ appears to be a ``disaster
problem'' due to its infinite-dimensional character, lack of clarity
about what is implementable, what is not, and the fact that this optimal $q$ must
depend in a complex way on the target $\pi$ to which we have only
a limited access. In particular MALA provides a specific approach to constructing
$\pi$-tailored proposals and HMC can be viewed as a combination of
Gibbs and special Metropolis moves for an extended
target.

In this optimisation context, it is thus
reasonable to restrict ourselves to some parametric family of
proposals $q_{\xi}$, or more generally of Markov transition kernels
$P_{\xi},$ where $\xi\in \Xi$ is a tuning parameter, possibly high-dimensional.

The aim of adaptive Markov chain Monte Carlo is conceptually very
simple. One expects that there is a set $\Xi_{\pi} \subset \Xi $ of good
parameters $\xi$ for which the kernel $P_{\xi} $ converges quickly to
$\pi$, and one allows the algorithm to search for $\Xi_{\pi}$ ``on the
fly'' and redesign the transition kernel during the simulation as more
and more information about $\pi$ becomes available. Thus an adaptive
MCMC algorithm would apply the kernel $P_{\xi^{(n)}}$ to sample $\theta^{(n)}$
given $\theta^{(n-1)}$, where the tuning parameter $\xi^{(n)}$ is itself a
random variable which may depend on the whole history $\theta^{(0)},
\dots, \theta^{(n-1)}$ and on $\xi^{(n-1)}$. Adaptive MCMC rests on the
hope that the adaptive parameter $\xi^{(n)}$ will find $\Xi_{\pi}$, stay
there essentially forever and inherit good convergence properties.

There are at least two fundamental
difficulties in executing this strategy in practice. First, standard measures of efficiency of
Markovian kernels, like the total variation convergence rate (c.f. \cite{MT2009,roberts2004general}),
$L^2(\pi)$ spectral gap
(\cite{diaconis1991geometric,roberts1996markov,MR1490046,levin2009markov}) or
asymptotic variance (\cite{peskun1973optimum,Geyer1992,tierney1998note}) in the Markov chain central limit
theorem will not be available explicitly, and their estimation from a
Markov chain trajectory is often an even more challenging task than the underlying MCMC estimation
problem itself.

 Secondly, when executing an adaptive strategy and trying to improve
the transition kernel on the fly, the Markov property of the process
is violated, therefore standard theoretical tools do not apply, and establishing validity of the approach becomes
significantly more difficult. While the approach has been successfully
applied in some very challenging practical problems
(\cite{solonen2012efficient,richardson2010bayesian,MarkJim}), there are examples
of seemingly reasonable adaptive algorithms that fail to converge to
the intended target distribution
(\cite{bai2011containment,latuszynski2010adaptive}), indicating that
compared to standard MCMC even more care must be taken to ensure
validity of inferential conclusions.

While heuristics-based adaptive algorithms have been considered
already in \cite{gilks1994adaptive}, a remarkable result providing a tool to address the difficulty of
optimising Markovian kernels coherently is the \cite{roberts1997weak}
paper on scaling the proposal variance.  It considers settings of increasing dimensionality and 
investigates efficiency of the random walk Metropolis algorithm as a
function of its average acceptance rate. More specifically, given a
sequence of targets $\pi_d$ on the product state space $\Theta^d$ with iid components constructed from conveniently
smooth marginal $f,$
\begin{eqnarray}
  \label{iidtarget}
  \pi_d(\theta) & := & \prod_{i=1}^{d} f(\theta_i), \qquad
                       \textrm{for}\quad d=1,2,\dots
\end{eqnarray}
consider a sequence of Markov chains $\theta_d,$ $d=1,2,\dots,$
where the chain $\theta_d  = (\theta_d^{(n)})_{n=0,1,...}$ is a random
walk Metropolis targeting $\pi_d$ 
with proposal increments distributed as $N(0, \sigma^2_d I_{d\times
  d})$.

It then turns out that the only sensible scaling of the proposal as dimensionality
increases is to take $\sigma_d^2 = l^2 d^{-1}$. In this regime the
sequence of time-rescaled first coordinate processes 
\begin{eqnarray*}
Z_d^{(t)} & := & \theta_{d, 1}^{(\lfloor td \rfloor)}, \qquad \textrm{for} \quad d=1,2,\dots 
\end{eqnarray*}
converges in a suitable sense to the solution $Z$ of a stochastic
differential equation
\begin{eqnarray*}
dZ_t & = & h(l)^{1/2}dB_t + \frac{1}{2}h(l) \nabla \log f(Z_t)dt.
\end{eqnarray*}
Hence maximising the speed of the above diffusion $h(l)$ is equivalent
to maximising the efficiency of the algorithm as the dimension goes to
infinity. Surprisingly, there is a one-to-one correspondence between
the value $l_{opt} = \argmax h(l)$ and the mean acceptance probability
of $0.234.$ 

The magic number $0.234$ does not depend on $f$ and gives a universal
tuning recipe to be used for example in adaptive algorithms: choose the scale of the increment so that approximately
$23\%$ of the proposals are accepted.

The result, established under restrictive assumptions, has been
empirically verified
to hold much more generally, for non iid targets and also in medium-
and even
low-dimensional examples with $d$ as small as $5$. It has been also combined
with relative efficiency loss due to mismatch between the proposal and
target covariance matrices (see \cite{roberts2001optimal}).

The simplicity of the result and easy access to the average acceptance
rate makes optimal scaling the main theoretical driver in development
of adaptive MCMC algorithms, and adaptive MCMC is the main application
and motivation for researching optimal scaling. 

A large body of theoretical work extends optimal scaling formally to
different and more general scenarios. For example Metropolis for
smooth non iid targets has
been addressed e.g. by \cite{MR2344305}, and in infinite
dimensional settings by \cite{MR2537193}. Discrete and other
discontinuous targets have been considered in \cite{Roberts1998} and \cite{MR3025684}. For MALA algorithms an optimal acceptance rate of 0.574 has been
established in \cite{robertsetrosenthal98b} and confirmed in infinite-dimensional settings in \cite{pillai2012optimal} along with the
stepsize $\sigma^2_d = l^2 d^{-1/3}$. Hybrid Monte Carlo
(see Section \ref{sub:hamish}) has been analysed in a similar spirit by
\cite{MR3129023} and \cite{betancourt2014optimizing} concluding that
any value $\in [0.6, 0.9]$ will be close to optimal
and the leapfrog step size should be taken as $h = l \times d^{-1/4}$. These results not only
inform about optimal tuning, but also provide an efficiency ordering
on the algorithms in $d-$dimensions. Metropolis algorithms need
$\mathcal{O}(d)$ steps to explore the state space, while MALA and HMC
need respectively $\mathcal{O}(d^{1/3})$ and  $\mathcal{O}(d^{1/4}).$ 

Further extensions include studying the transient phase before reaching
stationarity
(\cite{christensen2005scaling,jourdain2012optimal,MR3263094}),
the scaling of multiple-try MCMC (\cite{MR2891436}) and delayed rejection MCMC
(\cite{MR3270597}), and the temperature scale of parallel tempering type
algorithms (\cite{MR2826692,MR3161644}). Interestingly, the optimal scaling of the discussed in Section \ref{sub:psudos}
pseudo-marginal algorithms as obtained in
\cite{sherlock2014efficiency}, and extended to more general settings
in \cite{doucet2012efficient,sherlock2014optimal}, suggests an acceptance rate of just $0.07$.

While each of these numerous optimal scaling results gives rise, at least in
principle, to an adaptive MCMC design,
the pioneering and most successful algorithm is the Adaptive Metropolis of \cite{haario2001adaptive}. With its increasing popularity in applications, this has fuelled the
development of the field.

Here one considers a normal increment proposal that estimates the target
covariance matrix from past samples and applies appropriate dimension-dependent scaling
and covariance shrinkage. Precisely, the proposal takes the form
\begin{eqnarray}
  \label{AMprop}
  q(\cdot | \theta^{(n-1)}) & = & N( \theta^{(n-1)} , 
                                  C^{(n)}),
\end{eqnarray}
with the covariance matrix 
\begin{eqnarray}
C^{(n)} & = &  \frac{(2.38)^2}{d} \Big(\hat{\textrm{cov}}(\theta^{(0)}, \dots,
                                  \theta^{(n-1)}) + \varepsilon
                                  I_{d\times d}\Big)
\end{eqnarray}
which is efficiently computed using
a recursive formula.

Versions and refinements of the adaptive Metropolis algorithm \citep{roberts2009examples,andrieu2008tutorial}  have served well in
applications and motivated much of the theoretical development. These include, among
many other contributions,
adaptive Metropolis, delayed rejection adaptive
Me\-tro\-polis (\cite{haario2006dram}), regional adaptation and parallel
chains \citep{craiu:rosenthal:yang:2009}, and the robust version of
\cite{vihola2012robust} estimating the shape of the distribution
rather than its covariance matrix and hence suitable for heavy tailed targets.

Analogous development of adaptive MALA algorithms in
\cite{Atchade2006,marshall2012adaptive} and of adaptive Hamiltonian and
Riemannian Manifold Monte Carlo in \cite{wang2013adaptive} building on the adaptive
scaling theory, resulted in a similar drastic mixing improvement as the
original Adaptive Metropolis.

Another substantial and still unexplored area where adaptive
algorithms are applied for very high dimensional and multimodal problems is model and
variable selection
(\cite{nott2005adaptive,richardson2010bayesian,lamnisos2013adaptive,ji2013adaptive,MarkJim}). These
algorithms can incorporate reversible jump moves \citep{green:1995}
and are guided by scaling limits for discrete distributions as well as
temperature spacing of
parallel tempering to address multimodality. Successful
implementations allow for fully Bayesian variable selection in models
with over 20 000 variables for
which otherwise only ad hoc heuristic approaches have been used in the
literature.

To address the second difficulty with adaptive algorithms, several
approaches have been developed to establish their theoretical
underpinning. While for
standard MCMC, convergence in total variation and law of large numbers are obtained almost
trivially, and the effort concentrates on stronger results, like CLTs,
geometric convergence, nonasymptotic analysis, and, maybe most
importantly, comparison and ordering of algorithms, adaptive samplers
are intrinsically difficult.  The most elegant
and theoretically-valid strategy is to change the underlying
Markovian kernel at regeneration times only
(\cite{gilks1998adaptive}). Unfortunately, this is not very appealing for practitioners since regenerations are difficult to identify in more
complex settings and are essentially impractically rare in high
dimensions. The original Adaptive Metropolis of \cite{haario2001adaptive} has been validated (under
some restrictive additional conditions) by controlling the dependencies introduced by
the adaptation and using convergence results for mixingales. The
approach has been further developed in \cite{atchade2005adaptive} and
\cite{Atchade2006} to verify its ergodicity under
weaker assumptions and apply the mixingale approach to adaptive MALA. Another successful approach
(\cite{andrieu2006ergodicity} refined in \cite{saksman2008ergodicity}) rests on martingale difference
approximations and martingale limit theorems to obtain, under suitable
technical assumptions, versions of LLN and CLTs. There are close links
between analysing adaptive MCMC and stochastic approximation
algorithms and in particular the adaptation step can be often written
as a mean field of the stochastic approximation procedure; \citet{andrieu2001controlled,atchade2009adaptive,andrieu2015stability}
contribute to this direction of analysis. \citet{fort2011convergence}
develop an approach where both adaptive and interacting MCMC
algorithms can be treated in the same framework. This allows addressing
``external adaptation'' algorithms such as the interacting
tempering algorithm (a simplified version of the celebrated
equi-energy sampler of \citet{MR2283711}) or adaptive parallel tempering
in \citet{miasojedow2013adaptive}. 

We present here the rather general but fairly simple coupling approach
(\cite{roberts2007coupling}) to establishing convergence. Successfully
applied to a variety of adaptive Metropolis samplers under weak
regularity conditions (\cite{bai2011containment}), adaptive Gibbs and
adaptive Metropolis within adaptive Gibbs samplers (\cite{latuszynski2010adaptive}), it shows
that two properties \emph{Diminishing Adaptation} and
\emph{Containment} are sufficient to guarantee that an adaptive MCMC
algorithm will converge asymptotically to the correct target
distribution. To this end recall the total variation distance between
two measures defined as $\| \nu(\cdot) - \mu(\cdot)\|  := \sup_{A\in
  \mathcal{F}}|\nu(A) - \mu(A)|,$ and for every Markov transition
kernel $P_{\xi},$ $\xi \in \Xi$ and every starting point $\theta \in
\Theta$ define the $\varepsilon $ convergence function
$M_{\varepsilon}:\Theta \times \Xi \to \mathbb{N}$ as 
\begin{eqnarray*}
  M_{\varepsilon}(\theta, \xi) & :=& \inf\{n \geq 1: \|
                                     P^{(n)}_{\xi}(\theta, \cdot) -
                                     \pi(\cdot) \| \leq \varepsilon  \}.
\end{eqnarray*}
Let $\{(\theta^{(n)}, \xi^{(n)})\}_{n=0}^{\infty}$ be the
corresponding adaptive MCMC algorithm and by $A^{(n)}((\theta, \xi), \cdot)$
denote its marginal distribution at time $t,$ i.e.
\begin{eqnarray*}
  A^{(n)}((\theta, \xi), B) & := & \mathbb{P}(\theta^{(n)} \in B |
                                   \theta^{(0)}= \theta, \xi^{(0)} = \xi). 
\end{eqnarray*}
The adaptive algorithm is ergodic for every starting values of $\theta
 $ and $\xi $ if
 $
  \lim_{n\to \infty }\|
                                     A^{(n)}((\theta, \xi, \cdot) -
                                     \pi(\cdot) \|  =  0.
 $
The two conditions guaranteeing ergodicity are
\begin{definition}[Diminishing Adaptation]
The adaptive algorithm with starting values $\theta^{(0)}=\theta$ and
$\xi^{(0)} = \xi$ satisfies Diminishing Adaptation, if
\begin{eqnarray*}
  \lim_{n \to \infty} D^{(n)} & = & 0 \quad \textrm{in probability, where} \\
 D^{(n)} &:=& \sup_{\theta \in \Theta} \| P_{\xi^{(n+1)}}(\theta,
              \cdot) - P_{\xi^{(n)}}(\theta,
              \cdot)   \|.
\end{eqnarray*}
\end{definition}
\begin{definition}[Containment]
The adaptive algorithm with starting values $\theta^{(0)}=\theta$ and
$\xi^{(0)} = \xi$ satisfies Containment, if for all $\varepsilon > 0$
the sequence $\{M_{\varepsilon}(\theta^{(n)},
\xi^{(n)})\}_{n=0}^{\infty}$ is bounded in probability.
\end{definition}
While diminishing adaptation is a standard requirement, Containment
is subject to some discussion. On one hand, it may seem difficult to verify in practice; on
the other, it may appear restrictive in the context of
ergodicity results under some weaker conditions
(c.f. \cite{fort2011convergence}). However, it turns out
(\cite{latuszynski2014containment}) that if Containment is not
satisfied, then the algorithm may still converge, but with positive
probability it will be
asymptotically less efficient than any nonadaptive ergodic MCMC
scheme. Hence algorithms that do not satisfy Containment are termed
AdapFail and are best avoided. Containment has been further studied
in \cite{bai2011containment} and is in particular implied by
simultaneous geometric or polynomial drift conditions of the adaptive kernels.

   Given that adaptive algorithms may be incorporated in essentially
   any sampling scheme, their introduction seems to be one of the most
   important innovations of the last two decades. However, despite
   substantial effort and many ingenious contributions, the theory of
   adaptive MCMC lags behind practice even more than may be the
   case in other
   computational areas. While theory always matters, the numerous
   unexpected and
   counterintuitive examples of transient adaptive algorithms suggest
   that in this area theory matters even more for healthy development.

For adaptive MCMC to become a routine tool, a clear-cut result is needed
saying that under some easily verifiable conditions these algorithms
are valid and perform not much worse than their nonadaptive counterpart with
fixed parameters. Such a result is yet to be established and may
require deeper understanding of how to construct stable adaptive
MCMC, rather than aiming heavy technical artillery at algorithms
currently in use without modifying them.

\subsection{Estimated likelihoods and pseudo-marginals}\label{sub:psudos}

There are numerous settings of interest where the target density $\pi(\cdot|y)$ is not available in closed form. For
instance, in latent variable models, the likelihood function $\ell(\theta|y)$ is often only available as an intractable
integral
$$
\ell(\theta|y) = \int_\mathcal{Z} g(z,y|\theta)\,\text{d}z\,,
$$
which leads to
$$
\pi(\theta|y) \propto \pi(\theta) \int_\mathcal{Z} g(z,y|\theta)\,\text{d}z
$$
being equally intractable. A solution proposed from the early days of MCMC \citep{tanner:wong:1987} is to consider $z$
as an auxiliary variable and to simulate the joint distribution
$\pi(\theta,z|y)$ on $\Theta \times \mathcal{Z} $ by a standard method, leading to
simulating the marginal density $\pi(\cdot|y)$ as a by-product. However, when the dimension of the auxiliary variable $z$
grows with the sample size, this technique may run into difficulties as induced MCMC algorithms are more and more
likely to have convergence issues. An illustration of this case is provided by hidden Markov models, which have eventually to
resort to particle filters as Markov chain algorithms become ineffective \citep{chopin:2007,fearnhead:clifford:2003}.
Another situation where the target density $\pi(\cdot|y)$ cannot be directly computed is the case of the ``doubly
intractable" likelihood \citep{murray:etal:2006}, when the likelihood function $\ell(\theta|y)\propto g(y|\theta)$ itself contains a term that is intractable,
in that it makes the normalising constant
$$
\mathfrak{Z}(\theta) = \int_\mathcal{Y}  g(y|\theta) \,\text{d}y
$$
impossible to compute. The resulting posterior writes
\begin{eqnarray*}
\pi(\theta|y) & = & \frac{ \pi(\theta) g(y|\theta) }{ \mathfrak{Z}(\theta)
  p(y)  }, \qquad \textrm{where} \\
p(y) &=& \int_{\Theta} \frac{ \pi(\theta) g(y|\theta) }{
         \mathfrak{Z}(\theta)} \text{d}\theta,
\end{eqnarray*}
and consequently the Metropolis-Hastings acceptance rate becomes
$$
\alpha(\theta, \theta') =
 \min\left\{ 
1,
\frac{ \pi(\theta') g(y,\theta') q(\theta'|\theta) }{
  \pi(\theta) g(y,\theta) q(\theta|\theta') }  
\times 
\frac{
  \mathfrak{Z}(\theta) }{ \mathfrak{Z}(\theta')}  \right\},
$$
and cannot be evaluated exactly for algorithmic purposes.

Examples of this kind abound in Markov random fields models, as for instance for the Ising model
\citep{murray:nested:potts,moller:etal:2006}.

Both the approaches of 
\cite{murray:etal:2006} and \cite{moller:etal:2006} require sampling data
from the likelihood $\ell(\theta|y)$, which limits their applicability.
The latter uses in addition an importance
sampling function and may suffer from poor acceptance rates. 
%and retrospectively can be reinterpreted as Grouped Independence
%Metropolis-Hastings (GIMH of \cite{andrieu:roberts:2009}, see below)
%with sample size 1. When perfect sampling from the likelihood is
%impossible, \cite{girolami2013playing} develop an approach, also in
%the framework of GIMH, where the likelihoods are unbiasedly estimated
%by random truncation of their series expansions. 

\cite{andrieu:roberts:2009} propose a more
general resolution of such problems by designing a Metropolis--Hastings algorithm that
replaces the intractable target density $\pi(\cdot|y)$ with an unbiased estimator, following an idea of
\cite{beaumont:2003}. The approach is termed pseudo-marginal.
Rather than evaluating the posterior exactly, a positive unbiased
estimate $S_{\theta}$ of $\pi(\theta|y)$ is utilised. More formally, a
new Markov chain is constructed that evolves on the extended state
space $\Theta \times \mathbb{R}_+$, where at iteration $n$, given the pair
$(\theta^{(n-1)}, S_{\theta^{(n-1)}}^{(n-1)})$ of parameter value and
density estimate at this value, the proposal $(\theta', S_{\theta'})$ is
obtained by sampling $\theta' \sim q(\cdot|\theta^{(n-1)})$ and obtaining
$S_{\theta'}$, the estimate of $\pi(\theta'|y)$. Analogously to the
standard Metropolis-Hastings step, the pair  $(\theta', S_{\theta'})$ is accepted as $(\theta^{(n)}, S_{\theta^{(n)}}^{(n)})$ with probability 
$$
\min\left\{ 1, \frac{ S_{\theta'} q(\theta^{(n-1)}|\theta') }{ S_{\theta^{(n-1)}}^{(n-1)} q(\theta'|\theta^{(n-1)}) } \right\},
$$
otherwise the proposal is rejected and the new value set as
$(\theta^{(n)}, S_{\theta^{(n)}}^{(n)}) :=(\theta^{(n-1)}, S_{\theta^{(n-1)}}^{(n-1)}).$

It is not difficult to verify that the bivariate chain on extended
state space $\Theta \times \mathbb{R}_+$ enjoys the correct
$\pi(\theta|y)$ marginal on $\Theta$ and the approach is valid, see
\cite{andrieu:roberts:2009}) for details (and also
\cite{andrieu2012convergence} for an abstracted account).

One specific instance of constructing unbiased estimators of the
posterior is presented in \cite{girolami2013playing} and based on
random truncations of infinite series expansions. The paper also
offers an excellent overview of inference methods for intractable likelihoods.

The performance of the pseudo-marginal approach will depend on the quality of
the estimators $S_{\theta}$ and hence stabilising them as well as
understanding this relationship is an active area of current
development. Often $S_{\theta}$ is constructed as an importance sampler
based on an importance sample $z$.  Thus in particular, the improvements from using multiple
samples of $z$ to estimate $\pi$ are of interest and can be assessed from
\cite{andrieu2012convergence} where the efficiency of the algorithm is
studied in terms of its spectral gap and CLT asymptotic
variance. \citet{sherlock2014efficiency}, \citet{doucet2012efficient} and \citet{sherlock2014optimal}, on the other hand, investigate
the efficiency as a function of the acceptance rate and 
variance of the noise, deriving the optimal scaling, as discussed in
Section \ref{sub:adapmc}.

As an alternative to the above procedure of using estimates of the
intractable likelihood to design a new Markov chain
on an extended state space with correct marginal, one could naively use
these estimates to approximate the Metropolis-Hastings accept-reject
ratio and let the Markov chain evolve in the original state
space. This would amount to
dropping the current realisation of $S_{\theta}$ and obtaining a new
one in each accept-reject attempt. Such a procedure is termed Monte Carlo
within Metropolis \citep{andrieu:roberts:2009}. Unfortunately this approach does
not preserve the stationary distribution, and the resulting Markov
chain may even not be ergodic \citep{medina2015stability}. If ergodic,
the difference between stationary distribution, resulting from the
noisy acceptance must be quantified, which is a highly nontrivial task
and the bounds will rarely be tight (see also \cite{alquier2014noisy,
  pillai2014ergodicity,rudolf2015perturbation} for related methodology
and 
theory). The approach
is however an interesting avenue since at the price of being biased, it
overcomes mixing difficulties of the exact pseudo-marginal version. 

Design and understanding of pseudo-marginal algorithms
is a direction of dynamic methodological development that in the
coming years will be
further fuelled not only by complex models with intractable
likelihoods, but also by the need of  MCMC algorithms for Big Data. In
this context
the likelihood function cannot be evaluated for the whole
dataset even in the \emph{iid} case just because computing the long
product of individual likelihoods is infeasible. Several Big Data MCMC
approaches have been already considered in
\citet{welling2011bayesian,korattikara2013austerity,teh2014consistency,icml2014c1_bardenet14,maclaurin2014firefly,minsker2014robust,quiroz2014speeding,strathmann2015unbiased}.

\newcommand\kak{\mathfrak{h}}
\subsection{Particle MCMC}\label{sub:pmcmc}

While we refrain from covering particle filters here, since others \citep{BeskosJMS} in this volume are focussing on this
technique, a recent advance at the interface between MCMC, pseudo-marginals, and particle filtering is the notion
of particle MCMC (or {\em pMCMC}), developed by \cite{andrieu:doucet:holenstein:2010}. This innovation is indeed rather similar to the
pseudo-marginal algorithm approach, taking advantage of the state-space models and auxiliary variables used by particle filters. 
It differs from standard particle filters in that it targets (mostly) the marginal posterior distribution of the parameters.

The simplest setting in which pMCMC applies is one of a state-space model where a latent sequence $x_{0:T}$ is a Markov
chain with joint density
$$
p_0(x_0|\theta)p_1(x_1|x_0,\theta))\cdots p_T(x_T|x_{T-1},\theta)\,,
$$
and is associated with an observed sequence $y_{1:T}$ such that
$$
y_{1:T}|x_{1:T},\theta \sim \prod_{i=1}^T q_i(y_i|x_i,\theta)\,.
$$
The iterations of pMCMC are MCMC-like in that, at iteration $t$, a new value $\theta^\prime$ of $\theta$ is
proposed from an arbitrary transition kernel $\kak(\cdot|\theta^{(t)})$ and then a new value of the latent series
$x_{0:T}^\prime$ is generated from a particle filter approximation of $p(x_{0:T}|\theta^\prime,y_{1:T})$.
Since the particle filter returns as a by-product \citep{delmoral:doucet:jasra:2006} an unbiased estimator of the marginal posterior of $y_{1:T}$,
$\hat{q}(y_{1:T}|\theta^\prime)$, this estimator can be used as such in the Metropolis--Hastings ratio
$$
\dfrac{\hat{q}(y_{1:T}|\theta^\prime)\pi(\theta^\prime)\kak(\theta^{(t)}|\theta^\prime)}{\hat{q}(y_{1:T}|\theta)\pi(\theta^{(t)})\kak(\theta^\prime|\theta^{(t)})}
\wedge 1\,.
$$
Its validity follows from the general argument of \cite{andrieu:roberts:2009}, although some additional (notational)
effort is needed to demonstrate all random variables used therein are correctly assessed (see
\citealp{andrieu:doucet:holenstein:2010} and \citealp{wilkinson:2011}, the latter providing a very progressive introduction to
the notions of pMCMC and particle Gibbs, which helped greatly in composing this section). Note however that the general
validation of pMCMC as targetting the joint posterior of the states and parameters and of the parallel particle Gibbs
sampler does not follow from pseudo-marginal arguments.

This approach is being used increasingly in complex dynamic models like those found in signal processing
\citep{whiteley:andrieu:doucet:2010}, dynamical systems like the PDEs in biochemical kinetics \citep{wilkinson:2011b}
and probabilistic graphical models \citep{lindsen:jordan:schon:2014}. An extension to approximating the
sequential filtering distribution is found in \cite{chopin:jacob:papaspiliopoulos:2013}.

\subsection{Parallel MCMC}\label{sub:paramcee}

Since MCMC relies on local updating based on the current value of a Markov chain, opportunities for exploiting parallel resources,
either CPU or GPU, would seem quite limited, In fact, the possibilities reach far beyond the basic notion of running independent or coupled
MCMC chains on several processors. For instance, \cite{craiu:meng:2005} construct parallel antithetic coupling to
create negatively correlated MCMC chains (see also \citealp{frigessi:2000}), while \cite{craiu:rosenthal:yang:2009} use parallel exploration of the sample
space to tune an adaptive MCMC algorithm. \cite{jacob:robert:smith:2010} exploit GPU facilities to improve by
Rao-Blackwellisation the Monte Carlo approximations produced by a Markov chain, even though the parallelisation does not
improve the convergence of the chain. See also \cite{lee2009a} and \cite{suchard2010} for more detailed contributions on the
appeal of using GPUs towards massive parallelisation, and \cite{wilkinson:2005} for a general survey on the topic.

Another recently-explored direction  is ``prefetching". Based on \cite{brockwell:2006} this approach computes the $2^2,
2^3, \ldots, 2^k$ values of the posterior that will be needed $2, 3, \ldots, k$ sweeps ahead by simulating the possible
``futures" of the Markov chain, according to whether the next $k$ proposals are accepted or not, in parallel. Running a regular Metropolis--Hastings algorithm then means building a
decision tree back to the current iteration and drawing $2,3, \ldots,k$ uniform variates to go down the tree to the
appropriate branch. As noted by \cite{brockwell:2006}, ``in the case where one can guess whether or not acceptance
probabilities will be `high' or `low', the tree could be made deeper down `high' probability paths and shallower in the
`low' probability paths."  This idea is exploited in \cite{angelino:etal:2014}, by creating ``speculative moves" that
consider the reject branch of the prefetching tree more often than not, based on some preliminary or dynamic evaluation
of the acceptance rate. Using a fast but close-enough approximation to the true target (and a fixed sequence of
uniforms) may also produce a ``single most likely path" on which prefetched simulations can be run. The basic idea is
thus to run simulations and costly likelihood computations on many parallel processors along a prefetched path, a path
that has been prefetched for its high approximate likelihood. There are obviously instances where this speculative
simulation is not helpful because the actual chain with the genuine target ends up following another path.
\cite{angelino:etal:2014} actually go further by constructing sequences of approximations for the precomputations. The
proposition for the sequence found therein is to subsample the original data and use a normal approximation to the
difference of the log (sub-)likelihoods. See \cite{strid:2010} for related ideas.

A different use of parallel capabilities is found in \cite{calderhead:2014}. At each iteration of Calderhead's algorithm,
$N$ replicas are generated, rather than $1$ in traditional Metropolis--Hastings. The Markov chain actually consists of
$N$ components, from which one component is selected at random as a seed for the next proposal. This approach can
be seen as a special type of data augmentation \citep{tanner:wong:1987}, where the index of the selected component is an
auxiliary variable. The neat trick in the proposal (and the reason for its efficiency gain)
is that the stationary distribution of the auxiliary variable can be determined and hence used $N$ times
in updating the vector of $N$ components. An interesting feature of this approach is when the original Metropolis--Hastings algorithm is
expressed as a finite state space Markov chain on the set of indices $\{1,\ldots,N\}$. Conditional on the values of
the $N$ dimensional vector, the stationary distribution of that sub-chain is no longer uniform. Hence, picking
$N$ indices from the stationary helps in selecting the most appropriate images, which explains why the rejection
rate decreases. The paper indeed evaluates the impact of increasing the number of proposals in terms of effective sample
size (ESS), acceptance rate, and mean squared jump distance. Since this proposal is an
almost free bonus resulting from using $N$ processors, it sounds worth investigating and comparing with more complex parallel schemes.

\cite{neiswanger:wang:xing:2013} introduced the notion of embarrassingly parallel MCMC, where ``embarrassing" refers to
the ``embarrassingly simple" solution proposed therein, namely to solve the difficulty in handling very large datasets by running completely independent parallel MCMC samplers on parallel threads or computers and using the outcomes of those samplers as density estimates, pulled together as a product towards an approximation of the true posterior density. In other words, the idea is to break the posterior as
\begin{equation}\label{eq:partition}
p(\theta|y) \propto \prod_{i=1}^m p_i(\theta|y)
\end{equation}
and to use the estimate
$$
\hat p(\theta|y) \propto \prod_{i=1}^m \hat p_i(\theta|y)
$$
where the individual estimates are obtained, say, nonparametrically. The method is then ``asymptotically
exact" in the weak (and unsurprising) sense of converging in the number of MCMC iterations. Still, there is a theoretical
justification that is not found in previous parallel methods that mixed all resulting samples without accounting for the
subsampling. And the point is made that, in many cases, running MCMC samplers with subsamples produces faster convergence.
The decomposition of $p(\cdot)$ into its components is done by partitioning the iid data into $M$ subsets and taking a
power $1/m$ of the prior in each case. (This may induce issues about impropriety.) However, the subdivision is arbitrary and
can thus be implemented in cases other than the fairly restrictive iid setting. Because each (subsample)  nonparametric
estimate involves $T$ terms, the resulting overall estimate contains $Tm$ terms and the authors suggest using an independent
Metropolis sampler to handle this complexity. This is in fact necessary for
producing a final sample from the (approximate) true posterior distribution. 

In a closely related way, \cite{wang:dunson:2013} start from the same product representation of the target (posterior), namely,
\eqref{eq:partition}. However, they criticise the choice made by \cite{neiswanger:wang:xing:2013} to use MCMC approximations 
for each component of the product for the following reasons:
\begin{enumerate}
\item   Curse of dimensionality in the number of parameters $d$;
\item   Curse of dimensionality in the number of subsets $m$;
\item   Tail degeneration;
\item   Support inconsistency and mode misspecification.
\end{enumerate}
Point 1 is relevant, but there may be ways other than kernel estimation to mix samples from the terms in the
product. Point 2 is less of a clearcut drawback: while the $Tm$ terms corresponding to a product of $m$
sums of $T$ terms sounds self-defeating, \cite{neiswanger:wang:xing:2013} use a clever device to avoid the
combinatorial explosion, namely operating on one component at a time. Having non-manageable targets is not such an issue in
the post-MCMC era. Point 3 is formally correct, in that the kernel tail behaviour induces the kernel estimate tail
behaviour, most likely disconnected from the true target tail behaviour, but this feature is true for any non-parametric
estimate, even for the Weierstrass transform defined below, and hence maybe not so relevant in practice. In fact, by
lifting the tails up, the simulation from the subposteriors should help in visiting the tails of the true target.  
Finally, point 4 does not seem to be life-threatening. Assuming that the true target can be computed up to a normalising
constant, the value of the target for every simulated parameter could be computed, eliminating those outside the support
of the product and highlighting modal regions. 

The Weierstrass transform of a density $f$ is a convolution of $f$ and of an arbitrary kernel $K$. \cite{wang:dunson:2013}  propose to
simulate from the product of the Weierstrass transform, using a multi-tiered Gibbs sampler. Hence, the parameter is
only simulated once and from a controlled kernel, while the random effects from the convolution are related with each
subposterior. While the method requires coordination between the parallel threads, the components of the target are
separately computed on a single thread. The clearest perspective on the Weierstrass transform may possibly be the
rejection sampling version where simulations from the subpriors are merged together into a normal proposal
on $\theta$, to be accepted with a probability depending on the subprior simulations.

\cite{vanderwerken:schmidler:2013} keep with the spirit of parallel MCMC papers like consensus Bayes
\citep{scott:etal:2013},  embarrassingly parallel MCMC \citep{neiswanger:wang:xing:2013} and Weierstrass MCMC
\citep{wang:dunson:2013}, namely that the computation of the likelihood can be broken into batches and MCMC run over
those batches independently. The idea of the authors is to replace an exploration of the whole space operated via a
single Markov chain (or by parallel chains acting independently which all have to ``converge") with parallel and
independent explorations of parts of the space by separate Markov chains. The motivation is that ``Small is beautiful":
it takes a shorter while to explore each set of the partition, hence to converge, and, more importantly, each chain can work 
in parallel with the others. More specifically, given a partition of the space, into sets $A_i$ with posterior weights
$w_i$, parallel chains are associated with targets equal to the original target restricted to those $A_i$‘s. This is
therefore an MCMC version of partitioned sampling. With regard to the shortcomings listed in the quote above, the
authors consider that there does not need to be a bijection between the partition sets and the chains, in that a chain
can move across partitions and thus contribute to several integral evaluations simultaneously. It is somewhat unclear
(a) whether or not this impacts ergodicity (it all depends on the way the chain is constructed, i.e., against which
target) as it could lead to an over-representation of some boundary regions and (b) whether or not it improves the
overall convergence properties of the chain(s). A more delicate issue with the partitioned MCMC approach stands with the
partitioning. Indeed, in a complex and high-dimension model, the construction of the appropriate partition is a
challenge in itself as we often have no prior idea where the modal areas are. Waiting for a correct exploration of the
modes is indeed faster than waiting for crossing between modes, provided all modes are represented and the chain for
each partition set $A_i$ has enough energy to explore this set. It actually sounds unlikely that a target with huge gaps
between modes will see a considerable improvement from the partioned version when the partition sets $A_i$ are selected
on the go, because some of the boundaries between the partition sets may be hard to reach with an off-the-shelf
proposal. A last comment about this innovative paper is that the adaptive construction of the partition has much in
common with Wang-Landau schemes \citep{wangetlandau01,leeetal06,atchade:liu:2010,jacob:ryder:2014}.

\section{ABC and others, exactly delivering an approximation}\label{sec:abc}

Motivated by highly complex models where MCMC algorithms and other Monte Carlo methods were too inefficient by far,
approximate methods have emerged where the output cannot be considered as simulations from the genuine posterior, even under idealised situations of
infinite computing power. These methods include ABC techniques, described in
more details below, but also variational Bayes \citep{jaakkola:jordan:2000}, empirical likelihood \citep{owen:2001},
INLA \citep{rue:martino:chopin:2009} and other solutions that rely on pseudo-models, or on summarised versions of the data,
or both. It is quite important to signal this evolution as we think that it may be a central feature of
computational Bayesian statistics in the coming years. From a statistical perspective, it also induces a somewhat paradoxical
situation where loss of information is balanced by improvement in precision, for a given computational budget. This
perspective is not only interesting at the computational level but forces us (as statisticians) to re-evaluate in depth
the nature of a statistical model and could produce a paradigm shift in the near future by giving a brand new meaning to
George Box's motto that ``all models are wrong". 

\subsection{ABC per se}\label{sub:abc.0}

It seems important to discuss ABC (Approximate Bayesian computation) in this partial tour of Bayesian
computational techniques as (a) they provide the only approach to their model for some Bayesians, (b)
they deliver samples in the parameter space that are exact simulations from a posterior of some kind \citep{wilkinson:2013},
$\pi^{\text{ABC}}(\theta|\by_0)$ if not the original posterior $\pi(\theta|\by_0)$, where $\by_0$ denotes the data in
this section (c) they may be more intuitive to some
researchers outside statistics, as they entail simulating from the inferred model, i.e., going forward from parameter to
data, rather than backward, from data to parameter, as in traditional Bayesian inference, (d) they can be merged with
MCMC algorithms, and (e) they allow drawing inference directly from summaries of the data rather than the data
itself. 

ABC techniques play a role in the 2000s that MCMC methods did in the 1990s, in that they handle new models for
which earlier (e.g., MCMC) algorithms were at a loss, in the same way the latter (MCMC) were able to handle
models that regular Monte Carlo approaches could not reach, such as latent variable models \citep{tanner:wong:1987,
diebolt:robert:1994, richardson:green:1997}. New models for which ABC unlocked the gate include Markov random
fields, Kingman's coalescent for phylogeographical data, likelihood models with an intractable normalising constant, and
models defined by their quantile function or their characteristic function. While the ABC approach first appeared a
``quick-and-dirty" solution, to be considered only until more elaborate representations could be found,
those algorithms have been progressively incorporated into the statistician's toolbox as a novel form of generic
nonparametric inference handling partly-defined statistical models. They are therefore attractive as much for this
reason as for being handy computational solutions when everything else fails. 

A statistically intriguing feature of those methods is that they customarily require---for greater
efficiency---replacing the data with (much) smaller-dimension summaries\footnote{Maybe due to their initial introduction
in population genetics, the oxymoron `summary statistics' is now prevalent in descriptions of ABC algorithms, included
in the statistical literature, where the (linguistically sufficient) term `statistic' would suffice.} or summary statistics, 
because of the complexity of the former. In almost every case calling for ABC, those summaries are not sufficient
statistics and the method thus implies from the start a loss of statistical information, at
least at a formal level, since relying on the raw data is out of the question and therefore the additional information it
provides is moot. This imposed reduction of the statistical information raises many relevant questions, from the choice
of summary statistics \citep{blum:nunes:prangle:sisson:2013} to the consistency of the ensuing inference \citep{robert:cornuet:marin:pillai:2011}.

Although it has now diffused into a wide range of applications, the technique of Approximate Bayesian Computation (ABC) was
first introduced by and for population genetics \citep{tavare:balding:griffith:donnelly:1997,
pritchard:seielstad:perez:feldman:1999} to handle ancestry models driven by Kingman's coalescent and with strictly
intractable likelihoods \citep{beaumont:2010}. The likelihood function of such genetic models is indeed ``intractable" in
the sense that, while derived from a fully defined and parameterised probability model, this function cannot be computed
(at all or within a manageable time) for a single value of the parameter and for the given data. Bypassing the original
example to avoid getting mired into the details of population genetics, examples of intractable likelihoods include
densities with intractable normalising constants, i.e., $f (\by|\theta) = g(\by|\theta)/Z(\theta)$ such as in Potts
\citep{potts52} and auto-exponential \citep{besag:1972} models, and pseudo-likelihood models
\citep{cucala:marin:robert:titterington:2006}. 

\begin{example} A very simple illustration of an intractable likelihood is provided by Bayesian inference based on the
median and median absolute deviation statistics of a sample from an arbitrary location-scale family,
$y_1,\ldots,y_n\stackrel{\text{iid}}{\sim} \sigma^{-1} g(\sigma^{-1}\{y-\mu\})$, as the joint distribution of this
statistic is not available in closed form.
\findex\end{example}

The concept at the core of ABC methods can be seen as both very na{\"\i}ve and intrinsically related to the foundations
of Bayesian statistics as {\em inverse probability} \citep{rubin:1984}. This concept is that data $\by$ simulated conditional
on values of the parameter close to the ``true" value of the parameter should look more similar to the actual data
$\by_0$ than data $\by$ simulated conditional on values of the parameter far from the ``true" value. ABC actually
involves an acceptance/rejection step in that parameters simulated from the prior are accepted only when
$$
\rho(\by,\by_0) < \epsilon\,,
$$
where $\rho(\cdot,\cdot)$ is a distance and $\epsilon>0$ is called the tolerance. It can be shown that the algorithm 
exactly samples the posterior when $\epsilon=0$, but this is very rarely achievable in practice \citep{grelaud:marin:robert:rodolphe:tally:2009}.
An algorithmic representation is as follows:

\begin{footnotesize}
\begin{algorithm}[H]
\caption{ABC (basic version)}
\begin{algorithmic}
\FOR {$t=1$ to $N$}
\REPEAT
\STATE Generate $\theta^{*}$ from the prior $\pi(\cdot)$
\STATE Generate $\by^{*}$ from the model $f(\cdot|\theta^{*})$
\STATE Compute the distance $\rho(\mathbf{y}^0,\mathbf{y}^{*})$
\STATE Accept $\theta^{*}$ if $\rho(\mathbf{y}^0,\mathbf{y}^{*})<\epsilon$
\UNTIL acceptance
\ENDFOR
\RETURN $N$ accepted values of $\theta^*$
\end{algorithmic}
\label{algo:ABC.0}
\end{algorithm}
\end{footnotesize}

Calibration of the ABC method in Algorithm \ref{algo:ABC.0} involves selecting the distance
$\rho(\cdot,\cdot)$ and deducing the tolerance from computational cost constraints.
However, in realistic settings, ABC is never implemented as such because comparing raw data to simulated raw data is
rarely efficient, noise dominating signal (see, e.g., \cite{marin:pudlo:robert:ryder:2011} for toy examples). It is
therefore natural that one first considers dimension-reduction techniques to bypass this curse of dimensionality. For instance, if rudimentary estimates $S(y)$ of the parameter $\theta$ are available, they are
good candidates. In the ABC literature, they are called {\em summary statistics}, a term that does not impose
any constraint on their form and hence leaves open the question of performance,
as discussed in \cite{marin:pudlo:robert:ryder:2011,blum:nunes:prangle:sisson:2013}. A 
more practical version of the ABC algorithm is shown in Algorithm \ref{algo:ABC+sum} below, with a different output for
each choice of the summary statistic. We stress in this version of the algorithm the construction of the tolerance
$\epsilon$ as a quantile of the simulated distances $\rho(S(\mathbf{y}^0),S(\mathbf{y}^{(t)}))$, rather than an
additional parameter of the method.

\begin{footnotesize}
\begin{algorithm}[H]
\caption{ABC (version with summary)}
\begin{algorithmic}
\FOR {$t=1$ to $N_{ref}$}
\STATE Generate $\theta^{(t)}$ from the prior $\pi(\cdot)$
\STATE Generate $\by^{(t)}$ from the model $f(\cdot|\theta^{(t)})$
\STATE Compute $d_t=\rho(S(\mathbf{y}^0),S(\mathbf{y}^{(t)}))$
\ENDFOR \\
Order distances $d_{(1)}\leq d_{(2)}\leq \ldots \leq d_{(N_{ref})}$ \\
\RETURN the values $\theta^{(t)}$ associated with the $k$ smallest distances
\end{algorithmic}
\label{algo:ABC+sum}
\end{algorithm}
\end{footnotesize}

An immediate question about this approximate algorithm is how much it remains connected with the original posterior
distribution and in case it does not, where does it draw its legitimacy. A first remark in this connection
is that it constitutes at best a convergent approximation to the posterior distribution $\pi(\theta|S(y_0))$. It can
easily be seen that ABC generates outcomes from a genuine posterior distribution when the data is randomised with scale
$\epsilon$ \citep{wilkinson:2013,fearnhead:prangle:2012}. This interpretation indicates a decrease in the precision of
the inference but it does not provide a universal validation of the method. A second perspective on the ABC output is
that it is based on a nonparametric approximation of the sampling distribution \citep{blum:2010,blum:francois:2010}, connected
with both indirect inference \citep{drovandi:pettitt:faddy:2011} and $k$-nearest neighbour estimation
\citep{biau:etal:2014}. While a purely Bayesian nonparametric analysis of this aspect has not yet emerged, this
brings an additional if cautious support for the method.

\begin{figure}[H]
\includegraphics[width=.45\textwidth]{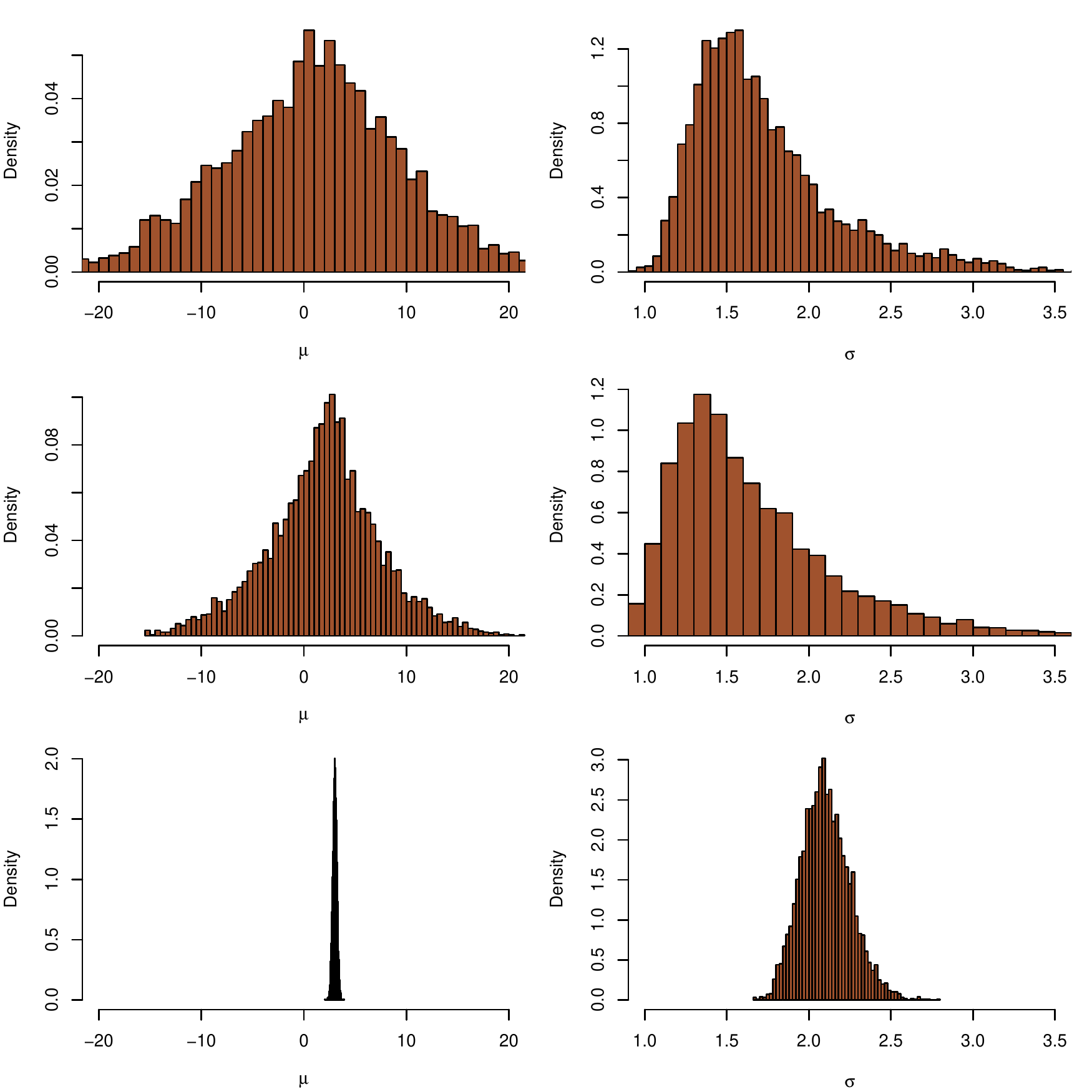}
\caption{\label{fig:medmad}
Comparison of the posterior distributions on $\mu$ {\em (left)} and $\sigma$ {\em (right)} when using an ABC algorithm
\ref{algo:ABC+sum} with distance $\rho_1$ {\em (top)} and $\rho_2$ {\em (central)}, and when using a standard Gibbs
sampler {\em (bottom)}. All three samples are based on the same number of subsampled parameters. The dataset is a
$\mathcal{N}(3,2^2)$ sample and the tolerance value $\epsilon$ corresponds to $\alpha=.5$\%~of the reference table.}
\end{figure}

\begin{figure}[H]
\includegraphics[width=.45\textwidth]{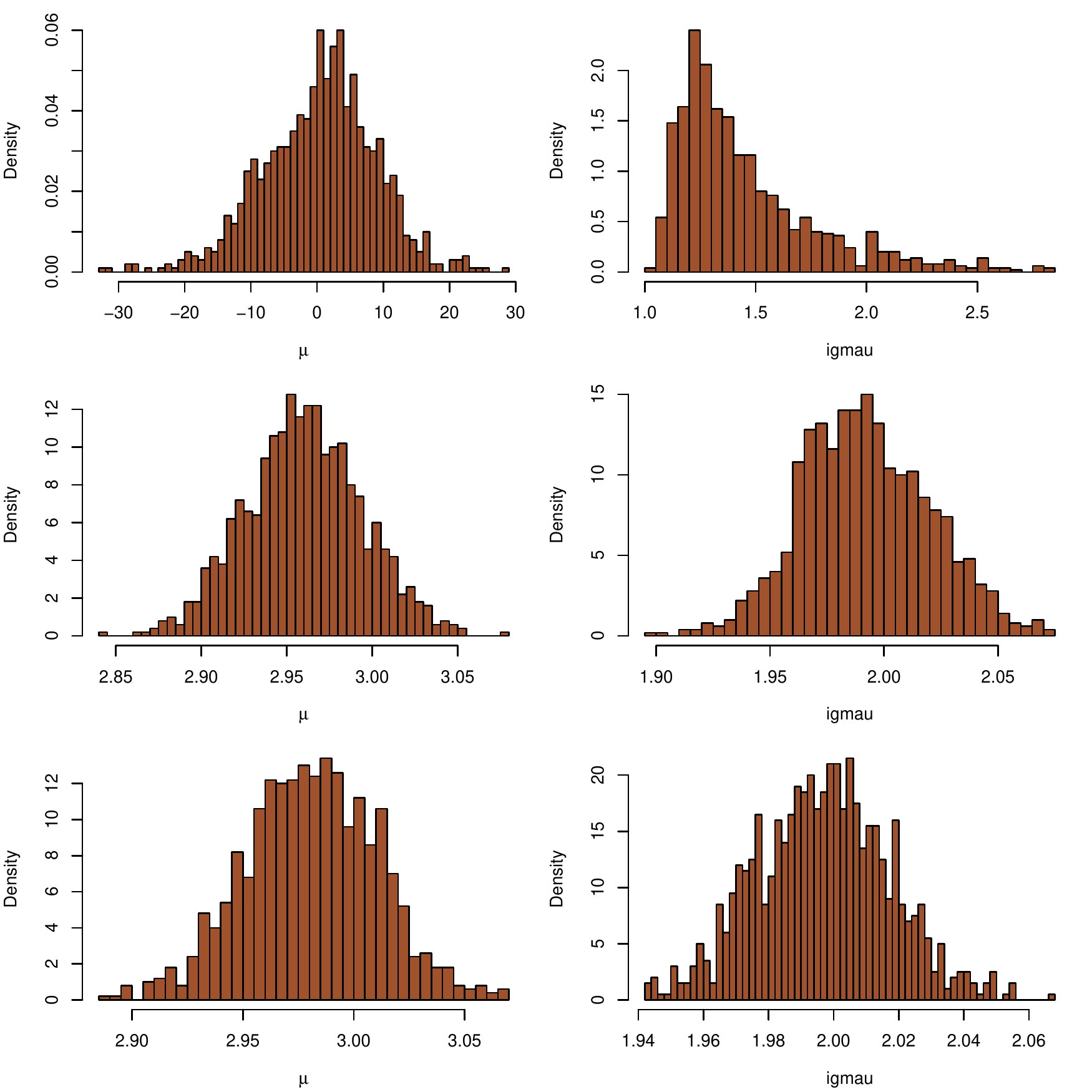}
\caption{\label{fig:beaumad}
Comparison of the posterior distributions on $\mu$ {\em (left)} and $\sigma$ {\em (right)} when using an ABC algorithm
\ref{algo:ABC+sum} with distance $\rho_1$ {\em (top)}, a post-processed version by Beaumont et al.'s
(\citeyear{beaumont:zhang:balding:2002}) local regression {\em (central)}, and when using a standard Gibbs
sampler {\em (bottom)}. The simulation setting is the same as in Figure \ref{fig:medmad}.}
\end{figure}

\begin{example} 
Continuing from the previous example of a location-scale sample only monitored through the pair median plus mad
statistic, we consider the special case of a normal sample $y_1,\ldots,y_n\sim\mathcal{N}(\mu,\sigma^2)$, with $n=100$. Using a
conjugate prior $\mu\sim\mathcal{N}(0,10)$, $\sigma^{-2}\sim\mathcal{G}a(2,5)$, we generated $10^6$ parameter values,
along with the corresponding pairs of summary statistics. When creating the distance $\rho(\cdot,\cdot)$, we used both
following versions:
\begin{align*}
\rho_1(S(\mathbf{y}^0),S(\mathbf{y})) &= \nicefrac{|\text{med}(\mathbf{y}^0)-\text{med}(\mathbf{y}|}{
\text{mad}(\text{med}(\mathbf{Y}))}\\
&\qquad + \nicefrac{|\text{mad}(\mathbf{y}^0)-\text{mad}(\mathbf{y}|}{
\text{mad}(\text{mad}(\mathbf{Y}))}\\
\rho_2(S(\mathbf{y}^0),S(\mathbf{y})) &= \nicefrac{|\text{med}(\mathbf{y}^0)-\text{med}(\mathbf{y}|}{
\text{mad}(\text{med}(\mathbf{Y}))} +\\
&\qquad \nicefrac{|\log\text{mad}(\mathbf{y}^0)-\log\text{mad}(\mathbf{y}|}{
\text{mad}(\log\text{mad}(\mathbf{Y}))}\\
\end{align*}
where the denominators are computed from the reference table in order to scale the components properly. Figure
\ref{fig:medmad} shows the impact of the choice of this distance, but even more clearly the discrepancy between
inference based on the ABC and the true inference on $(\mu,\sigma^2)$.

The discrepancy can however be completely eliminated by post-processing:  Figure \ref{fig:beaumad} reproduces Figure
\ref{fig:medmad} by comparing the histograms of an ABC sample with the version corrected by Beaumont et al.'s
(\citeyear{beaumont:zhang:balding:2002}) local regression, as the latter is essentially equivalent to a regular Gibbs 
output.
\findex\end{example}

\cite{barber:voss:webster:2013} studies the rate of convergence for ABC algorithms through the mean square error when
approximating a posterior moment. They show the convergence rate is of order $\text{O}(n^{\nicefrac{2}{q+4}})$, when $q$
is the dimension of the ABC summary statistic, associated with an optimal tolerance in $\text{O}(n^{\nicefrac{-1}{4}})$.
Those rates are connected with the nonparametric nature of ABC, as already suggested in the earlier literature: for instance, 
\cite{blum:2010}, who links ABC with standard kernel density non-parametric estimation and find a tolerance
(re-expressed as a bandwidth) of order $n^{\nicefrac{-1}{q+4}}$ and an rmse of order 
$\nicefrac{2}{q+4}$ as well, while \cite{fearnhead:prangle:2012} obtain similar rates, with a tolerance of order
$n^{\nicefrac{-1}{q+2}}$ for noisy ABC. See also \cite{calvet:czellar:2014}. Similarly, \cite{biau:etal:2014} obtain
precise convergence rates for ABC interpreted as a $k$-nearest-neighbour estimator. 

\cite{lee:latuszynski:2014} have also produced precise characterisations of the geometric ergodicity or lack thereof of four
ABC-MCMC algorithms:
\begin{enumerate}
\item the standard ABC-MCMC (with $N$ replicates of the
simulated pseudo-data to each simulated parameter value),
\item versions
involving simulations of the replicates repeated at the subsequent step,
\item use of a stopping rule in the generation
of the pseudo data, and 
\item a ``gold-standard” algorithm based on the (unavailable) measure of an $\epsilon$ ball around the
data. 
\end{enumerate}
Based a result by \cite{roberts:tweedie:1996}, also used in \cite{mengersen:tweedie:1996}, namely that an MCMC chain
cannot be geometrically ergodic when there exist almost-absorbing states, they derive that (under
some technical assumptions) the first two versions above cannot be variance-bounding (i.e., that the spectral gap is
zero),  while the last two versions can be both variance-bounding and geometrically ergodic under some appropriate 
conditions on the prior and the above ball measure. This result is thus rather striking in simulating a {\em random}
number of auxiliary variables is sufficient to produce geometric ergodicity. We note that this result does not
contradict the parallel result of \cite{bornn:etal:2014}, who establish that there is no efficiency gain in simulating
$N>1$ replicates of the pseudo-data, since there is no randomness involved in that approach.
However, the latter result only applies to functions with finite variances.

When testing hypotheses and selecting models, the Bayesian approach relies on modelling hypotheses and model indices as
part of the parameter and hence ABC naturally operates as this level as well, as demonstrated in Algorithm
\ref{algo:ABCMoo} following \cite{cornuet:etal:2008}, \cite{grelaud:marin:robert:rodolphe:tally:2009}
and \cite{toni:etal:2009}. In fields like population genetics, model choice and hypotheses validation is presumably the
primary motivation for using ABC methods as exemplified in \cite{belle:etal:2008,cornuet:ravigne:estoup:2010,
excoffier:leuenberger:wegman:2009,ghirotto:etal:2010,guillemaud:etal:2009, leuenberger:wegmann:2010,
patin:etal:2009,ramakrishnan:hadly:2009,verdu:etal:2009,wegmann:excoffier:2010}. It is also the area that attracts
most of the criticisms addressed against ABC: while some are easily dismissed \citep[see,
e.g.,][]{templeton:2008,templeton:2010,clade:2010,berger:fienberg:raftery:robert:2010}, the impact of the choice of the
summary statistics on the value of the posterior probability remains a delicate issue that prompted
\cite{pudlo:pnas:2014} to advocate the alternative use of a posterior predictive error. 

\begin{footnotesize}
\begin{algorithm}[H]
\caption{ABC (model choice)}
\begin{algorithmic}
\FOR {$i=1$ to $N$}
\STATE Generate $\modl$ from the prior $\pi(\mathcal{M}=m)$
\STATE Generate $\theta_{\modl}$ from the prior $\pi_{\modl}(\theta_\modl)$
\STATE Generate $\by$ from the model $f_{\modl}(\by|\theta_{\modl})$
\STATE Compute the distance $\rho\{S(\by),S(\by_0)\}$
\STATE Set $\modl^{(i)}=\modl$ and $\theta^{(i)}=\theta_\modl$
\ENDFOR
\RETURN the values $\modl^{(i)}$ associated with the $k$ smallest distances
\end{algorithmic}\label{algo:ABCMoo}
\end{algorithm}
\end{footnotesize}

Indeed, \cite{robert:cornuet:marin:pillai:2011} pointed out the potential irrelevance of ABC-based posterior probabilities, due to the possible ancilarity (for model choice) of summary statistics, as also explained in \cite{didelot:everitt:johansen:lawson:2011}. \cite{marin:pillai:robert:rousseau:2011} consider for instance the comparison of normal and Laplace fits on both normal and Laplace samples and show that using sample mean and sample variance as summary statistics produces Bayes factors converging to values near 1, instead of the consistent $0$ and $+\infty$. 

\cite{marin:pillai:robert:rousseau:2011} analyses this phenomenon with the aim of producing a necessary and sufficient
consistency condition on summary statistics. Quite naturally, the summaries that are acceptable must display different
behaviour under both models, in the guise of ranges of means $\mathbb{E}_\theta[S(\by_0)]$ that do not intersect for the two
models. (In the counter-example of the normal-Laplace test, the expectations of the sample mean and variance can be
recovered under both models.) This characterisation then leads to a practical asymptotic test validating summary
statistics and to the realisation that a larger number of summaries helps in achieving this goal (while degrading the
estimated tolerance). More importantly, it shows that the reduction of information represented by an ABC approach may
prevent discriminating between models, at least when trying to recover the Bayes factor. In the end, this is a natural consequence of simplifying the description of both the data and the model, and can be found in most limited information settings.

\subsection{More fish in the alphabet soup}\label{sub:def.1}
Besides ABC, approximation techniques have spread wide and far towards analysing more complex or less completely defined
models. Rather than a confusion, this multiplicity of available approximations is beneficial both to the understanding of
the underlying model and to the calibration of those different methods. 

Variational Bayes methods have been proposed for at least two decades to substitute exponential families
$q(\theta|\lambda)$ for complex posterior distributions $\pi(\theta)$ \citep{jordan:etal:1999,mackay:2002}. The central notion in those
methods is that the exponential family structure and a so-called mean-field representation of the approximation
$$
q(\theta|\lambda)=\prod_{i=1}^k q_i(\theta_i|\lambda_i)
$$
allows for a sometimes closed-form minimisation of the Kullback-Leibler distance $\text{KL}(q(\theta|\lambda),\pi(\theta))$
between the true target and its approximation. If not, the setting is quite congenial to the use of EM algorithms
\citep{paisley:blei:jordan:2012}. See \cite{salimans:knowles:2013} for a contemporary view on this approach, which
offers considerable gains in terms of computing time, while being difficult to assess in terms of discrepancy with the
``truth", i.e., the outcome that would result from using the genuine posterior.

Another approach that has met with considerable interest in the past five years is Integrated nested Laplace approximation (INLA) \citep{rue:martino:chopin:2009}.
The method operates on latent Gaussian random fields, with
likelihoods of the form
$$
\prod_{i=1}^n f(x_i|\eta_i,\theta)\,,
$$
where the $x_i$'s are the observables and the $\eta_i$'s are latent variables. Using Laplace approximations to the
marginal distributions $\pi(\theta|\mathbf{x}_0)$ and to $f(\mathbf{\eta}|\mathbf{x}_0)$, INLA produces fast and accurate
approximations of the true posterior distribution as well as of the marginal likelihood value. Thanks to the
availability of a well-constructed package called {\sf R-INLA}, this approach has gathered a large group of followers.

A somewhat exotic example of variational approximation is expectation-propagation (EP) \citep{minka:2001}, which starts from an
arbitrary decomposition of the target distribution 
$$
\pi(\theta) = \prod_{j=1}^k \pi_j(\theta)
$$
(often inspired by a likelihood decomposition into groups of observations)
and iteratively approximate each term $\pi_j$ in the product by a density member of an exponential family,
$\nu(\cdot|\lambda)$m using the
other approximations as a marginal. Given the current approximation of $\pi(\theta)$ at iteration $t$,
$$
\nu(\theta|\lambda_t) = \prod_{j=1}^k \nu_j(\theta|\lambda_t)\,,
$$
where $\lambda_t$ is the current value of the hyperparameter, the $t$-th step in the expectation-propagation (EP)
algorithm goes as follows:
\begin{enumerate}
\item Select $1\le j\le k$ at random
\item Define the marginal 
$$
\nu_{-j}(\theta|\lambda_t) \propto \frac{\nu(\theta|\lambda_t)}{\nu_j(\theta|\lambda_t)}\,;
$$
\item Update the hyperparameter $\lambda_t$ by solving
$$
\lambda_{t+1} = \argmin_\lambda \text{KL}\left\{\pi_j(\theta)\nu_{-j}(\theta|\lambda_t),\nu(\theta|\lambda)\right\}
$$
\item Update $\nu_j(\theta|\lambda_t)$ as 
$$
\nu_j(\theta|\lambda_{t+1}) \propto \frac{ \nu(\theta|\lambda_{t+1}) }{ \nu_{-j}(\theta|\lambda_t) }\,.
$$
\end{enumerate}
(In the above, KL denotes the Kullback-Leibler divergence.) The algorithm stops at stationarity. The convergence of this
approach is not yet fully understood, but \cite{barthelme:chopin:2014} consider expectation-propagation as a practical
substitute for ABC, avoiding the selection of summary statistics by using a local constraint
$$
||x_i-x^\text{obs}||\le \epsilon
$$
on each element of the simulated pseudo-data vector, $x^\text{obs}$ being the actual data.  In addition,
expectation-propagation provides an approximation of the evidence. In the ABC setting, when using a Normal distribution
as the exponential family default, implementing EP means computing empirical mean and empirical variance, one
observation at a time, under the above tolerance constraint. Obviously, using a Normal candidate means that the final
approximation will also look much like a Normal distribution, which both links with other Normal approximations like
INLA and variational methods, and signals a difficulty with EP in less smooth cases, such as ridge-like or multimodal
posteriors.

While different approximations keep being developed and tested, with arguments ranging from efficient
programming, to avoiding simulations, to having an ability to deal with more complex structures, their drawback
is the overall incapacity to assess the amount of approximation involved. Bootstrap evaluations can be
attempted in the simplest cases but cannot be extended to more realistic situations.

%\input{proximal.tex}
% !TEX root = GLPR14S.tex
\section{Optimisation in modern Bayesian computation}\label{sec:opt}
Optimisation methodology for high-dimensional ma\-xi\-mum-a-pos\-te\-riori (MAP) estimation is another area of Bayesian computation that has received a lot of attention over the last years, particularly for problems related to machine learning, signal processing and computer vision. One reason for this is that for many Bayesian models optimisation is significantly more computationally tractable than integration. This has generated a lot of interest in MAP estimators, especially for applications involving very high-dimensional parameter spaces or tight computing time constraints, for which calculating other summaries of the posterior distribution is not feasible. Here we review some of the major breakthroughs in this topic, which originated mainly outside the statistics community. We focus on developments related to high-dimensional convex optimisation, though many of the techniques discussed below are also useful for non-convex optimisation. In particular, in Section \ref{proximal} we concentrate on \emph{proximal optimisation algorithms}, a powerful class of iterative methods that exploit tools from convex analysis, monotone operator theory and theory of non-expansive mappings to construct carefully designed fixed-point schemes. We refer the reader to the excellent book by \citet{combettes:2011} for the mathematics underpinning proximal optimisation algorithms, and to the recent tutorial papers by \citet{pesquet:2011}, \citet{Cevher:2014} and \citet{BoydBook} for an overview of the field and applications to signal processing and machine learning.

However, we do think it is vital to insist that, at the same time as asserting that modern optimisation methodology represents a
much-underused opportunity in Bayesian inference, in its raw form it inevitably fails to deliver essential elements of the Bayesian paradigm. The vision is not to deliver a point estimate of an unknown structure, but the full
richness of Bayesian inference in its coherence, its proper treatment of uncertainty, its intrinsic treatment of model
uncertainty, and so on. Bayesian statistics does not boil down to optimisation with penalisation
\citep{lange:chi:zhou:2014}. We need to express the uncertainty associated with decisions and estimation, stemming from
the stochastic nature of the data, and our lack of knowledge about relevant mechanisms. 

The challenge is to use the awesome capacity of fast optimisation in a high-dimensional parameter space to focus on
local regions of that space where a combination of analytic and numerical investigation can deliver at least
approximations to full posterior distributions and derived quantities. The community has barely risen to this challenge,
with only isolated examples such as the discussion in \cite{green:mardiafest} of a problem in unlabelled shape analysis.
However, the growing community of INLA \citep{rue:martino:chopin:2009} users may bring an heightened awareness of such
possibilities, along with its efficient code \citep{schrodle:held:2011,muif:etal:2013}. Another promising research area is to use mathematical and algorithmic tools from convex optimisation to design more efficient high-dimensional MCMC algorithms \citep{pereyra:2014}.

\subsection{Proximal algorithms}\label{proximal}
Similarly to many other computational methodologies that are widely used nowadays, proximal algorithms were first proposed several decades ago by \citet{moreau:1962}, \citet{martinet:1970} and \citet{rockafellar:1976}, and regained attention recently in the context of large-scale inverse problems and ``big data".

We consider the computation of maximisers of posterior densities $\pi(\btheta) = \exp{\{-g(\btheta)\}}/\kappa$ that are high-dimensional and log-concave, which we formulate as
\begin{equation}\label{primal}
\hat{\btheta}_{MAP} = \argmin_{\btheta \in \mathbb{R}^n} g(\btheta)
\end{equation}
where $g$ belongs to the class $\Gamma_0(\mathbb{R}^n)$ of lower semicontinuous convex functions from $\mathbb{R}^n \rightarrow (-\infty,+\infty]$. Notice that $g$ may be non-differentiable and take value $g(\btheta) = +\infty$, reflecting constraints in the parameter space. In order to introduce proximal algorithms we first recall the following standard definitions and results from convex analysis: We say that $\bvphi \in \mathbb{R}^n$ is a subgradient of $g$ at $\btheta \in \mathbb{R}^n$ if it satisfies $(\bu-\btheta)^T\bvphi + g(\btheta) \leq g(\bu), \forall \bu \in \mathbb{R}^n$. The set of all such subgradients defines the subdifferential set $\partial g(\btheta)$, and $\hat{\btheta}_{MAP}$ is a minimiser of $g$ if and only if $\boldsymbol{0} \in \partial g(\hat{\btheta}_{MAP})$. The (convex) conjugate of $g \in \Gamma_0(\mathbb{R}^n)$ is the function $g^* \in \Gamma_0(\mathbb{R}^n)$ defined as  $g^*(\bvphi) = \sup_{\bu \in \mathbb{R}^n} \bu^T\bvphi - g(\bu)$. The subgradients of $g$ and $g^*$ satisfy the property $\bvphi \in \partial g(\btheta) \Leftrightarrow \btheta \in \partial g^*(\bvphi)$.

Proximal algorithms take their name from the proximity mapping, defined for $g \in \Gamma_0(\mathbb{R}^n)$ and $\lambda > 0$ as \citep{moreau:1962}
\begin{equation}\label{proxMap}
\prox^\lambda_{g}(\btheta) = \argmin_{\bu \in \mathbb{R}^n}\, g(\bu) + \|\bu - \btheta\|^2/2\lambda.
\end{equation}
In order to gain intuition about this mapping it is useful to analyse its behaviour when $\lambda \in \mathbb{R}^+$ is
either very small or very large. In the limit $\lambda \rightarrow \infty$, the quadratic penalty term vanishes and
\eqref{proxMap} maps all points to $\hat{\btheta}_{MAP}$. In the opposite limit $\lambda \rightarrow 0$, \eqref{proxMap}
becomes the identity operator and maps $\btheta$ to itself. For finite values of $\lambda$, $\prox^\lambda_{g}(\btheta)$
behaves similarly to a gradient mapping and moves points in the direction of $\hat{\btheta}_{MAP}$. Like gradients,
proximity mappings have several properties that are useful for devising fixed-point methods \citep{combettes:2011}.\\
\noindent \textbf{Property 1}: The proximity mapping of $g$ is related to its subdifferential by the inclusion $\{\btheta - \prox^\lambda_{g}(\btheta)\}/\lambda \in \partial g\{\prox^\lambda_{g}(\btheta)\}$, which collapses to $\nabla g\{\prox^\lambda_{g}(\btheta)\}$ when $g \in \mathcal{C}^1$. As a result, for any $\lambda > 0$, the minimiser of $g$ verifies the fixed-point equation $\btheta = \prox^\lambda_g(\btheta)$.\\
\noindent \textbf{Property 2}: Proximity mappings are firmly non-expansive; that is, $\|\prox^\lambda_{g}(\btheta)-\prox^\lambda_{g}(\bu)\|^2 \leq (\btheta-\bu)^T\{\prox^\lambda_{g}(\btheta)-\prox^\lambda_{g}(\bu)\}, \forall \btheta,\bu \in \mathbb{R}^n$.\\
\noindent \textbf{Property 3}: The proximity mappings of $g$ and its conjugate $g^*$ are related by Moreau's decomposition formula: $\btheta = \prox_{g}^\lambda(\btheta) + \lambda \prox^{1/\lambda}_{g^*}(\btheta/\lambda)$.

The simplest proximal method to solve \eqref{primal} is the \emph{proximal point algorithm} given by the iteration
\begin{equation}\label{proxPoint}
\btheta^{k+1} = \prox_{g}^{\lambda}(\btheta^{k}).
\end{equation}
Every sequence $\{\btheta^{k}\}_{k\in\mathbb{N}}$ produced by this algorithm converges to $\hat{\btheta}_{MAP}$, even if proximity mappings are evaluated inexactly, as long as the errors are of certain types (e.g., summable). A more general proximal point algorithm includes relaxation, i.e., 
$$
\btheta^{k+1} = (1-\alpha_k) \btheta^{k} + \alpha_k\prox_{g}^{\lambda}(\btheta^{k}), \quad \alpha_k \in (0,2),
$$
and with over-relaxation (i.e., $\alpha_k \in (1,2)$) often converges faster than \eqref{proxPoint}. Notice from Property 1 that \eqref{proxPoint} can be interpreted as an implicit (backward) subgradient steepest descent to minimise $g$, i.e., $\btheta^{k+1} = \btheta^{k} - \lambda\bvphi$, with $\bvphi \in \partial g\left(\btheta^{k+1}\right)$. Alternatively, proximal point algorithms can also be interpreted as explicit (forward) gradient steepest descent to minimise the \emph{Moreau envelope} of $g$, $e_\lambda (\btheta) = \inf_{\bu \in \mathbb{R}^n}\, g(\bu) + \|\bu - \btheta\|^2/2\lambda$, a convex lower bound on $g$ that by construction is continuously differentiable and has the same minimiser as $g$.

Proximal point algorithms may appear of little relevance because evaluating $\prox^\lambda_g$ can be as difficult as solving \eqref{primal} in the first place (notice that \eqref{proxMap} is a convex minimisation problem similar to \eqref{primal}). Surprisingly, many advanced proximal optimisation methods can in fact be shown to be either applications of this simple algorithm, or closely related to it.

Most proximal methods operate by splitting $g$, e.g.,
\begin{equation}\label{primal_split1}
\hat{\btheta}_{MAP}=\argmin_{\btheta \in \mathbb{R}^n} \{g_1(\btheta) + g_2(\btheta)\},
\end{equation}
such that $g_1 \in \Gamma_0(\mathbb{R}^n)$ and $g_2 \in \Gamma_0(\mathbb{R}^n)$ have gradients or proximity mappings that are easy to compute or approximate. For example, for many Bayesian models it is possible to find a decomposition $g(\btheta) = g_1(\btheta) + g_2(\btheta)$ such that $g_1$ is $\beta$-Lipschitz\footnote{$g_1 \in \mathcal{C}^1$ has $\beta$-Lipschitz continuous gradient if $\| \nabla g_1(\btheta) - \nabla g_1(\bu)\| \leq \beta \|\btheta-\bu\|, \quad \forall (\btheta,\bu) \in \mathbb{R}^N \times \mathbb{R}^N$} differentiable and $g_2 \in \Gamma_0(\mathbb{R}^n)$, possibly non-differentiable, has a proximity mapping that can be computed efficiently with a specialised algorithm. This decomposition is useful for instance in linear inverse problems, where $g_1$ is often related to a Gaussian observation model involving linear operators and $g_2$ to a log-prior promoting a parsimonious representation (e.g., sparsity on some appropriate dictionary, low-rankness) or enforcing convex constraints (e.g., positivity, positive definiteness). For models that admit this decomposition, it is possible to compute $\hat{\btheta}_{MAP}$ efficiently with a \textit{forward-backward} algorithm, also known as the proximal gradient algorithm
\begin{equation}\label{fb_primal}
\btheta^{k+1} = \prox_{g_2}^{\lambda_n}(\btheta^{k} - \lambda_n\nabla g_1(\btheta^{k})).
\end{equation}
For $\lambda_n = \lambda \in (0,1/\beta)$ the objective function $g(\btheta^k)$ converges to $g(\hat{\btheta}_{MAP})$ with rate $O(1/k)$. If the value of the Lipschitz constant $\beta$ is unknown $\lambda_n$ can be found by line-search.

A remarkable property of \eqref{fb_primal} is that it can be accelerated to converge with rate $O(1/k^2)$, which is optimal for this class of problems \citep{nesterov:2004}. This can be achieved for instance by introducing an extrapolation step
\begin{equation}\label{fb_primal_accelerated}
\begin{split}
\btheta^{+} &= \btheta^{k} + \omega_k (\btheta^{k} - \btheta^{k-1}),\\
\btheta^{k+1} &= \prox_{g_2}^{\beta^{-1}}(\btheta^{+} - \beta^{-1}\nabla g_1(\btheta^{+})),
\end{split}
\end{equation}
where $\{\omega_k\}_{k \in \mathbb{N}}$ is an appropriate sequence of extrapolation parameters. It was noticed by \citet{pesquet:2011} that several important convex optimisation algorithms can be derived as applications of the forward-backward algorithm, for example the projected gradient algorithm for minimising a Lipschitz differentiable function subject to a convex constraint (in this case the proximity mapping reduces to a projection onto the convex set). Notice that \eqref{fb_primal} can be interpreted as an implementation of the proximal point iteration \eqref{proxPoint} where $\prox_{g}^{\lambda}(\btheta^{k})$ is approximated by replacing $g_1$ with its first order Taylor series approximation around the point $\btheta^{k}$.

Moreover, in some cases it may be more efficient to compute $\hat{\btheta}_{MAP}$ by solving the dual of \eqref{primal_split1}, for instance if $g$ admits a decomposition $g(\btheta) = g_1(\btheta) + g_2(L\btheta)$ for some linear operator $L \in \mathbb{R}^{n\times p}$, $g_1 \in \Gamma_0(\mathbb{R}^n)$ strongly convex and $g_2 \in \Gamma_0(\mathbb{R}^p)$ with efficient proximity mapping. In this case, the Fenchel--Rockafellar theorem states that $\hat{\btheta}_{MAP}$ can be computed by solving the dual problem \citep[ch. 19]{combettes:2011}
\begin{equation}\label{dual}
\bpsi^* = \argmin_{\bpsi \in \mathbb{R}^p} g_1^*(-L^T\bpsi) + g_2^*(\bpsi)
\end{equation}
and setting $\hat{\btheta}_{MAP} = \nabla g_1^*(-L^T\bpsi^*)$. This $p$-dimensional problem can be solved iteratively with a forward-backward algorithm
$
\bpsi^{k+1} = \prox^{\lambda_n}_{g^*_2}(\bpsi^{k} - \lambda_n \nabla g^*_1(-L^T\bpsi^{k}))%, \quad \btheta^{k+1} = \nabla g_1^*(-L^T \bpsi^{k+1})
$
that can also be accelerated to converge with rate $O(1/k^2)$, and where we note that the proximity mapping of $g_2^*$ is typically evaluated by using Property 3, and that the strong convexity of $g_1$ implies Lipschitz differentiability of $g_1^*$. Computing $\hat{\btheta}_{MAP}$ via \eqref{dual} can lead to important computational savings, in particular if $p \ll n$ or if $g_2$ is separable and has a proximity mapping that can be computed in parallel for each element of $\btheta$ (this is generally not possible for $g_2 \circ L$). We refer the reader to \cite{komodakis:2014} for an overview of recent dual and primal-dual algorithms and guidelines for parallel implementations.

Another important proximal optimisation method is the Douglas--Rachford splitting algorithm given by
\begin{equation}\label{DouglasRachford}
\begin{split}
\btheta^{k+\frac{1}{2}} &= \prox_{g_1}^\lambda(\btheta^{k}),\\
\btheta^{k+1} &= \btheta^{k} - \btheta^{k+\frac{1}{2}} + \prox_{g_2}^\lambda (2\btheta^{k+\frac{1}{2}} - \btheta^{k}).
\end{split}
\end{equation}
From a theoretical viewpoint this algorithm is more general than the forward-backward algorithm because it does not require $g_1$ or $g_2$ to be continuously differentiable. However, its practical application is limited to problems for which both $g_1$ and $g_2$ have efficient proximity mappings. Similarly to the forward-backward algorithm, \eqref{DouglasRachford} includes many proximal algorithms that been proposed in the literature for specific models, and can also be interpreted as an application of the proximal point algorithm.

The proximal method that is arguably most widely used in Bayesian inference is the \emph{alternating direction method of multipliers} (ADMM), which operates by formulating \eqref{primal_split1} as a constrained optimisation problem
\begin{equation}\label{primal_constrained}
\begin{split}
\argmin_{\btheta \in \mathbb{R}^n, \,\bz \in \mathbb{R}^n} g_1(\btheta) + g_2(\bz)\\
\textrm{subject to}\quad \boldsymbol \btheta = \bz,
\end{split}
\end{equation}
and then using augmented Langrangian techniques to express \eqref{primal_constrained} as an unconstrained saddle point problem with saddle function
$ g_1(\btheta) + g_2(\bz) + \lambda\bvphi^T(\btheta-\bz) + ||\btheta-\bz||^2/{2\lambda}$ \citep{boyd:2011}. ADMM solves this problem with the iteration
\begin{equation}\label{ADMM}
\begin{split}
\btheta^{k+1} &= \prox_{g_1}^\lambda(\bz^{k} - \bvphi^{k}),\\
\bz^{k+1} &= \prox_{g_2}^\lambda (\btheta^{k+1} + \bvphi^{k}),\\
\bvphi^{k+1} &= \bvphi^{k} + \btheta^{k+1} - \bz^{k+1},
\end{split}
\end{equation}
that also involves the proximity mappings of $g_1$ and $g_2$. This basic ADMM iteration can be tailored to specific models in many ways (e.g., to exploit decompositions of the form $g_1 = \tilde{g}_1 \circ L_1$ and $g_2 = \tilde{g}_2 \circ L_2$ so that proximal updates can be performed in parallel for all components of $\btheta$, $\bz$ and $\bvphi$). Interestingly, ADMM can be interpreted as an application of the Douglas--Rachford algorithm to the dual of \eqref{primal_constrained}, and is therefore also a special case of the proximal point algorithm. For more details about the ADMM algorithm, see the recent tutorial by \citet{boyd:2011}.

Furthermore, an important characteristic of proximal optimisation algorithms is that they can be massively parallelised to take advantage of parallel computer architectures. Suppose for instance that $g$ admits the decomposition $g(\btheta) = \sum_{m = 1}^M g_m(L_m \btheta)$ with $g_m \in \Gamma(\mathbb{R}^{p_m})$ and $L_m \in \mathbb{R}^{n \times {p_m}}$ such that the mappings of $g_m$ are easy to compute and $Q = \sum_{m = 1}^M L_m^T L_m$ is invertible. Then, in a manner akin to \eqref{primal_constrained}, we express \eqref{primal} as
\begin{equation}\label{primal_constrained_parallel}
\begin{split}
\argmin_{\bz_1 \in \mathbb{R}^n, \ldots,\,\bz_M \in \mathbb{R}^n} \sum\nolimits_{m = 1}^M g_m(\bz_m)\\
\textrm{subject to}\quad \bz_m = L_m\btheta, \, \forall m = 1,\ldots,M,
\end{split}
\end{equation}
and compute $\hat{\btheta}_{MAP}$ with the following iteration
\begin{equation}
\begin{split}
\btheta^{k+1} &= Q^{-1} \sum\nolimits_{m = 1}^M L_m^T (\bz_m^{k} - \bvphi_m^{k}),\\
\bz_m^{k+1} &= \prox^\lambda_{g_m}(L_m\btheta^{k+1} - \bvphi_m^{k}), \, \forall m = 1,\ldots,M,\\
\bvphi_m^{k+1} &= \bvphi_m^{k} + L_m\btheta^{k+1} - \bz_m^{k+1}, \,\,\, \forall m = 1,\ldots,M,
\end{split}
\end{equation}
that can be parallelised with factor $M$ at a coarse level (e.g., on a multi-processor system). Further parallelisation may be possible at a finer scale (e.g., on a vectorial processor such as GPU or FPGA) by taking advantage of the structure of $\prox^\lambda_{g_m}$ or by using specialised algorithms. This algorithm, known as the \emph{simultaneous direction method of multipliers}, is also closely related to the ADMM, Douglas--Rachford and proximal point algorithms. Notice that splitting $g$ not only allows the exploitation of parallel computer architectures, but may also significantly simplify the computation of proximity mappings; often $\prox^\lambda_{g_m}$ has a closed-form expression. Lastly, it is worth mentioning that there are  other modern proximal optimisation algorithms that can be massively parallelised, for example the \emph{generalised forward backward} algorithm \citep{raguet:2013}, the \emph{parallel proximal} algorithms \citep{combettes:2008, pesquet:2012a}, and the parallel primal-dual algorithm \citep{pesquet:2012b}.

Finally, main current topics of research in proximal optimisation include theory and methodology for: 1) randomised and stochastic algorithms that operate with estimators of gradients and proximity mappings to reduce computational complexity and allow for errors in the update rules, 2) adaptive and variable metric algorithms (e.g. Riemannian and Newton-type) that exploit the model's geometry to improve convergence speed, and 3) proximal methods for non-convex problems. We anticipate that in the future new and stronger connections will emerge between proximal optimisation and stochastic simulation, in particular through developments in stochastic optimisation and high-dimensional MCMC sampling. For example, one connection is through the integration of modern stochastic convex optimisation and Markovian stochastic approximation \citep{combettes:2014, andrieu2015stability}, and of proximal optimisation and high-dimensional MCMC sampling \citep{pereyra:2014}.

%Finally, it is worth mentioning that proximal optimisation methods have been greatly motivated by theoretical results related to compressive sensing and low-rank matrix recovery stating that it is possible to recover 

%high-dimensional statistics results stating that it is possible to approximate 
%We anticipate that  To conclude, we highlight some connections that are emerging between  Presently most stochastic proximal optimisation algorithms only consider the case where estimators are computed by using a random subset of the data, rather than the entire set of observations.

%For example, \citet{atchade:2014} describe a stochastic forward-backward algorithm where gradients are estimated using an auxiliary MCMC algorithm. We wonder to use MCMC-based estimators for the class of stochastic algorithms studied in \citet{combettes:2014}.
\subsection{Convex relaxations}\label{sub:opt.1}
Modern proximal optimisation was greatly motivated by important theoretical results on the recovery of partially-observed sparse vectors and low-rank matrices through convex minimisation \citep{candes:2006,candes:2009} and on \emph{compressive sensing} \citep{candes:2008}. A key idea underlying these works is that of approximating a combinatorial optimisation problem, whose solution is NP-hard, with a ``relaxed'' convex problem that is computationally tractable, and whose solution is in some sense close to the solution of the original problem. Reciprocally, the development of modern convex optimisation has in turn generated much interest in log-concave models, convex regularisers, and ``convexifications'' (i.e., convex relaxations for intractable or poorly tractable models) for statistical inference problems involving high-dimensionality, large datasets and computing time constraints \citep{chandrasekaran:2012,chandrasekaran:2013}.

\subsection{Illustrative example}\label{sec:sub.opt3}
For illustration, we show an application of proximal optimisation to Bayesian image resolution enhancement. The goal is
to recover a high-resolution image $\btheta \in \mathbb{R}^n$ from a blurred and noisy observed image $\by \sim
\mathcal{N}(H\btheta,\allowbreak\sigma^2\boldsymbol{I}_n)$, where $H \in \mathbb{R}^{n\times n}$ is a linear operator representing the blur point spread function of the low resolution acquisition system and $\sigma^2$ is the system's noise power. This inverse problem is ill-posed, a difficulty that Bayesian image processing methods address by exploiting prior knowledge about $\btheta$. Here we use the following hierarchical Bayesian model \citep{oliveira:2009}
\begin{equation}
\begin{split}
f(\by|\btheta) &= (2\pi\sigma^2)^{-n/2}\exp\{-\|\by - H\btheta\|^2_2/2\sigma^2\},\\
\pi(\btheta|\alpha) &  \propto \alpha^{-n}\exp{\left(-\alpha\|\nabla_d \btheta\|_{1-2}\right)},\\
\pi(\alpha) & = e^{-\alpha}\boldsymbol{1}_{\mathbb{R}^+}(\alpha),%\alpha^{\gamma-1}\exp\{-\beta \alpha\}\beta^\gamma/\Gamma(\gamma),\\
\end{split}
\end{equation}
where $\pi(\btheta|\alpha)$ is the (improper) \emph{total-variation} Markov random field, $\| \cdot \|_{1-2}$ denotes the composite $\ell_1 - \ell_2$ norm and $\nabla_d$ is the discrete gradient operator that computes the vertical and horizontal differences between neighbour image pixels. This prior is log-concave and models the fact that differences between neighbouring image pixels are usually very small but occasionally take large values; it is arguably the most widely used prior in modern statistical image processing. The values of $H$ and $\sigma^2$ are typically determined during the system's calibration process and are here assumed known.

We compute the MAP estimator of $\btheta$ associated with the marginal posterior $\pi(\btheta|\by) = \int_0^\infty \pi(\btheta,\alpha | \by) \textrm{d}\alpha$, which is unimodal but not log-concave,
\begin{equation}\label{TVdeconvolution}
\begin{split}
\hat{\btheta}_{MAP} = \argmin_{\btheta \in \mathbb{R}^n} \quad &\|\by - H\btheta\|^2_2/2\sigma^2 \\
&+ (n+1)\log \left(\|\nabla_d \btheta\|_{1-2} + 1\right).
\end{split}
\end{equation}
Problem \eqref{TVdeconvolution} is not convex, but can nevertheless be solved efficiently with proximal algorithms by using a \emph{majorisation--minimisation} strategy. To be precise, starting from some initial condition $\btheta^{(0)}$, e.g.,  $\btheta^{(0)} = \by$, we iteratively minimise the following sequence of strictly convex majorants \citep{oliveira:2009}
\begin{equation}\label{TVdeconvolution2}
\begin{split}
&\btheta^{(t+1)} =\argmin_{\btheta \in \mathbb{R}^n} \|\by - H\btheta\|^2_2/2\sigma^2 + \alpha_{\textrm{eff}}^{(t)}\|\nabla_d \btheta\|_{1-2},\\
&\textrm{with}\quad  \alpha_{\textrm{eff}}^{(t)} = (n+1)(\|\nabla_d \btheta^{(t)}\|_{1-2} + 1).
\end{split}
\end{equation}
Iteration \eqref{TVdeconvolution2} involves a convex subproblem that can easily be solved using most modern proximal optimisation techniques. For example, here we use the state-of-the-art ADMM algorithm \emph{SALSA} \citep{figueiredo:2011} implemented with $g_1 (\btheta) =  \|\by - H\btheta\|^2_2/2\sigma^2$, $g_2(\bu) = \alpha_{eff}^{(t)}\|\nabla_d \bu\|_{1-2}$, and the constraint $\btheta = \bu$ [though we could have also used other modern algorithms \citep{pesquet:2012a, pesquet:2012b, raguet:2013}]. To compute the proximity mapping of $g_1$ we use the fact that $H$ is block-circulant to compute matrix products and pseudo-inverses with the FFT algorithm. We compute the proximity mapping of $g_2$ with a highly parallelised implementation of the specialised algorithm of \citet{chambolle:2004}. 

\begin{figure}[H]
\includegraphics[width=.5\textwidth]{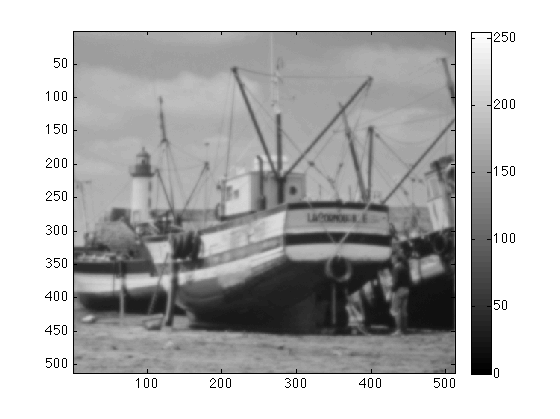}
\caption{\label{fig:boat_observed} Observed blurred noisy image $\by$.}
\includegraphics[width=.5\textwidth]{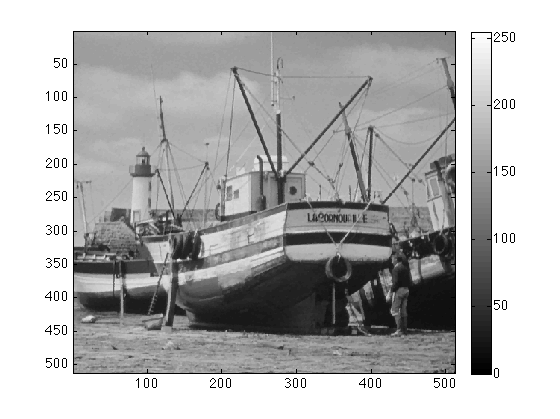}
\caption{\label{fig:boat_restored} Resolution enhanced image $\hat{\btheta}_{MAP}$ obtained by solving \eqref{TVdeconvolution} with the majorisation-minimisation strategy \eqref{TVdeconvolution2}.}
\includegraphics[width=.5\textwidth]{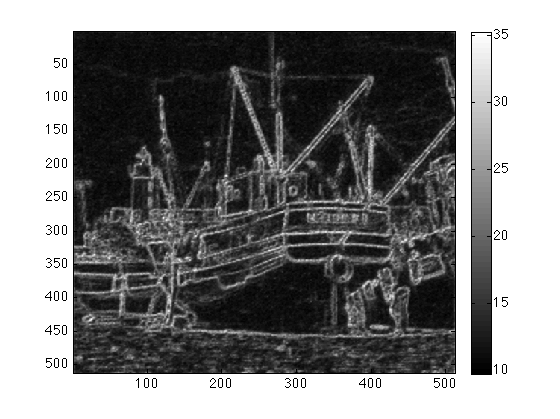}
\caption{\label{fig:boat_confidence} Widths of pixel-wise $90\%$ marginal credibility intervals estimated with the proximal MCMC algorithm of \citet{pereyra:2014}.}
\end{figure}
\begin{figure}[H]
\includegraphics[width=.5\textwidth]{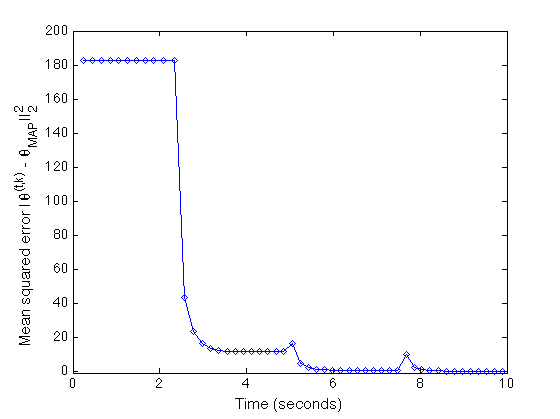}
\caption{\label{fig:boat_mseplot} Convergence of the estimate $\btheta$ to $\hat{\btheta}_{MAP}$ vs computing time (seconds).}
%\includegraphics[width=.45\textwidth]{logPi.png}
%\caption{$g(\btheta^{(k)},\alpha^{(k)} | \by, H, \sigma^2)$ vs Time (seconds)}
\end{figure}

Figure \ref{fig:boat_observed} presents a blurred and noisy observation $\by$ of the popular ``boats'' image of size $512 \times 512$ pixels, generated with a uniform $9\times9$ blur and a noise power of $\sigma^2 = 0.5^2$ (blurred-signal-to-noise ratio $BRSN = 10\log_{10}\{\|H{\btheta}_0\|^2_2/\sigma^2\} = 40$dB). Figure \ref{fig:boat_restored} below shows the MAP estimate $\hat{\btheta}_{MAP}$ obtained by solving \eqref{TVdeconvolution} using $4$ iterations of \eqref{TVdeconvolution2} and a total of $51$ ADMM iterations. We observe that this resolution enhancement process has produced a remarkably sharp image with very noticeable fine detail. Moreover, Figure \ref{fig:boat_confidence} shows the magnitude of the marginal $90\%$ credibility regions for each pixel, as measured by the distance between the $5\%$ and $95\%$ quantile estimates. These estimates were computed using the proximal Metropolis-adjusted Langevin algorithm \citep{pereyra:2014}, which is appropriate for high-dimensional densities that are not continuously differentiable. We observe in Figure \ref{fig:boat_confidence} that the uncertainty is mainly concentrated at the contours and object boundaries, revealing that model is able to accurately detect the presence of sharp edges in the image but with some uncertainty about their exact location. Finally, Figure \ref{fig:boat_mseplot} shows the convergence of the estimates $\btheta^{(t,k)}$ produced by each ADMM iteration to $\hat{\btheta}_{MAP}$ (as measured by the mean squared error $\| \btheta^{(t,k)} - \hat{\btheta}_{MAP} \|_2^2$) . Notice that computing $\hat{\btheta}_{MAP}$ only required $10$ seconds (experiment conducted on an Apple Macbook Pro computer running {\sc Matlab} 2013, a C++ implementation would certainly produce even faster results). This is remarkably fast given the high dimensionality of the problem ($n = 262\,144$). The computation of the credibility regions by MCMC sampling ($20\,000$ samples with a thinning factor of $1\,000$ to reduce the algorithm's memory foot-print) required $75$ hours.

%(Here I should mention that proximal optimisation algorithms have allowed to tackle very large inference problems in signal and image processing, machine learning, and other applications domains. In particular, they are widely used for problems related to sparsity and variable/model selection that involve priors with combinatorial terms. Instead of performing inference with this "true" model, the log-prior is first "relaxed" or approximated by a concave tight bound yielding a posterior distribution that is log-concave. Maximum a posteriori estimates can then be computed using state of the art convex optimisation algorithms. In a certain way, this approach also seeks to address inference problems that are intractable, in the sense that computing the MAP of the original model would be NP-hard).

%\input{BigData}
\section{Discussion}\label{sec:discn}
\subsection{Bayesian computation in the era of data science}\label{sub:discn.1}

Is there a revolution taking place right now and have we missed the train,
standing on the platform, only concerned with small-print on the train
schedules -- apart, that is, from the obvious but not-so-new requirement to handle massive
datasets (and the mistakes that come with them)?!

%Machine-learners insist on the `inevitability' of turning statistics into an
%`optimize+penalize' exercise. This is hopefully tunnel vision and we can continue to
%present a broader perspective on inference, while accepting the limitations
%of the representation.

As with other areas of statistical science, the Bayesian computation community has to decide whether data science is an opportunity or a threat. Inevitably if we do not treat it as an opportunity, it will become a threat. Thanks to the ubiquity of ``big data'' (as an over-hyped phrase mostly useful for attracting research funding, but also to at least some extent in reality), a new potentially multi-disciplinary field of data science is rapidly opening up. This field is attracting huge material resources, and will absorb much human talent. Statistical science has to be a part of this, for its own survival, but also for the sake of society. As Tim Harford has cogently argued \citep{harford}:

\begin{quotation}
Recall big data's four articles of faith. Uncanny accuracy is easy to overrate if we simply ignore false positives
[\ldots]. The claim that causation has been ``knocked off its pedestal'' is fine if we are making predictions in a
stable environment but not if the world is changing [\ldots] or if we ourselves hope to change it. The promise that ``N
= All'', and therefore that sampling bias does not matter, is simply not true in most cases that count. As for the idea
that ``with enough data, the numbers speak for themselves'' -- that seems hopelessly na{\"\i}ve in data sets where spurious patterns vastly outnumber genuine discoveries.

``Big data'' has arrived, but big insights have not. The challenge now is to solve new problems and gain new answers -- without making the same old statistical mistakes on a grander scale than ever.
\end{quotation}

It is a mistake to think that Bayes has no part to play in these developments, but more of us need to get more involved,
and learn new tools, as in the way the Consensus Monte Carlo algorithm \citep{scott:etal:2013} exploits the Hadoop
environment \citep{white2012} and the MapReduce programming model \citep{dean:ghemawat}. Another direction that can
prevent a potential schism between Bayesian modelling and highly complex models is to aim for modularity and local
learning, that it, to abandon the goal of modelling big universes for analysing a series of small worlds, in spite of the loss of coherence, amd hence compromise to the Bayesian  paradigm, that this entails. The curious
case of the cut models presented in \cite{plummer:2014} is an illustration of the potential for developing partial-information Bayesian inference tools where ``small is beautiful" because this is the only viable solution.

\subsection{Do we care enough about applications?}\label{sub:discn.3}

Bayesian computation began in order to answer rather practical problems -- how can we perform a Bayesian analysis of
these data using this model? -- or the corresponding meta-problems -- how can Bayesian analysis be performed generally
and reliably for this class of models? The focus was applied methodology (although since the methods were new, they
tended to be published in premier theory/methodology journals). Because the research community wanted to understand (the
advantages, performance and limitations of) the methods they were advocating, more theoretical work started to be
conducted, and, for example, many probabilists were attracted to study the Markov chains that MCMC methodologists
created. The centre of mass of research activity drifted away from the original motivations, just as has happened in
other areas of mathe\-ma\-ti\-cal\-ly-rigorous computation. 

At the same time, those working with data became more ambitious with regard to the scale of data, the complexity of modelling and the sophistication of analysis, all factors that have in principle (and often in fact) stimulated new developments in Bayesian computation. But to a large extent this is a rich, self-stimulating and self-supporting area of research; new applications may or may not need new computational techniques, but new techniques don't seem to need applications to justify themselves. It is apposite to ask to what extent is cutting-edge computational methodology research really delivering answers to questions that application domains are posing. And to what extent is cutting-edge computational methodology research successfully answering real questions?

We may not be unanimous about answers to these questions, except we can probably all agree they are ``not entirely''.
We will also disagree about how much this matters, but again there may be something to agree about, that we have failed
if methodological innovations disconnect completely from applications. Legitimate differences in research goals
partially explain the trend in this direction, but it is fair to say that there is a big communication problem between
the computational statistics community and many of the communities where Bayesian computational methods are applied.
Unfortunately people in these communities do not always keep up with the state of the art in computational statistics.
At the same time, statisticians are often not aware of important developments arising in other fields. (ABC is a good
illustration: it took more than five years of development within the population genetics community before statisticians
became aware the technique existed and a few more years before they realised this was proper Bayesian inference applied
on approximate models.) We can perhaps blame the fact that there are not enough people working at the interface of the different communities, but life at the interface is not easy because multidisciplinary and interdisciplinary research is often seen as ``marginal" by both communities and is thus difficult to publish, communicate, etc. Then there are of course problems in dissemination, related to the different writing styles, journals, computing languages, software, etc. of each community. 

We strongly encourage those developing new techniques always to find a way to disseminate them in such a way that at least \emph{somebody} else could use them, preferably someone without the ability to have invented the technique for themselves! -- and advocate, of course, that successful dissemination be properly rewarded in our career structures.

In a somewhat parallel path, we have seen over the past decades the emergence of new languages and meta-languages intended to handle complexity
both of problems and of solutions towards a wider audience of users. BUGS \citep{lunn:bugs:2012} is the archetypal
example of such languages and it has been successful to the extent that a large proportion of the users has a fairly
limited statistical background and often even less of a computational background. However, the population of BUGS users
and sympathisers is tiny compared to that of SAS or other corporate statistical systems. In this respect, we have failed to
disseminate concepts like Bayesian analysis and wonderful tools like MCMC algorithms, because most people are unable to
turn them into codes by themselves. (Perusing one of the numerous statistics and machine-learning on-line forums
like Cross Validated quickly exposes the methodological gap between academics and the masses!) It is unclear how novel
programming developments like STAN \citep{stan-software:2014} are going to modify this picture, in that they still
assume a decent understanding of both modelling and simulation issues. In that respect, network-based approaches as
those covered by BUGS sound more promising towards ``modelling locally to learn globally". Similarly, ABC software is
either too specific, like DIYABC \citep{cornuet:etal:2008} which addresses only population genetic questions, or too
dependent on the ability of the modeller to program simulated outcomes from the model under study.

\subsection{Anticipating the future}\label{sub:discn.4}
In which of the areas we discuss do we expect a particular emphasis of effort, or significant progress, or do we see particular needs for new efforts or new directions?

One expectation is that in the future computational methodologies will be more flexible and malleable. Over the past 25 years Bayesian modelling and inference techniques have been applied successfully to thousands of problems across a wide range application domains. Each application brings its own constraints in terms of model dimensionality and complexity, data, inferences, accuracy and computing times. These constraints also vary significantly within specific applications. For example, in hyperspectral remote sensing, when a new Bayesian model is introduced it is often first explored and validated by MCMC sampling, then approximated with a variational Bayes method, and then approximated again so that it can be applied to gigabyte-large datasets by using optimisation techniques. Similarly, an interesting result revealed by a fast inference technique can be analysed more deeply with more reliable and accurate methods. Therefore we expect that in the future the different main computational methodologies will become more adaptable and that the boundaries between them will be less well defined, with many algorithms developed that combine simulation, variational approximations and optimisation. These will be able to handle a wide spectrum of models, degrees of accuracy and computing times, as well as models that have some parts that are simple but high-dimensional and others that are more complex but that only involve low-dimensional components. This can be achieved by using approximations and optimisation to improve stochastic sampling, by using simulation within deterministic algorithms to handle specific parts of the model that are difficult to compute analytically, or in completely new and original ways.

We also anticipate that computational methodologies will continue to be challenged by larger and larger datasets. There is of course a threat that the whole field turns into a library of machine-learning techniques, with limited validation on reference learning sets and a quick turnover of methods, which would both impoverish the field and fail to reach a general audience of practitioners. We must retain a sense of the stochastic elements in data collection, data analysis, and
inference, recognising uncertainty in data and models, to preserve the inductive strength of data science -- seeing beyond the data we have to what it might have been, what it might be next time, and where it came from.

\input GLPR14S.bbl

\end{document}

%% file: GLPR14S.bbl
\hyphenation{Post-Script Sprin-ger}

%% file: GLPR14Xiv.bbl
\begin{thebibliography}{251}
\expandafter\ifx\csname natexlab\endcsname\relax\def\natexlab#1{#1}\fi
\expandafter\ifx\csname url\endcsname\relax
  \def\url#1{\texttt{#1}}\fi
\expandafter\ifx\csname urlprefix\endcsname\relax\def\urlprefix{URL }\fi
\providecommand{\eprint}[2][]{\url{#2}}

\bibitem[{Afonso et~al.(2011)Afonso, Bioucas-Dias and
  Figueiredo}]{figueiredo:2011}
\textsc{Afonso, M.}, \textsc{Bioucas-Dias, J.} and \textsc{Figueiredo, M.}
  (2011).
\newblock An augmented {L}agrangian approach to the constrained optimization
  formulation of imaging inverse problems.
\newblock \textit{IEEE. Trans. on Image Process.}, \textbf{20} 681--695.

\bibitem[{Albert(1988)}]{albert:1988}
\textsc{Albert, J.} (1988).
\newblock Computational methods using a {B}ayesian hierarchical generalized
  linear model.
\newblock \textit{J. American Statist. Assoc.}, \textbf{83} 1037--1044.

\bibitem[{Aldous et~al.(2008)Aldous, Krikun and Popovic}]{aldkripop2008a}
\textsc{Aldous, D.}, \textsc{Krikun, M.} and \textsc{Popovic, L.} (2008).
\newblock Stochastic models for phylogenetic trees on higher-order taxa.
\newblock \textit{J. Math. Biology}, \textbf{56} 525--557.

\bibitem[{Alquier et~al.(2014)Alquier, Friel, Everitt and
  Boland}]{alquier2014noisy}
\textsc{Alquier, P.}, \textsc{Friel, N.}, \textsc{Everitt, R.} and
  \textsc{Boland, A.} (2014).
\newblock {Noisy Monte Carlo: Convergence of Markov chains with approximate
  transition kernels}.
\newblock \textit{Statistics and Computing} 1--19.

\bibitem[{Andrieu et~al.(2011)Andrieu, Doucet and
  Holenstein}]{andrieu:doucet:holenstein:2010}
\textsc{Andrieu, C.}, \textsc{Doucet, A.} and \textsc{Holenstein, R.} (2011).
\newblock Particle {M}arkov chain {M}onte {C}arlo (with discussion).
\newblock \textit{J. Royal Statist. Society Series B}, \textbf{72 (2)}
  269--342.

\bibitem[{Andrieu and Moulines(2006)}]{andrieu2006ergodicity}
\textsc{Andrieu, C.} and \textsc{Moulines, {\'E}.} (2006).
\newblock {On the ergodicity properties of some adaptive MCMC algorithms}.
\newblock \textit{The Annals of Applied Probability}, \textbf{16} 1462--1505.

\bibitem[{Andrieu and Robert(2001)}]{andrieu2001controlled}
\textsc{Andrieu, C.} and \textsc{Robert, C.} (2001).
\newblock {Controlled MCMC for optimal sampling}.
\newblock Tech. rep., Cahiers du Ceremade.

\bibitem[{Andrieu and Roberts(2009)}]{andrieu:roberts:2009}
\textsc{Andrieu, C.} and \textsc{Roberts, G.} (2009).
\newblock The pseudo-marginal approach for efficient {M}onte {C}arlo
  computations.
\newblock \textit{Ann. Statist.}, \textbf{37} 697--725.

\bibitem[{Andrieu et~al.(2015)Andrieu, Tadi{\'c} and
  Vihola}]{andrieu2015stability}
\textsc{Andrieu, C.}, \textsc{Tadi{\'c}, V.~B.} and \textsc{Vihola, M.} (2015).
\newblock On the stability of some controlled {M}arkov chains and its
  applications to stochastic approximation with {M}arkovian dynamic.
\newblock \textit{The Annals of Applied Probability}, \textbf{25} 1--45.

\bibitem[{Andrieu and Thoms(2008)}]{andrieu2008tutorial}
\textsc{Andrieu, C.} and \textsc{Thoms, J.} (2008).
\newblock {A tutorial on adaptive MCMC}.
\newblock \textit{Statistics and Computing}, \textbf{18} 343--373.

\bibitem[{Andrieu and Vihola(2015)}]{andrieu2012convergence}
\textsc{Andrieu, C.} and \textsc{Vihola, M.} (2015).
\newblock Convergence properties of pseudo-marginal markov chain monte carlo
  algorithms.
\newblock \textit{Annals of Applied Probability}, \textbf{25} 1030--1077.

\bibitem[{Angelino et~al.(2014)Angelino, Kohler, Waterland, Seltzer and
  Adams}]{angelino:etal:2014}
\textsc{Angelino, E.}, \textsc{Kohler, E.}, \textsc{Waterland, A.},
  \textsc{Seltzer, M.} and \textsc{Adams, R.} (2014).
\newblock Accelerating {MCMC} via parallel predictive prefetching.
\newblock \textit{arXiv preprint arXiv:1403.7265}.

\bibitem[{Atchad{\'e}(2006)}]{Atchade2006}
\textsc{Atchad{\'e}, Y.} (2006).
\newblock An adaptive version for the {Metropolis adjusted Langevin} algorithm
  with a truncated drift.
\newblock \textit{Methodology and Computing in Applied Probability}, \textbf{8}
  235--254.

\bibitem[{Atchad{\'e} et~al.(2011{\natexlab{a}})Atchad{\'e}, Fort, Moulines and
  Priouret}]{atchade2009adaptive}
\textsc{Atchad{\'e}, Y.}, \textsc{Fort, G.}, \textsc{Moulines, E.} and
  \textsc{Priouret, P.} (2011{\natexlab{a}}).
\newblock {Adaptive Markov chain Monte Carlo}: theory and methods.
\newblock In \textit{Bayesian Time Series Models}. Cambridge University Press.

\bibitem[{Atchad{\'e} and Rosenthal(2005)}]{atchade2005adaptive}
\textsc{Atchad{\'e}, Y.} and \textsc{Rosenthal, J.} (2005).
\newblock {On adaptive Markov chain Monte Carlo algorithms}.
\newblock \textit{Bernoulli}, \textbf{11} 815--828.

\bibitem[{Atchad{\'e} and Liu(2010)}]{atchade:liu:2010}
\textsc{Atchad{\'e}, Y.~F.} and \textsc{Liu, J.~S.} (2010).
\newblock The {Wang-Landau} algorithm in general state spaces: applications and
  convergence analysis.
\newblock \textit{Statistica Sinica}, \textbf{20} 209--233.

\bibitem[{Atchad{\'e} et~al.(2011{\natexlab{b}})Atchad{\'e}, Roberts and
  Rosenthal}]{MR2826692}
\textsc{Atchad{\'e}, Y.~F.}, \textsc{Roberts, G.~O.} and \textsc{Rosenthal,
  J.~S.} (2011{\natexlab{b}}).
\newblock Towards optimal scaling of {M}etropolis-coupled {M}arkov chain
  {M}onte {C}arlo.
\newblock \textit{Stat. Comput.}, \textbf{21} 555--568.

\bibitem[{Bai et~al.(2011)Bai, Roberts and Rosenthal}]{bai2011containment}
\textsc{Bai, Y.}, \textsc{Roberts, G.} and \textsc{Rosenthal, J.} (2011).
\newblock On the containment condition for adaptive {Markov chain Monte Carlo}
  algorithms.
\newblock \textit{Advances and Applications in Statistics}, \textbf{21} 1--54.

\bibitem[{{Barber} et~al.(2015){Barber}, {Voss} and
  {Webster}}]{barber:voss:webster:2013}
\textsc{{Barber}, S.}, \textsc{{Voss}, J.} and \textsc{{Webster}, M.} (2015).
\newblock The rate of convergence for {A}pproximate {B}ayesian computation.
\newblock \textit{Electronic Journal of Statistics}, \textbf{9} 80--105.

\bibitem[{Bardenet et~al.(2014)Bardenet, Doucet and
  Holmes}]{icml2014c1_bardenet14}
\textsc{Bardenet, R.}, \textsc{Doucet, A.} and \textsc{Holmes, C.} (2014).
\newblock Towards scaling up {Markov chain Monte Carlo}: an adaptive
  subsampling approach.
\newblock In \textit{Proceedings of the 31st International Conference on
  Machine Learning (ICML-14)} (T.~Jebara and E.~P. Xing, eds.). JMLR Workshop
  and Conference Proceedings, 405--413.

\bibitem[{Barthelm{\'e} and Chopin(2014)}]{barthelme:chopin:2014}
\textsc{Barthelm{\'e}, S.} and \textsc{Chopin, N.} (2014).
\newblock Expectation propagation for likelihood-free inference.
\newblock \textit{Journal of the American Statistical Association},
  \textbf{109} 315--333.

\bibitem[{Bauschke and Combettes(2011)}]{combettes:2011}
\textsc{Bauschke, H.~H.} and \textsc{Combettes, P.~L.} (2011).
\newblock \textit{Convex Analysis and Monotone Operator Theory in {H}ilbert
  Spaces}.
\newblock Springer New York.

\bibitem[{Beaumont(2003)}]{beaumont:2003}
\textsc{Beaumont, M.} (2003).
\newblock Estimation of population growth or decline in genetically monitored
  populations.
\newblock \textit{Genetics}, \textbf{164} 1139--1160.

\bibitem[{Beaumont(2010)}]{beaumont:2010}
\textsc{Beaumont, M.} (2010).
\newblock Approximate {B}ayesian computation in evolution and ecology.
\newblock \textit{Annual Review of Ecology, Evolution, and Systematics},
  \textbf{41} 379--406.

\bibitem[{Beaumont et~al.(2010)Beaumont, Nielsen, Robert, Hey, Gaggiotti,
  Knowles, Estoup, Mahesh, Coranders, Hickerson, Sisson, Fagundes, Chikhi,
  Beerli, Vitalis, Cornuet, Huelsenbeck, Foll, Yang, Rousset, Balding and
  Excoffier}]{clade:2010}
\textsc{Beaumont, M.}, \textsc{Nielsen, R.}, \textsc{Robert, C.}, \textsc{Hey,
  J.}, \textsc{Gaggiotti, O.}, \textsc{Knowles, L.}, \textsc{Estoup, A.},
  \textsc{Mahesh, P.}, \textsc{Coranders, J.}, \textsc{Hickerson, M.},
  \textsc{Sisson, S.}, \textsc{Fagundes, N.}, \textsc{Chikhi, L.},
  \textsc{Beerli, P.}, \textsc{Vitalis, R.}, \textsc{Cornuet, J.-M.},
  \textsc{Huelsenbeck, J.}, \textsc{Foll, M.}, \textsc{Yang, Z.},
  \textsc{Rousset, F.}, \textsc{Balding, D.} and \textsc{Excoffier, L.} (2010).
\newblock In defense of model-based inference in phylogeography.
\newblock \textit{Molecular Ecology}, \textbf{19(3)} 436--446.

\bibitem[{Beaumont et~al.(2002)Beaumont, Zhang and
  Balding}]{beaumont:zhang:balding:2002}
\textsc{Beaumont, M.}, \textsc{Zhang, W.} and \textsc{Balding, D.} (2002).
\newblock Approximate {B}ayesian computation in population genetics.
\newblock \textit{Genetics}, \textbf{162} 2025--2035.

\bibitem[{B{\'e}dard(2007)}]{MR2344305}
\textsc{B{\'e}dard, M.} (2007).
\newblock Weak convergence of {M}etropolis algorithms for non-i.i.d. target
  distributions.
\newblock \textit{Ann. Appl. Probab.}, \textbf{17} 1222--1244.

\bibitem[{B{\'e}dard et~al.(2012)B{\'e}dard, Douc and Moulines}]{MR2891436}
\textsc{B{\'e}dard, M.}, \textsc{Douc, R.} and \textsc{Moulines, E.} (2012).
\newblock Scaling analysis of multiple-try {MCMC} methods.
\newblock \textit{Stochastic Process. Appl.}, \textbf{122} 758--786.

\bibitem[{B{\'e}dard et~al.(2014)B{\'e}dard, Douc and Moulines}]{MR3270597}
\textsc{B{\'e}dard, M.}, \textsc{Douc, R.} and \textsc{Moulines, E.} (2014).
\newblock Scaling analysis of delayed rejection {MCMC} methods.
\newblock \textit{Methodol. Comput. Appl. Probab.}, \textbf{16} 811--838.

\bibitem[{Belle et~al.(2008)Belle, Benazzo, Ghirotto, Colonna and
  Barbujani}]{belle:etal:2008}
\textsc{Belle, E.}, \textsc{Benazzo, A.}, \textsc{Ghirotto, S.},
  \textsc{Colonna, V.} and \textsc{Barbujani, G.} (2008).
\newblock {Comparing models on the genealogical relationships among Neandertal,
  Cro-Magnoid and modern Europeans by serial coalescent simulations}.
\newblock \textit{Heredity}, \textbf{102} 218--225.

\bibitem[{Bennett et~al.(1996)Bennett, Racine-Poon and
  Wakefield}]{bennett:racine-poon:wakefield:1996}
\textsc{Bennett, J.}, \textsc{Racine-Poon, A.} and \textsc{Wakefield, J.}
  (1996).
\newblock {MCMC} for nonlinear hierarchical models.
\newblock In \textit{{M}arkov chain {M}onte {C}arlo in Practice} (W.~Gilks,
  S.~Richardson and D.~Spiegelhalter, eds.). Chapman and Hall, New York,
  339--358.

\bibitem[{Berger et~al.(2010)Berger, Fienberg, Raftery and
  Robert}]{berger:fienberg:raftery:robert:2010}
\textsc{Berger, J.}, \textsc{Fienberg, S.}, \textsc{Raftery, A.} and
  \textsc{Robert, C.} (2010).
\newblock Incoherent phylogeographic inference.
\newblock \textit{Proc. National Academy Sciences}, \textbf{107} E57.

\bibitem[{Besag(1994)}]{besag1994comments}
\textsc{Besag, J.} (1994).
\newblock Comments on ``{R}epresentations of knowledge in complex systems'' by
  {U}. {G}renander and {M}.{I}. {M}iller.
\newblock \textit{J. Roy. Statist. Soc. Ser. B}, \textbf{56} 591--592.

\bibitem[{Besag and Green(1993)}]{besag:green:1993}
\textsc{Besag, J.} and \textsc{Green, P.} (1993).
\newblock Spatial statistics and {B}ayesian computation (with discussion).
\newblock \textit{J. Royal Statist. Society Series B}, \textbf{55} 25--38.

\bibitem[{Besag et~al.(1995)Besag, Green, Higdon and Mengersen}]{bghm:1995}
\textsc{Besag, J.}, \textsc{Green, P.~J.}, \textsc{Higdon, D.} and
  \textsc{Mengersen, K.} (1995).
\newblock Bayesian computation and stochastic systems (with discussion).
\newblock \textit{Statistical Science}, \textbf{10} 3--66.

\bibitem[{Besag(1972)}]{besag:1972}
\textsc{Besag, J.~E.} (1972).
\newblock Nearest-neighbour systems and the auto-logistic model for binary
  data.
\newblock \textit{J. Roy. Statist. Soc. Ser. B}, \textbf{34} 75--83.

\bibitem[{Beskos et~al.(2015)Beskos, Jasra, Muzaffer and Stuart}]{BeskosJMS}
\textsc{Beskos, A.}, \textsc{Jasra, A.}, \textsc{Muzaffer, E.~A.} and
  \textsc{Stuart, A.~M.} (2015).
\newblock Sequential {Monte Carlo methods for Bayesian} elliptic inverse
  problems.
\newblock \textit{Statistics and Computing}, \textbf{25}.
\newblock In press.

\bibitem[{Beskos et~al.(2006)Beskos, Papaspiliopoulos, Roberts and
  Fearnhead}]{beskos:papaspiliopoulos:roberts:fearnhead:2006}
\textsc{Beskos, A.}, \textsc{Papaspiliopoulos, O.}, \textsc{Roberts, G.} and
  \textsc{Fearnhead, P.} (2006).
\newblock Exact and computationally efficient likelihood-based estimation for
  discretely observed diffusion processes (with discussion).
\newblock \textit{J. Royal Statist. Society Series B}, \textbf{68} 333--382.

\bibitem[{Beskos et~al.(2013)Beskos, Pillai, Roberts, Sanz-Serna and
  Stuart}]{MR3129023}
\textsc{Beskos, A.}, \textsc{Pillai, N.}, \textsc{Roberts, G.},
  \textsc{Sanz-Serna, J.-M.} and \textsc{Stuart, A.} (2013).
\newblock Optimal tuning of the hybrid {M}onte {C}arlo algorithm.
\newblock \textit{Bernoulli}, \textbf{19} 1501--1534.

\bibitem[{Beskos et~al.(2009)Beskos, Roberts and Stuart}]{MR2537193}
\textsc{Beskos, A.}, \textsc{Roberts, G.} and \textsc{Stuart, A.} (2009).
\newblock Optimal scalings for local {M}etropolis-{H}astings chains on
  nonproduct targets in high dimensions.
\newblock \textit{Ann. Appl. Probab.}, \textbf{19} 863--898.

\bibitem[{Betancourt(2013)}]{betancourt2013general}
\textsc{Betancourt, M.} (2013).
\newblock A general metric for {R}iemannian manifold {H}amiltonian {M}onte
  {C}arlo.
\newblock In \textit{National Conference on the Geometric Science of
  Information} (F.~Nielsen and F.~Barbaresco, eds.). Lecture Notes in Computer
  Science 8085, Springer, 327--334.

\bibitem[{Betancourt et~al.(2014)Betancourt, Byrne and
  Girolami}]{betancourt2014optimizing}
\textsc{Betancourt, M.}, \textsc{Byrne, S.} and \textsc{Girolami, M.} (2014).
\newblock Optimizing the integrator step size for {H}amiltonian {M}onte
  {C}arlo.
\newblock \textit{arXiv preprint arXiv:1411.6669}.

\bibitem[{{Betancourt} et~al.(2014){Betancourt}, {Byrne}, {Livingstone} and
  {Girolami}}]{betancourt:byrne:linvingstone:girolami:2014}
\textsc{{Betancourt}, M.~J.}, \textsc{{Byrne}, S.}, \textsc{{Livingstone}, S.}
  and \textsc{{Girolami}, M.} (2014).
\newblock The geometric foundations of {H}amiltonian {M}onte {C}arlo.
\newblock \textit{ArXiv e-prints}.
\newblock \eprint{1410.5110}.

\bibitem[{Biau et~al.(2014)Biau, C\'erou and Guyader}]{biau:etal:2014}
\textsc{Biau, G.}, \textsc{C\'erou, F.} and \textsc{Guyader, A.} (2014).
\newblock New insights into {A}pproximate {B}ayesian computation.
\newblock \textit{Annales de l'IHP (Probability and Statistics)}, \textbf{51}
  376--403.

\bibitem[{Blum(2010)}]{blum:2010}
\textsc{Blum, M.} (2010).
\newblock {Approximate {B}ayesian computation: a non-parametric perspective}.
\newblock \textit{J. American Statist. Assoc.}, \textbf{105} 1178--1187.

\bibitem[{Blum and Fran{\c c}ois(2010)}]{blum:francois:2010}
\textsc{Blum, M.} and \textsc{Fran{\c c}ois, O.} (2010).
\newblock Non-linear regression models for approximate {B}ayesian computation.
\newblock \textit{Statist. Comput.}, \textbf{20} 63--73.

\bibitem[{Blum et~al.(2013)Blum, Nunes, Prangle and
  Sisson}]{blum:nunes:prangle:sisson:2013}
\textsc{Blum, M.}, \textsc{Nunes, M.}, \textsc{Prangle, D.} and \textsc{Sisson,
  S.} (2013).
\newblock A comparative review of dimension reduction methods in {A}pproximate
  {B}ayesian computation.
\newblock \textit{Stat Sci}, \textbf{28} 189--208.

\bibitem[{{Bornn} et~al.(2014){Bornn}, {Pillai}, {Smith} and
  {Woodard}}]{bornn:etal:2014}
\textsc{{Bornn}, L.}, \textsc{{Pillai}, N.}, \textsc{{Smith}, A.} and
  \textsc{{Woodard}, D.} (2014).
\newblock A pseudo-marginal perspective on the {ABC} algorithm.
\newblock \textit{ArXiv e-prints}.
\newblock \eprint{1404.6298}.

\bibitem[{Bou-Rabee and Hairer(2012)}]{bou2012nonasymptotic}
\textsc{Bou-Rabee, N.} and \textsc{Hairer, M.} (2012).
\newblock Nonasymptotic mixing of the {MALA} algorithm.
\newblock \textit{IMA Journal of Numerical Analysis} drs003.

\bibitem[{Boyd et~al.(2011)Boyd, Parikh, Chu, Peleato and Eckstein}]{boyd:2011}
\textsc{Boyd, S.}, \textsc{Parikh, N.}, \textsc{Chu, E.}, \textsc{Peleato, B.}
  and \textsc{Eckstein, J.} (2011).
\newblock Distributed optimization and statistical learning via the alternating
  direction method of multipliers.
\newblock \textit{Found. Trends Mach. Learn.}, \textbf{3} 1--122.

\bibitem[{Brockwell(2006)}]{brockwell:2006}
\textsc{Brockwell, A.} (2006).
\newblock Parallel {Markov} chain {Monte Carlo} simulation by pre-fetching.
\newblock \textit{J. Comput. Graphical Stat.}, \textbf{15} 246--261.

\bibitem[{Calderhead(2014)}]{calderhead:2014}
\textsc{Calderhead, B.} (2014).
\newblock A general construction for parallelizing {M}etropolis--{H}astings
  algorithms.
\newblock \textit{Proceedings of the National Academy of Sciences},
  \textbf{111} 17408--17413.

\bibitem[{Calvet and Czellar(2014)}]{calvet:czellar:2014}
\textsc{Calvet, C.} and \textsc{Czellar, V.} (2014).
\newblock Accurate methods for {A}pproximate {B}ayesian computation filtering.
\newblock \textit{J. Financial Econometrics}.
\newblock (to appear).

\bibitem[{Cand\`es et~al.(2006)Cand\`es, Romberg and Tao}]{candes:2006}
\textsc{Cand\`es, E.~J.}, \textsc{Romberg, J.~K.} and \textsc{Tao, T.} (2006).
\newblock Stable signal recovery from incomplete and inaccurate measurements.
\newblock \textit{Communications on Pure and Applied Mathematics}, \textbf{59}
  1207--1223.

\bibitem[{Cand\`es and Tao(2009)}]{candes:2009}
\textsc{Cand\`es, E.~J.} and \textsc{Tao, T.} (2009).
\newblock The power of convex relaxation: {N}ear-optimal matrix completion.
\newblock \textit{IEEE Trans. Inform. Theory}, \textbf{56} 2053--2080.

\bibitem[{Cand\`es and Wakin(2008)}]{candes:2008}
\textsc{Cand\`es, E.~J.} and \textsc{Wakin, M.~B.} (2008).
\newblock An introduction to compressive sampling.
\newblock \textit{IEEE Signal Process. Mag.}, \textbf{25} 21--30.

\bibitem[{Capp\'e et~al.(2002)Capp\'e, Robert and
  Ryd\'en}]{cappe:robert:ryden:2002}
\textsc{Capp\'e, O.}, \textsc{Robert, C.} and \textsc{Ryd\'en, T.} (2002).
\newblock Reversible jump {MCMC} converging to birth-and-death {MCMC} and more
  general continuous time samplers.
\newblock \textit{J. Royal Statist. Society Series B}, \textbf{65} 679--700.

\bibitem[{Carlin and Gelfand(1991)}]{carlin:gelfand:1991b}
\textsc{Carlin, B.} and \textsc{Gelfand, A.} (1991).
\newblock An iterative {M}onte {C}arlo method for nonconjugate {B}ayesian
  analysis.
\newblock \textit{Statistics and Computing}, \textbf{1} 119--28.

\bibitem[{Carlin et~al.(1992)Carlin, Gelfand and
  Smith}]{carlin:gelfand:smith:1992}
\textsc{Carlin, B.}, \textsc{Gelfand, A.} and \textsc{Smith, A.} (1992).
\newblock Hierarchical {B}ayesian analysis of change point problems.
\newblock \textit{Applied Statistics (Series C)}, \textbf{41} 389--405.

\bibitem[{Cevher et~al.(2014)Cevher, Becker and Schmidt}]{Cevher:2014}
\textsc{Cevher, V.}, \textsc{Becker, S.} and \textsc{Schmidt, M.} (2014).
\newblock Convex optimization for big data: Scalable, randomized, and parallel
  algorithms for big data analytics.
\newblock \textit{Signal Processing Magazine, IEEE}, \textbf{31} 32--43.

\bibitem[{Chambolle(2004)}]{chambolle:2004}
\textsc{Chambolle, A.} (2004).
\newblock An algorithm for total variation minimization and applications.
\newblock \textit{J. Math. Imaging Vis.}, \textbf{20} 89--97.

\bibitem[{Chandrasekaran and Jordan(2013)}]{chandrasekaran:2013}
\textsc{Chandrasekaran, V.} and \textsc{Jordan, M.~I.} (2013).
\newblock Computational and statistical tradeoffs via convex relaxation.
\newblock \textit{PNAS}, \textbf{110} 1181--1190.

\bibitem[{Chandrasekaran et~al.(2012)Chandrasekaran, Recht, Parrilo and
  Willsky}]{chandrasekaran:2012}
\textsc{Chandrasekaran, V.}, \textsc{Recht, B.}, \textsc{Parrilo, P.} and
  \textsc{Willsky, A.} (2012).
\newblock The convex geometry of linear inverse problems.
\newblock \textit{Found. Comput. Math.}, \textbf{12} 805--849.

\bibitem[{Chipman et~al.(2008)Chipman, George and
  McCulloch}]{chipman:george:mcculoch:2006}
\textsc{Chipman, H.}, \textsc{George, E.} and \textsc{McCulloch, R.} (2008).
\newblock {BART}: {B}ayesian additive regression trees.
\newblock Tech. rep., Acadia University.
\newblock ArXiv:0806.3286v1.

\bibitem[{Chopin(2007)}]{chopin:2007}
\textsc{Chopin, N.} (2007).
\newblock Inference and model choice for time-ordered hidden {M}arkov models.
\newblock \textit{J. Royal Statist. Society Series B}, \textbf{69(2)} 269--284.

\bibitem[{Chopin et~al.(2013)Chopin, Jacob and
  Papaspiliopoulos}]{chopin:jacob:papaspiliopoulos:2013}
\textsc{Chopin, N.}, \textsc{Jacob, P.~E.} and \textsc{Papaspiliopoulos, O.}
  (2013).
\newblock {SMC}2: an efficient algorithm for sequential analysis of state space
  models.
\newblock \textit{J. Royal Statist. Society Series B}, \textbf{75} 397--426.

\bibitem[{Christensen et~al.(2005)Christensen, Roberts and
  Rosenthal}]{christensen2005scaling}
\textsc{Christensen, O.}, \textsc{Roberts, G.} and \textsc{Rosenthal, J.}
  (2005).
\newblock {Scaling limits for the transient phase of local Metropolis--Hastings
  algorithms}.
\newblock \textit{Journal of the Royal Statistical Society: Series B
  (Statistical Methodology)}, \textbf{67} 253--268.

\bibitem[{Combettes and Pesquet(2008)}]{combettes:2008}
\textsc{Combettes, P.~L.} and \textsc{Pesquet, J.-C.} (2008).
\newblock A proximal decomposition method for solving convex variational
  inverse problems.
\newblock \textit{Inverse Problems}, \textbf{24} 065014.

\bibitem[{Combettes and Pesquet(2011)}]{pesquet:2011}
\textsc{Combettes, P.~L.} and \textsc{Pesquet, J.-C.} (2011).
\newblock Proximal splitting methods in signal processing.
\newblock In \textit{Fixed-Point Algorithms for Inverse Problems in Science and
  Engineering} (H.~H. Bauschke, R.~S. Burachik, P.~L. Combettes, V.~Elser,
  D.~R. Luke and H.~Wolkowicz, eds.). Springer {N}ew {Y}ork, 185--212.

\bibitem[{Combettes and Pesquet(2012)}]{pesquet:2012b}
\textsc{Combettes, P.~L.} and \textsc{Pesquet, J.-C.} (2012).
\newblock Primal-dual splitting algorithm for solving inclusions with mixtures
  of composite, {L}ipschitzian, and parallel-sum type monotone operators.
\newblock \textit{Set-Valued A.}, \textbf{20} 307--330.

\bibitem[{{Combettes} and {Pesquet}(2014)}]{combettes:2014}
\textsc{{Combettes}, P.~L.} and \textsc{{Pesquet}, J.-C.} (2014).
\newblock Stochastic {Quasi-Fej\'er} block-coordinate fixed point iterations
  with random sweeping.
\newblock \textit{ArXiv e-prints}.
\newblock \eprint{1404.7536}.

\bibitem[{Cornuet et~al.(2010)Cornuet, Ravign\'e and
  Estoup}]{cornuet:ravigne:estoup:2010}
\textsc{Cornuet, J.-M.}, \textsc{Ravign\'e, V.} and \textsc{Estoup, A.} (2010).
\newblock {Inference on population history and model checking using DNA
  sequence and microsatellite data with the software {DIYABC} (v1.0)}.
\newblock \textit{BMC Bioinformatics}, \textbf{11} 401.

\bibitem[{Cornuet et~al.(2008)Cornuet, Santos, Beaumont, Robert, Marin,
  Balding, Guillemaud and Estoup}]{cornuet:etal:2008}
\textsc{Cornuet, J.-M.}, \textsc{Santos, F.}, \textsc{Beaumont, M.},
  \textsc{Robert, C.}, \textsc{Marin, J.-M.}, \textsc{Balding, D.},
  \textsc{Guillemaud, T.} and \textsc{Estoup, A.} (2008).
\newblock Inferring population history with {DIYABC}: a user-friendly approach
  to {A}pproximate {B}ayesian computation.
\newblock \textit{Bioinformatics}, \textbf{24} 2713--2719.

\bibitem[{Cotter et~al.(2013)Cotter, Roberts, Stuart, White
  et~al.}]{cotter2013mcmc}
\textsc{Cotter, S.}, \textsc{Roberts, G.}, \textsc{Stuart, A.}, \textsc{White,
  D.} \textsc{et~al.} (2013).
\newblock {MCMC} methods for functions: modifying old algorithms to make them
  faster.
\newblock \textit{Statistical Science}, \textbf{28} 424--446.

\bibitem[{Craiu et~al.(2009)Craiu, Rosenthal and
  Yang}]{craiu:rosenthal:yang:2009}
\textsc{Craiu, R.}, \textsc{Rosenthal, J.} and \textsc{Yang, C.} (2009).
\newblock Learn from thy neighbour: Parallel-chain and regional adaptive
  {MCMC}.
\newblock \textit{J. American Statist. Assoc.}, \textbf{104} 1454--1466.

\bibitem[{Craiu and Meng(2005)}]{craiu:meng:2005}
\textsc{Craiu, R.~V.} and \textsc{Meng, X.-L.} (2005).
\newblock Multiprocess parallel antithetic coupling for backward and forward
  {M}arkov chain {M}onte {C}arlo.
\newblock \textit{Ann. Statist.}, \textbf{33} 661--697.

\bibitem[{Cucala et~al.(2009)Cucala, Marin, Robert and
  Titterington}]{cucala:marin:robert:titterington:2006}
\textsc{Cucala, L.}, \textsc{Marin, J.-M.}, \textsc{Robert, C.} and
  \textsc{Titterington, D.} (2009).
\newblock {B}ayesian inference in $k$-nearest-neighbour classification models.
\newblock \textit{J. American Statist. Assoc.}, \textbf{104 (485)} 263--273.

\bibitem[{Dean and Ghemawat(2008)}]{dean:ghemawat}
\textsc{Dean, J.} and \textsc{Ghemawat, S.} (2008).
\newblock Map{R}educe: simplified data processing on large clusters.
\newblock \textit{Communications of the ACM}, \textbf{51} 107--113.

\bibitem[{Del~Moral et~al.(2006)Del~Moral, Doucet and
  Jasra}]{delmoral:doucet:jasra:2006}
\textsc{Del~Moral, P.}, \textsc{Doucet, A.} and \textsc{Jasra, A.} (2006).
\newblock Sequential {M}onte {C}arlo samplers.
\newblock \textit{J. Royal Statist. Society Series B}, \textbf{68} 411--436.

\bibitem[{Dellaportas et~al.(2004)Dellaportas, Papaspiliopoulos and
  Roberts}]{roberts:papaspiliopoulos:dellaportas:2001}
\textsc{Dellaportas, P.}, \textsc{Papaspiliopoulos, O.} and \textsc{Roberts,
  G.} (2004).
\newblock {B}ayesian inference for non-{G}aussian {O}rnstein-uhlenbeck
  stochastic volatility processes.
\newblock \textit{J. Royal Statist. Society Series B}, \textbf{66} 369--393.

\bibitem[{Dellaportas and Wright(1991)}]{dellaportas:wright:2001}
\textsc{Dellaportas, P.} and \textsc{Wright, D.} (1991).
\newblock Positive embedded integration in {B}ayesian analysis.
\newblock \textit{Statistics and Computing}, \textbf{1} 1--12.

\bibitem[{Diaconis and Stroock(1991)}]{diaconis1991geometric}
\textsc{Diaconis, P.} and \textsc{Stroock, D.} (1991).
\newblock Geometric bounds for eigenvalues of {M}arkov chains.
\newblock \textit{The Annals of Applied Probability} 36--61.

\bibitem[{Didelot et~al.(2011)Didelot, Everitt, Johansen and
  Lawson}]{didelot:everitt:johansen:lawson:2011}
\textsc{Didelot, X.}, \textsc{Everitt, R.}, \textsc{Johansen, A.} and
  \textsc{Lawson, D.} (2011).
\newblock Likelihood-free estimation of model evidence.
\newblock \textit{Bayesian Analysis}, \textbf{6} 48--76.

\bibitem[{Diebolt and Robert(1994)}]{diebolt:robert:1994}
\textsc{Diebolt, J.} and \textsc{Robert, C.} (1994).
\newblock Estimation of finite mixture distributions by {B}ayesian sampling.
\newblock \textit{J. Royal Statist. Society Series B}, \textbf{56} 363--375.

\bibitem[{Doll and Dion(1976)}]{doll1976generalized}
\textsc{Doll, J.} and \textsc{Dion, D.} (1976).
\newblock Generalized {L}angevin equation approach for atom/solid--surface
  scattering: Numerical techniques for {G}aussian generalized {L}angevin
  dynamics.
\newblock \textit{The Journal of Chemical Physics}, \textbf{65} 3762--3766.

\bibitem[{Douc and Robert(2011)}]{douc:robert:2010}
\textsc{Douc, R.} and \textsc{Robert, C.} (2011).
\newblock A vanilla {R}ao--{B}lackwellization of {M}etropolis--{H}astings
  algorithms.
\newblock \textit{Ann. Statist.}, \textbf{39} 261--277.

\bibitem[{Doucet et~al.(2012)Doucet, Pitt, Deligiannidis and
  Kohn}]{doucet2012efficient}
\textsc{Doucet, A.}, \textsc{Pitt, M.}, \textsc{Deligiannidis, G.} and
  \textsc{Kohn, R.} (2012).
\newblock Efficient implementation of {M}arkov chain {M}onte {C}arlo when using
  an unbiased likelihood estimator.
\newblock \textit{arXiv preprint arXiv:1210.1871 Biometrika to appear 2015}.

\bibitem[{Drovandi et~al.(2011)Drovandi, Pettitt and
  Fddy}]{drovandi:pettitt:faddy:2011}
\textsc{Drovandi, C.}, \textsc{Pettitt, A.} and \textsc{Fddy, M.} (2011).
\newblock Approximate {B}ayesian computation using indirect inference.
\newblock \textit{J. Royal Statist. Society Series A}, \textbf{60} 503--524.

\bibitem[{Duane et~al.(1987)Duane, Kennedy, Pendleton,  and
  Roweth}]{duane:etal:1987}
\textsc{Duane, S.}, \textsc{Kennedy, A.~D.}, \textsc{Pendleton, B.~J.},  and
  \textsc{Roweth, D.} (1987).
\newblock Hybrid {M}onte {C}arlo.
\newblock \textit{Phys. Lett. B}, \textbf{195} 216--222.

\bibitem[{Ermak(1975)}]{ermak1975computer}
\textsc{Ermak, D.} (1975).
\newblock {A computer simulation of charged particles in solution. I. Technique
  and equilibrium properties}.
\newblock \textit{The Journal of Chemical Physics}, \textbf{62} 4189--4196.

\bibitem[{Excoffier et~al.(2009)Excoffier, Leuenberger and
  Wegmann}]{excoffier:leuenberger:wegman:2009}
\textsc{Excoffier, C.}, \textsc{Leuenberger, D.} and \textsc{Wegmann, L.}
  (2009).
\newblock {Bayesian computation and model selection in population genetics}.
\newblock ArXiv:0901.2231.

\bibitem[{Fearnhead and Clifford(2003)}]{fearnhead:clifford:2003}
\textsc{Fearnhead, P.} and \textsc{Clifford, P.} (2003).
\newblock On-line inference for hidden {M}arkov models via particle filters.
\newblock \textit{J. Royal Statist. Society Series B}, \textbf{65} 887--899.

\bibitem[{Fearnhead and Prangle(2012)}]{fearnhead:prangle:2012}
\textsc{Fearnhead, P.} and \textsc{Prangle, D.} (2012).
\newblock Constructing summary statistics for {A}pproximate {B}ayesian
  computation: semi-automatic {A}pproximate {B}ayesian computation.
\newblock \textit{Journal of the Royal Statistical Society: Series B
  (Statistical Methodology)}, \textbf{74} 419--474.
\newblock (With discussion.).

\bibitem[{Fort et~al.(2011)Fort, Moulines and Priouret}]{fort2011convergence}
\textsc{Fort, G.}, \textsc{Moulines, E.} and \textsc{Priouret, P.} (2011).
\newblock Convergence of adaptive and interacting markov chain monte carlo
  algorithms.
\newblock \textit{The Annals of Statistics}, \textbf{39} 3262--3289.

\bibitem[{Frigessi et~al.(2000)Frigessi, Gasemyr and Rue}]{frigessi:2000}
\textsc{Frigessi, A.}, \textsc{Gasemyr, J.} and \textsc{Rue, H.} (2000).
\newblock Antithetic coupling of two {G}ibbs sampler chains.
\newblock \textit{The Annals of Statistics}, \textbf{28} 1128--1149.

\bibitem[{Geman and Geman(1984)}]{geman:geman:1984}
\textsc{Geman, S.} and \textsc{Geman, D.} (1984).
\newblock Stochastic relaxation, {G}ibbs distributions and the {B}ayesian
  restoration of images.
\newblock \textit{IEEE Trans. Pattern Anal. Mach. Intell.}, \textbf{6}
  721--741.

\bibitem[{Geyer(1992)}]{Geyer1992}
\textsc{Geyer, C.~J.} (1992).
\newblock {Practical Markov chain Monte Carlo}.
\newblock \textit{Statistical Science}, \textbf{7} 473--483.

\bibitem[{Ghirotto et~al.(2010)Ghirotto, Mona, Benazzo, Paparazzo, Caramelli
  and Barbujani}]{ghirotto:etal:2010}
\textsc{Ghirotto, S.}, \textsc{Mona, S.}, \textsc{Benazzo, A.},
  \textsc{Paparazzo, F.}, \textsc{Caramelli, D.} and \textsc{Barbujani, G.}
  (2010).
\newblock {Inferring genealogical processes from patterns of bronze-age and
  modern DNA variation in {S}ardinia}.
\newblock \textit{Mol. Biol. Evol.}, \textbf{27} 875--886.

\bibitem[{Gilks et~al.(1994)Gilks, Roberts and George}]{gilks1994adaptive}
\textsc{Gilks, W.}, \textsc{Roberts, G.} and \textsc{George, E.} (1994).
\newblock Adaptive direction sampling.
\newblock \textit{Journal of the Royal Statistical Society. Series D (The
  Statistician)}, \textbf{43} 179--189.

\bibitem[{Gilks et~al.(1998)Gilks, Roberts and Sahu}]{gilks1998adaptive}
\textsc{Gilks, W.}, \textsc{Roberts, G.} and \textsc{Sahu, S.} (1998).
\newblock {Adaptive Markov chain Monte Carlo through regeneration}.
\newblock \textit{Journal of the American Statistical Association}, \textbf{93}
  1045--1054.

\bibitem[{Girolami and Calderhead(2011)}]{girolami:2011}
\textsc{Girolami, M.} and \textsc{Calderhead, B.} (2011).
\newblock Riemann manifold {L}angevin and {H}amiltonian {M}onte {C}arlo
  methods.
\newblock \textit{Journal of the Royal Statistical Society: Series B
  (Statistical Methodology)}, \textbf{73} 123--214.

\bibitem[{Girolami et~al.(2013)Girolami, Lyne, Strathmann, Simpson and
  Atchad{\'e}}]{girolami2013playing}
\textsc{Girolami, M.}, \textsc{Lyne, A.-M.}, \textsc{Strathmann, H.},
  \textsc{Simpson, D.} and \textsc{Atchad{\'e}, Y.} (2013).
\newblock {Playing Russian roulette with intractable likelihoods}.
\newblock \textit{arXiv preprint arXiv:1306.4032}.

\bibitem[{Green(1995)}]{green:1995}
\textsc{Green, P.} (1995).
\newblock Reversible jump {MCMC} computation and {B}ayesian model
  determination.
\newblock \textit{Biometrika}, \textbf{82} 711--732.

\bibitem[{Green(2015)}]{green:mardiafest}
\textsc{Green, P.~J.} (2015).
\newblock {MAD-B}ayes matching and alignment for labelled and unlabelled
  configurations.
\newblock In \textit{Geometry driven statistics} (I.~L. Dryden and J.~T. Kent,
  eds.), chap.~19. Wiley, Chichester, 365--375.

\bibitem[{Grelaud et~al.(2009)Grelaud, Marin, Robert, Rodolphe and
  Tally}]{grelaud:marin:robert:rodolphe:tally:2009}
\textsc{Grelaud, A.}, \textsc{Marin, J.-M.}, \textsc{Robert, C.},
  \textsc{Rodolphe, F.} and \textsc{Tally, F.} (2009).
\newblock Likelihood-free methods for model choice in {G}ibbs random fields.
\newblock \textit{Bayesian Analysis}, \textbf{3(2)} 427--442.

\bibitem[{Grenander and Miller(1994)}]{MR1293234}
\textsc{Grenander, U.} and \textsc{Miller, M.~I.} (1994).
\newblock Representations of knowledge in complex systems.
\newblock \textit{J. Roy. Statist. Soc. Ser. B}, \textbf{56} 549--603.
\newblock With discussion and a reply by the authors.

\bibitem[{Griffin et~al.(2014)Griffin, {\L}atuszy\'nski and Steel}]{MarkJim}
\textsc{Griffin, J.}, \textsc{{\L}atuszy\'nski, K.} and \textsc{Steel, M.}
  (2014).
\newblock {Individual adaptation: an adaptive MCMC scheme for variable
  selection problems}.
\newblock \textit{submitted}.

\bibitem[{Guillemaud et~al.(2009)Guillemaud, Beaumont, Ciosi, Cornuet and
  Estoup}]{guillemaud:etal:2009}
\textsc{Guillemaud, T.}, \textsc{Beaumont, M.}, \textsc{Ciosi, M.},
  \textsc{Cornuet, J.-M.} and \textsc{Estoup, A.} (2009).
\newblock {Inferring introduction routes of invasive species using approximate
  {B}ayesian computation on microsatellite data}.
\newblock \textit{Heredity}, \textbf{104} 88--99.

\bibitem[{Haario et~al.(2006)Haario, Laine, Mira and Saksman}]{haario2006dram}
\textsc{Haario, H.}, \textsc{Laine, M.}, \textsc{Mira, A.} and \textsc{Saksman,
  E.} (2006).
\newblock {DRAM: efficient adaptive MCMC}.
\newblock \textit{Statistics and Computing}, \textbf{16} 339--354.

\bibitem[{Haario et~al.(2001)Haario, Saksman and Tamminen}]{haario2001adaptive}
\textsc{Haario, H.}, \textsc{Saksman, E.} and \textsc{Tamminen, J.} (2001).
\newblock {An adaptive Metropolis algorithm}.
\newblock \textit{Bernoulli}, \textbf{7} 223--242.

\bibitem[{Hairer et~al.(2006)Hairer, Lubich and Wanner}]{MR2221614}
\textsc{Hairer, E.}, \textsc{Lubich, C.} and \textsc{Wanner, G.} (2006).
\newblock \textit{Geometric numerical integration: {S}tructure-preserving
  algorithms for ordinary differential equations}, vol.~31 of \textit{Springer
  Series in Computational Mathematics}.
\newblock 2nd ed. Springer-Verlag, Berlin.

\bibitem[{Harford(2014)}]{harford}
\textsc{Harford, T.} (2014).
\newblock Big data: Are we making a big mistake?
\newblock \textit{Significance}, \textbf{11} 14--19.

\bibitem[{Hastings(1970)}]{hastings:1970}
\textsc{Hastings, W.} (1970).
\newblock {M}onte {C}arlo sampling methods using {M}arkov chains and their
  application.
\newblock \textit{Biometrika}, \textbf{57} 97--109.

\bibitem[{Huelsenbeck and Ronquist(2001)}]{hueron2001a}
\textsc{Huelsenbeck, J.~P.} and \textsc{Ronquist, F.} (2001).
\newblock {MRBAYES}: Bayesian inference of phylogenetic trees.
\newblock \textit{Bioinformatics}.

\bibitem[{Jaakkola and Jordan(2000)}]{jaakkola:jordan:2000}
\textsc{Jaakkola, T.} and \textsc{Jordan, M.} (2000).
\newblock Bayesian parameter estimation via variational methods.
\newblock \textit{Statistics and Computing}, \textbf{10} 25--37.

\bibitem[{Jacob et~al.(2011)Jacob, Robert and Smith}]{jacob:robert:smith:2010}
\textsc{Jacob, P.}, \textsc{Robert, C.} and \textsc{Smith, M.} (2011).
\newblock Using parallel computation to improve independent
  {M}etropolis--{H}astings based estimation.
\newblock \textit{J. Comput. Graph. Statist.}, \textbf{20} 616--635.

\bibitem[{Jacob and Ryder(2014)}]{jacob:ryder:2014}
\textsc{Jacob, P.~E.} and \textsc{Ryder, R.~J.} (2014).
\newblock The {W}ang--{L}andau algorithm reaches the flat histogram criterion
  in finite time.
\newblock \textit{Ann. Appl. Probab.}, \textbf{24} 34--53.

\bibitem[{Ji and Schmidler(2013)}]{ji2013adaptive}
\textsc{Ji, C.} and \textsc{Schmidler, S.~C.} (2013).
\newblock Adaptive {Markov chain Monte Carlo for Bayesian} variable selection.
\newblock \textit{Journal of Computational and Graphical Statistics},
  \textbf{22} 708--728.

\bibitem[{Jordan et~al.(1999)Jordan, Ghahramani, Jaakkola and
  Saul}]{jordan:etal:1999}
\textsc{Jordan, M.~I.}, \textsc{Ghahramani, Z.}, \textsc{Jaakkola, T.~S.} and
  \textsc{Saul, L.~K.} (1999).
\newblock An introduction to variational methods for graphical models.
\newblock \textit{Machine learning}, \textbf{37} 183--233.

\bibitem[{Jourdain et~al.(2012)Jourdain, Leli{\`e}vre and
  Miasojedow}]{jourdain2012optimal}
\textsc{Jourdain, B.}, \textsc{Leli{\`e}vre, T.} and \textsc{Miasojedow, B.}
  (2012).
\newblock Optimal scaling for the transient phase of the random walk
  {M}etropolis algorithm: the mean-field limit.
\newblock \textit{arXiv preprint arXiv:1210.7639; Annals of Applied
  Probability, to appear}.

\bibitem[{Jourdain et~al.(2014)Jourdain, Leli{\`e}vre and
  Miasojedow}]{MR3263094}
\textsc{Jourdain, B.}, \textsc{Leli{\`e}vre, T.} and \textsc{Miasojedow, B.}
  (2014).
\newblock Optimal scaling for the transient phase of {M}etropolis {H}astings
  algorithms: the longtime behavior.
\newblock \textit{Bernoulli}, \textbf{20} 1930--1978.

\bibitem[{Kent(1978)}]{Kent1978}
\textsc{Kent, J.} (1978).
\newblock Time-reversible diffusions.
\newblock \textit{Adv. in Appl. Probab.}, \textbf{10, no. 4} 819--835.

\bibitem[{Kloeden and Platen(1992)}]{MR1214374}
\textsc{Kloeden, P.~E.} and \textsc{Platen, E.} (1992).
\newblock \textit{Numerical solution of stochastic differential equations},
  vol.~23 of \textit{Applications of Mathematics (New York)}.
\newblock Springer-Verlag, Berlin.

\bibitem[{{Komodakis} and {Pesquet}(2014)}]{komodakis:2014}
\textsc{{Komodakis}, N.} and \textsc{{Pesquet}, J.-C.} (2014).
\newblock Playing with duality: An overview of recent primal-dual approaches
  for solving large-scale optimization problems.
\newblock \textit{ArXiv e-prints}.
\newblock \eprint{1406.5429}.

\bibitem[{Korattikara et~al.(2013)Korattikara, Chen and
  Welling}]{korattikara2013austerity}
\textsc{Korattikara, A.}, \textsc{Chen, Y.} and \textsc{Welling, M.} (2013).
\newblock Austerity in {MCMC} land: Cutting the {Metropolis--Hastings} budget.
\newblock \textit{arXiv preprint arXiv:1304.5299}.

\bibitem[{Kou et~al.(2006)Kou, Zhou and Wong}]{MR2283711}
\textsc{Kou, S.~C.}, \textsc{Zhou, Q.} and \textsc{Wong, W.~H.} (2006).
\newblock Equi-energy sampler with applications in statistical inference and
  statistical mechanics.
\newblock \textit{Ann. Statist.}, \textbf{34} 1581--1652.
\newblock With discussions and a rejoinder by the authors.

\bibitem[{Lamnisos et~al.(2013)Lamnisos, Griffin and
  Steel}]{lamnisos2013adaptive}
\textsc{Lamnisos, D.}, \textsc{Griffin, J.~E.} and \textsc{Steel, M.~F.}
  (2013).
\newblock Adaptive {Monte Carlo for Bayesian} variable selection in regression
  models.
\newblock \textit{Journal of Computational and Graphical Statistics},
  \textbf{22} 729--748.

\bibitem[{Lange et~al.(2014)Lange, Chi and Zhou}]{lange:chi:zhou:2014}
\textsc{Lange, K.}, \textsc{Chi, E.~C.} and \textsc{Zhou, H.} (2014).
\newblock A brief survey of modern optimization for statisticians.
\newblock \textit{International Statistical Review}, \textbf{82} 46--70.

\bibitem[{Larget and Simon(1999)}]{largetetsimon99}
\textsc{Larget, B.} and \textsc{Simon, D.~L.} (1999).
\newblock {M}arkov chain {M}onte {C}arlo algorithms for the {B}ayesian analysis
  of phylogenetic trees.
\newblock \textit{Mol. Biol. Evol.}, \textbf{16} 750--759.

\bibitem[{{\L}atuszy{\'n}ski et~al.(2011){\L}atuszy{\'n}ski, Kosmidis,
  Papaspiliopoulos and Roberts}]{latuszynski:kosmidis:papa:roberts:2010}
\textsc{{\L}atuszy{\'n}ski, K.}, \textsc{Kosmidis, I.},
  \textsc{Papaspiliopoulos, O.} and \textsc{Roberts, G.~O.} (2011).
\newblock Simulating events of unknown probabilities via reverse time
  martingales.
\newblock \textit{Random Structures \& Algorithms}, \textbf{38} 441--452.

\bibitem[{{\L}atuszy\'{n}ski et~al.(2013){\L}atuszy\'{n}ski, Roberts and
  Rosenthal}]{latuszynski2010adaptive}
\textsc{{\L}atuszy\'{n}ski, K.}, \textsc{Roberts, G.} and \textsc{Rosenthal,
  J.} (2013).
\newblock {Adaptive Gibbs samplers and related MCMC methods}.
\newblock \textit{Ann. Appl. Probab.}, \textbf{23(1)} 66--98.

\bibitem[{{\L}atuszy{\'n}ski and Rosenthal(2014)}]{latuszynski2014containment}
\textsc{{\L}atuszy{\'n}ski, K.} and \textsc{Rosenthal, J.~S.} (2014).
\newblock The containment condition and {AdapFail} algorithms.
\newblock \textit{Journal of Applied Probability}, \textbf{51} 1189--1195.

\bibitem[{{Lee} and {{\L}atuszy{\'n}ski}(2014)}]{lee:latuszynski:2014}
\textsc{{Lee}, A.} and \textsc{{{\L}atuszy{\'n}ski}, K.} (2014).
\newblock {Variance bounding and geometric ergodicity of {M}arkov chain {M}onte
  {C}arlo kernels for {A}pproximate {B}ayesian computation}.
\newblock \textit{Biometrika}, \textbf{101} 655--671.

\bibitem[{Lee et~al.(2009)Lee, Yau, Giles, Doucet and Holmes}]{lee2009a}
\textsc{Lee, A.}, \textsc{Yau, C.}, \textsc{Giles, M.}, \textsc{Doucet, A.} and
  \textsc{Holmes, C.} (2009).
\newblock {On the utility of graphics cards to perform massively parallel
  simulation with advanced Monte Carlo methods}.
\newblock \textit{Arxiv preprint arXiv:0905.2441}.

\bibitem[{Lee et~al.(2005)Lee, Okabe and Landau}]{leeetal06}
\textsc{Lee, H.~K.}, \textsc{Okabe, Y.} and \textsc{Landau, D.~P.} (2005).
\newblock Convergence and refinement of the {W}ang-{L}andau algorithm.
\newblock \textit{Technical Report}.

\bibitem[{Leuenberger and Wegmann(2010)}]{leuenberger:wegmann:2010}
\textsc{Leuenberger, C.} and \textsc{Wegmann, D.} (2010).
\newblock Bayesian computation and model selection without likelihoods.
\newblock \textit{Genetics}, \textbf{184} 243--252.

\bibitem[{Levin et~al.(2009)Levin, Peres and Wilmer}]{levin2009markov}
\textsc{Levin, D.~A.}, \textsc{Peres, Y.} and \textsc{Wilmer, E.~L.} (2009).
\newblock \textit{Markov chains and mixing times}.
\newblock American Mathematical Soc.

\bibitem[{{Lindsten} et~al.(2014){Lindsten}, {Jordan} and
  {Sch{\"o}n}}]{lindsen:jordan:schon:2014}
\textsc{{Lindsten}, F.}, \textsc{{Jordan}, M.~I.} and \textsc{{Sch{\"o}n},
  T.~B.} (2014).
\newblock Particle {G}ibbs with ancestor sampling.
\newblock \textit{ArXiv e-prints}.
\newblock \eprint{1401.0604}.

\bibitem[{Lunn et~al.(2010)Lunn, Thomas, Best and
  Spiegelhalter}]{lunn:bugs:2012}
\textsc{Lunn, D.}, \textsc{Thomas, A.}, \textsc{Best, N.} and
  \textsc{Spiegelhalter, D.} (2010).
\newblock \textit{The {BUGS} Book: A Practical Introduction to {B}ayesian
  Analysis}.
\newblock Chapman \& Hall/CRC Press.

\bibitem[{MacKay(2002)}]{mackay:2002}
\textsc{MacKay, D. J.~C.} (2002).
\newblock \textit{Information Theory, Inference \& Learning Algorithms}.
\newblock Cambridge University Press, Cambridge, UK.

\bibitem[{Maclaurin and Adams(2014)}]{maclaurin2014firefly}
\textsc{Maclaurin, D.} and \textsc{Adams, R.~P.} (2014).
\newblock Firefly {Monte Carlo: Exact MCMC} with subsets of data.
\newblock \textit{arXiv preprint arXiv:1403.5693}.

\bibitem[{Marin et~al.(2014)Marin, Pillai, Robert and
  Rousseau}]{marin:pillai:robert:rousseau:2011}
\textsc{Marin, J.}, \textsc{Pillai, N.}, \textsc{Robert, C.} and
  \textsc{Rousseau, J.} (2014).
\newblock Relevant statistics for {B}ayesian model choice.
\newblock \textit{J. Royal Statist. Society Series B}.
\newblock (to appear).

\bibitem[{Marin et~al.(2011)Marin, Pudlo, Robert and
  Ryder}]{marin:pudlo:robert:ryder:2011}
\textsc{Marin, J.}, \textsc{Pudlo, P.}, \textsc{Robert, C.} and \textsc{Ryder,
  R.} (2011).
\newblock Approximate {B}ayesian computational methods.
\newblock \textit{Statistics and Computing} 1--14.

\bibitem[{Marshall and Roberts(2012)}]{marshall2012adaptive}
\textsc{Marshall, T.} and \textsc{Roberts, G.} (2012).
\newblock An adaptive approach to {Langevin MCMC}.
\newblock \textit{Statistics and Computing}, \textbf{22} 1041--1057.

\bibitem[{Martinet(1970)}]{martinet:1970}
\textsc{Martinet, B.} (1970).
\newblock Regularisation d'in\'equations variationelles par approximations
  successives.
\newblock \textit{Revue {F}ran. d'{A}utomatique et {I}nfomatique {R}ech.
  Op\'erationelle}, \textbf{4} 154--159.

\bibitem[{Medina-Aguayo et~al.(2015)Medina-Aguayo, Lee and
  Roberts}]{medina2015stability}
\textsc{Medina-Aguayo, F.~J.}, \textsc{Lee, A.} and \textsc{Roberts, G.~O.}
  (2015).
\newblock {Stability of noisy Metropolis--Hastings}.
\newblock \textit{arXiv preprint arXiv:1503.07066}.

\bibitem[{Mengersen and Tweedie(1996)}]{mengersen:tweedie:1996}
\textsc{Mengersen, K.} and \textsc{Tweedie, R.} (1996).
\newblock Rates of convergence of the {H}astings and {M}etropolis algorithms.
\newblock \textit{Ann. Statist.}, \textbf{24} 101--121.

\bibitem[{Metropolis(1987)}]{metropolis:1987}
\textsc{Metropolis, N.} (1987).
\newblock The beginning of the {M}onte {C}arlo method.
\newblock \textit{Los Alamos Science}, \textbf{15} 125--130.

\bibitem[{Metropolis et~al.(1953)Metropolis, Rosenbluth, Rosenbluth, Teller and
  Teller}]{metropolis:1953}
\textsc{Metropolis, N.}, \textsc{Rosenbluth, A.~W.}, \textsc{Rosenbluth,
  M.~N.}, \textsc{Teller, A.~H.} and \textsc{Teller, E.} (1953).
\newblock Equations of state calculations by fast computing machines.
\newblock \textit{J. Chem. Phys.}, \textbf{21} 1087--1092.

\bibitem[{Meyn and Tweedie(2009)}]{MT2009}
\textsc{Meyn, S.} and \textsc{Tweedie, R.} (2009).
\newblock \textit{Markov chains and stochastic stability}.
\newblock Cambridge University Press.

\bibitem[{Miasojedow et~al.(2013)Miasojedow, Moulines and
  Vihola}]{miasojedow2013adaptive}
\textsc{Miasojedow, B.}, \textsc{Moulines, E.} and \textsc{Vihola, M.} (2013).
\newblock An adaptive parallel tempering algorithm.
\newblock \textit{Journal of Computational and Graphical Statistics},
  \textbf{22} 649--664.

\bibitem[{Minka(2001)}]{minka:2001}
\textsc{Minka, T.} (2001).
\newblock Expectation propagation for approximate {B}ayesian inference.
\newblock In \textit{UAI '01: Proceedings of the 17th Conference in Uncertainty
  in Artificial Intelligence} (D.~K. Jack S.~Breese, ed.). University of
  Washington, Seattle, 362--369.

\bibitem[{Minsker et~al.(2014)Minsker, Srivastava, Lin and
  Dunson}]{minsker2014robust}
\textsc{Minsker, S.}, \textsc{Srivastava, S.}, \textsc{Lin, L.} and
  \textsc{Dunson, D.~B.} (2014).
\newblock Robust and scalable {B}ayes via a median of subset posterior
  measures.
\newblock \textit{arXiv preprint arXiv:1403.2660}.

\bibitem[{M{\o}ller et~al.(2006)M{\o}ller, Pettitt, Reeves and
  Berthelsen}]{moller:etal:2006}
\textsc{M{\o}ller, J.}, \textsc{Pettitt, A.~N.}, \textsc{Reeves, R.} and
  \textsc{Berthelsen, K.~K.} (2006).
\newblock An efficient {Markov chain Monte Carlo} method for distributions with
  intractable normalising constants.
\newblock \textit{Biometrika}, \textbf{93} 451--458.

\bibitem[{Moreau(1962)}]{moreau:1962}
\textsc{Moreau, J.-J.} (1962).
\newblock Fonctions convexes duales et points proximaux dans un espace
  {H}ilbertien.
\newblock \textit{C. R. Acad. Sci. Paris S\'er. A Math.}, \textbf{255}
  2897--2899.

\bibitem[{{Muff} et~al.(2013){Muff}, {Riebler}, {Rue}, {Saner} and
  {Held}}]{muif:etal:2013}
\textsc{{Muff}, S.}, \textsc{{Riebler}, A.}, \textsc{{Rue}, H.},
  \textsc{{Saner}, P.} and \textsc{{Held}, L.} (2013).
\newblock {Bayesian analysis of measurement error models using {INLA}}.
\newblock \textit{ArXiv e-prints}.
\newblock \eprint{1302.3065}.

\bibitem[{Murray et~al.(2006{\natexlab{a}})Murray, Ghahramani,  and
  MacKay}]{murray:etal:2006}
\textsc{Murray, I.}, \textsc{Ghahramani, Z.},  and \textsc{MacKay, D.}
  (2006{\natexlab{a}}).
\newblock {MCMC} for doubly-intractable distributions.
\newblock In \textit{Uncertainty in Artificial Intelligence}. UAI-2006.

\bibitem[{Murray et~al.(2006{\natexlab{b}})Murray, MacKay, Ghahramani and
  Skilling}]{murray:nested:potts}
\textsc{Murray, I.}, \textsc{MacKay, D.~J.}, \textsc{Ghahramani, Z.} and
  \textsc{Skilling, J.} (2006{\natexlab{b}}).
\newblock Nested sampling for {P}otts models.
\newblock In \textit{Advances in Neural Information Processing Systems 18}
  (Y.~Weiss, B.~Sch\"{o}lkopf and J.~Platt, eds.). MIT Press, Cambridge, MA,
  947--954.

\bibitem[{Naylor and Smith(1982)}]{naylor:smith:1982}
\textsc{Naylor, J.} and \textsc{Smith, A.} (1982).
\newblock Application of a method for the efficient computation of posterior
  distributions.
\newblock \textit{Applied {S}tatistics}, \textbf{31} 214--225.

\bibitem[{Neal et~al.(2012)Neal, Roberts and Yuen}]{MR3025684}
\textsc{Neal, P.}, \textsc{Roberts, G.} and \textsc{Yuen, W.~K.} (2012).
\newblock Optimal scaling of random walk {M}etropolis algorithms with
  discontinuous target densities.
\newblock \textit{Ann. Appl. Probab.}, \textbf{22} 1880--1927.

\bibitem[{Neal(1999)}]{neal:1996}
\textsc{Neal, R.} (1999).
\newblock \textit{{B}ayesian Learning for Neural Networks}, vol. 118.
\newblock Springer--Verlag, New York.
\newblock Lecture Notes.

\bibitem[{Neal(2013)}]{neal:2013}
\textsc{Neal, R.} (2013).
\newblock {MCMC} using {H}amiltonian dynamics.
\newblock In \textit{Handbook of Markov Chain Monte Carlo} (S.~Brooks,
  A.~Gelman, G.~Jones and X.-L. Meng, eds.). Chapman \& Hall/CRC Press,
  113--162.

\bibitem[{Neiswanger et~al.(2013)Neiswanger, Wang and
  Xing}]{neiswanger:wang:xing:2013}
\textsc{Neiswanger, W.}, \textsc{Wang, C.} and \textsc{Xing, E.} (2013).
\newblock Asymptotically exact, embarrassingly parallel {MCMC}.
\newblock \textit{arXiv preprint arXiv:1311.4780}.

\bibitem[{Nesterov(2004)}]{nesterov:2004}
\textsc{Nesterov, Y.} (2004).
\newblock \textit{Introductory lectures on convex optimization: A basic
  course}, vol.~87 of \textit{Applied optimization.}
\newblock Kluwer Academic Publishers.

\bibitem[{Nott and Kohn(2005)}]{nott2005adaptive}
\textsc{Nott, D.} and \textsc{Kohn, R.} (2005).
\newblock Adaptive sampling for {B}ayesian variable selection.
\newblock \textit{Biometrika}, \textbf{92} 747--763.

\bibitem[{Oliveira et~al.(2009)Oliveira, Bioucas-Dias and
  Figueiredo}]{oliveira:2009}
\textsc{Oliveira, J.}, \textsc{Bioucas-Dias, J.} and \textsc{Figueiredo, M.}
  (2009).
\newblock Adaptive total variation image deblurring: {A}
  majorization-minimization approach.
\newblock \textit{Signal Process.}, \textbf{89} 1683--1693.

\bibitem[{Owen(2001)}]{owen:2001}
\textsc{Owen, A.~B.} (2001).
\newblock \textit{Empirical Likelihood}.
\newblock Chapman \& Hall.

\bibitem[{Paisley et~al.(2012)Paisley, Blei and
  Jordan}]{paisley:blei:jordan:2012}
\textsc{Paisley, J.}, \textsc{Blei, D.~M.} and \textsc{Jordan, M.~I.} (2012).
\newblock Variational {B}ayesian inference with stochastic search.
\newblock In \textit{Proceedings of the 29th International Conference on
  Machine Learning (ICML-12)}. 1367--1374.

\bibitem[{Papaspiliopoulos et~al.(2007)Papaspiliopoulos, Roberts and
  Sk\"{o}ld}]{papa:robe:skol:2007}
\textsc{Papaspiliopoulos, O.}, \textsc{Roberts, G.~O.} and \textsc{Sk\"{o}ld,
  M.} (2007).
\newblock A general framework for the parametrization of hierarchical models.
\newblock \textit{Statistical Science}, \textbf{22} 59--73.

\bibitem[{Parikh and Boyd(2014)}]{BoydBook}
\textsc{Parikh, N.} and \textsc{Boyd, S.} (2014).
\newblock Proximal algorithms.
\newblock \textit{Foundations and Trends in Optimization}, \textbf{1} 123--231.

\bibitem[{Patin et~al.(2009)Patin, Laval, Barreiro, Salas, Semino,
  Santachiara-Benerecetti, Kidd, Kidd, Van Der~Veen, Hombert
  et~al.}]{patin:etal:2009}
\textsc{Patin, E.}, \textsc{Laval, G.}, \textsc{Barreiro, L.}, \textsc{Salas,
  A.}, \textsc{Semino, O.}, \textsc{Santachiara-Benerecetti, S.}, \textsc{Kidd,
  K.}, \textsc{Kidd, J.}, \textsc{Van Der~Veen, L.}, \textsc{Hombert, J.}
  \textsc{et~al.} (2009).
\newblock {Inferring the demographic history of {A}frican farmers and {P}ygmy
  hunter-gatherers using a multilocus resequencing data set}.
\newblock \textit{PLoS Genetics}, \textbf{5} e1000448.

\bibitem[{Pearson(1894)}]{pearson:1894}
\textsc{Pearson, K.} (1894).
\newblock Contribution to the mathematical theory of evolution.
\newblock \textit{Proc. Trans. Royal Society A}, \textbf{185} 71--110.

\bibitem[{Pereyra(2015)}]{pereyra:2014}
\textsc{Pereyra, M.} (2015).
\newblock Proximal {M}arkov chain {M}onte {C}arlo algorithms.
\newblock \textit{Statist. Comp.}
\newblock (to appear).

\bibitem[{Peskun(1973)}]{peskun1973optimum}
\textsc{Peskun, P.} (1973).
\newblock Optimum {M}onte {C}arlo sampling using {M}arkov chains.
\newblock \textit{Biometrika}, \textbf{60} 607--612.

\bibitem[{Pesquet and Pustelnik(2012)}]{pesquet:2012a}
\textsc{Pesquet, J.-C.} and \textsc{Pustelnik, N.} (2012).
\newblock A parallel inertial proximal optimization method.
\newblock \textit{Pac. J. Optim.}, \textbf{8} 273--305.

\bibitem[{Pillai and Smith(2014)}]{pillai2014ergodicity}
\textsc{Pillai, N.~S.} and \textsc{Smith, A.} (2014).
\newblock {Ergodicity of approximate MCMC chains with applications to large
  data sets}.
\newblock \textit{arXiv preprint arXiv:1405.0182}.

\bibitem[{Pillai et~al.(2012)Pillai, Stuart and Thi{\'e}ry}]{pillai2012optimal}
\textsc{Pillai, N.~S.}, \textsc{Stuart, A.~M.} and \textsc{Thi{\'e}ry, A.~H.}
  (2012).
\newblock Optimal scaling and diffusion limits for the {L}angevin algorithm in
  high dimensions.
\newblock \textit{The Annals of Applied Probability}, \textbf{22} 2320--2356.

\bibitem[{Plummer(2014)}]{plummer:2014}
\textsc{Plummer, M.} (2014).
\newblock Cuts in {B}ayesian graphical models.
\newblock \textit{Statistics and Computing} to appear.

\bibitem[{Potts(1952)}]{potts52}
\textsc{Potts, R.~B.} (1952).
\newblock Some generalized order-disorder transitions.
\newblock \textit{Proceedings of Cambridge Philosophical Society}, \textbf{48}
  106--109.

\bibitem[{Pritchard et~al.(1999)Pritchard, Seielstad, Perez-Lezaun and
  Feldman}]{pritchard:seielstad:perez:feldman:1999}
\textsc{Pritchard, J.}, \textsc{Seielstad, M.}, \textsc{Perez-Lezaun, A.} and
  \textsc{Feldman, M.} (1999).
\newblock Population growth of human {Y} chromosomes: a study of {Y} chromosome
  microsatellites.
\newblock \textit{Mol. Biol. Evol.}, \textbf{16} 1791--1798.

\bibitem[{{Pudlo} et~al.(2014){Pudlo}, {Marin}, {Estoup}, {Cornuet}, {Gautier}
  and {Robert}}]{pudlo:pnas:2014}
\textsc{{Pudlo}, P.}, \textsc{{Marin}, J.-M.}, \textsc{{Estoup}, A.},
  \textsc{{Cornuet}, J.-M.}, \textsc{{Gautier}, M.} and \textsc{{Robert},
  C.~P.} (2014).
\newblock {ABC model choice via random forests}.
\newblock \textit{ArXiv e-prints}.
\newblock \eprint{1406.6288}.

\bibitem[{Quiroz et~al.(2014)Quiroz, Villani and Kohn}]{quiroz2014speeding}
\textsc{Quiroz, M.}, \textsc{Villani, M.} and \textsc{Kohn, R.} (2014).
\newblock {Speeding up MCMC by efficient data subsampling}.
\newblock \textit{arXiv preprint arXiv:1404.4178}.

\bibitem[{Raguet et~al.(2013)Raguet, Fadili and Peyr\'e}]{raguet:2013}
\textsc{Raguet, H.}, \textsc{Fadili, J.} and \textsc{Peyr\'e, G.} (2013).
\newblock A generalized forward-backward splitting.
\newblock \textit{SIAM Journal on Imaging Sciences}, \textbf{6} 1199--1226.

\bibitem[{Ramakrishnan and Hadly(2009)}]{ramakrishnan:hadly:2009}
\textsc{Ramakrishnan, U.} and \textsc{Hadly, E.} (2009).
\newblock {Using phylochronology to reveal cryptic population histories: review
  and synthesis of 29 ancient DNA studies}.
\newblock \textit{Molecular Ecology}, \textbf{18} 1310--1330.

\bibitem[{Richardson et~al.(2010)Richardson, Bottolo and
  Rosenthal}]{richardson2010bayesian}
\textsc{Richardson, S.}, \textsc{Bottolo, L.} and \textsc{Rosenthal, J.}
  (2010).
\newblock {Bayesian models for sparse regression analysis of high dimensional
  data}.
\newblock \textit{Bayesian Statistics}, \textbf{9}.

\bibitem[{Richardson and Green(1997)}]{richardson:green:1997}
\textsc{Richardson, S.} and \textsc{Green, P.} (1997).
\newblock On {B}ayesian analysis of mixtures with an unknown number of
  components (with discussion).
\newblock \textit{J. Royal Statist. Society Series B}, \textbf{59} 731--792.

\bibitem[{Robert and Casella(2011)}]{robert:casella:2011}
\textsc{Robert, C.} and \textsc{Casella, G.} (2011).
\newblock A short history of {M}arkov chain {M}onte {C}arlo: Subjective
  recollections from incomplete data.
\newblock \textit{Statist. Science}, \textbf{26} 102--115.

\bibitem[{Robert et~al.(2011)Robert, Cornuet, Marin and
  Pillai}]{robert:cornuet:marin:pillai:2011}
\textsc{Robert, C.}, \textsc{Cornuet, J.-M.}, \textsc{Marin, J.-M.} and
  \textsc{Pillai, N.} (2011).
\newblock Lack of confidence in {ABC} model choice.
\newblock \textit{Proceedings of the National Academy of Sciences},
  \textbf{108(37)} 15112--15117.

\bibitem[{Roberts et~al.(1997)Roberts, Gelman and Gilks}]{roberts1997weak}
\textsc{Roberts, G.}, \textsc{Gelman, A.} and \textsc{Gilks, W.} (1997).
\newblock {Weak convergence and optimal scaling of random walk Metropolis
  algorithms}.
\newblock \textit{The Annals of Applied Probability}, \textbf{7} 110--120.

\bibitem[{Roberts and Rosenthal(1998)}]{robertsetrosenthal98b}
\textsc{Roberts, G.} and \textsc{Rosenthal, J.} (1998).
\newblock Optimal scaling of discrete approximations to {L}angevin diffusions.
\newblock \textit{J. Roy. Stat. Soc. B}, \textbf{60} 255--268.

\bibitem[{Roberts and Rosenthal(2001)}]{roberts2001optimal}
\textsc{Roberts, G.} and \textsc{Rosenthal, J.} (2001).
\newblock {Optimal scaling for various Metropolis--Hastings algorithms}.
\newblock \textit{Statistical Science}, \textbf{16} 351--367.

\bibitem[{Roberts and Rosenthal(2004)}]{roberts2004general}
\textsc{Roberts, G.} and \textsc{Rosenthal, J.} (2004).
\newblock {General state space Markov chains and MCMC algorithms}.
\newblock \textit{Probability Surveys}, \textbf{1} 20--71.

\bibitem[{Roberts and Rosenthal(2007)}]{roberts2007coupling}
\textsc{Roberts, G.} and \textsc{Rosenthal, J.} (2007).
\newblock {Coupling and ergodicity of adaptive Markov chain Monte Carlo
  algorithms}.
\newblock \textit{Journal of Applied Probability}, \textbf{44} 458.

\bibitem[{Roberts and Rosenthal(2009)}]{roberts2009examples}
\textsc{Roberts, G.} and \textsc{Rosenthal, J.} (2009).
\newblock {Examples of adaptive MCMC}.
\newblock \textit{Journal of Computational and Graphical Statistics},
  \textbf{18} 349--367.

\bibitem[{Roberts and Stramer(2002)}]{roberts2002langevin}
\textsc{Roberts, G.} and \textsc{Stramer, O.} (2002).
\newblock {Langevin diffusions and Metropolis--Hastings algorithms}.
\newblock \textit{Methodology and Computing in Applied Probability}, \textbf{4}
  337--357.

\bibitem[{Roberts and Tweedie(1996{\natexlab{a}})}]{roberts1996exponential}
\textsc{Roberts, G.} and \textsc{Tweedie, R.} (1996{\natexlab{a}}).
\newblock {Exponential convergence of Langevin distributions and their discrete
  approximations}.
\newblock \textit{Bernoulli}, \textbf{2} 341--363.

\bibitem[{Roberts and Tweedie(1996{\natexlab{b}})}]{roberts:tweedie:1996}
\textsc{Roberts, G.} and \textsc{Tweedie, R.} (1996{\natexlab{b}}).
\newblock Geometric convergence and central limit theorems for multidimensional
  {H}astings and {M}etropolis algorithms.
\newblock \textit{Biometrika}, \textbf{83} 95--110.

\bibitem[{Roberts(1996)}]{roberts1996markov}
\textsc{Roberts, G.~O.} (1996).
\newblock Markov chain concepts related to sampling algorithms.
\newblock \textit{Markov chain Monte Carlo in practice}, \textbf{57}.

\bibitem[{Roberts(1998)}]{Roberts1998}
\textsc{Roberts, G.~O.} (1998).
\newblock Optimal {M}etropolis algorithms for product measures on the vertices
  of a hypercube.
\newblock \textit{Stochastics and Stochastic Reports}, \textbf{62} 275--283.

\bibitem[{Roberts and Rosenthal(2014)}]{MR3161644}
\textsc{Roberts, G.~O.} and \textsc{Rosenthal, J.~S.} (2014).
\newblock Minimising {MCMC} variance via diffusion limits, with an application
  to simulated tempering.
\newblock \textit{Ann. Appl. Probab.}, \textbf{24} 131--149.

\bibitem[{Roberts and Stramer(2001)}]{roberts2001inference}
\textsc{Roberts, G.~O.} and \textsc{Stramer, O.} (2001).
\newblock On inference for partially observed nonlinear diffusion models using
  the {Metropolis--Hastings} algorithm.
\newblock \textit{Biometrika}, \textbf{88} 603--621.

\bibitem[{Rockafellar(1976)}]{rockafellar:1976}
\textsc{Rockafellar, R.~T.} (1976).
\newblock Monotone operators and the proximal point algorithm.
\newblock \textit{{SIAM} J. Control Optim.}, \textbf{14} 877--898.

\bibitem[{Rossky et~al.(1978)Rossky, Doll and Friedman}]{Rossky1978}
\textsc{Rossky, P.}, \textsc{Doll, J.} and \textsc{Friedman, H.} (1978).
\newblock Brownian dynamics as smart {Monte Carlo} simulation.
\newblock \textit{The Journal of Chemical Physics}, \textbf{69} 4628--4633.

\bibitem[{Rubin(1984)}]{rubin:1984}
\textsc{Rubin, D.} (1984).
\newblock Bayesianly justifiable and relevant frequency calculations for the
  applied statistician.
\newblock \textit{Ann. Statist.}, \textbf{12} 1151--1172.

\bibitem[{Rubinstein(1981)}]{rubinstein81}
\textsc{Rubinstein, R.~Y.} (1981).
\newblock \textit{Simulation and the {M}onte {C}arlo {M}ethod}.
\newblock J. Wiley, New York.

\bibitem[{Rudolf and Schweizer(2015)}]{rudolf2015perturbation}
\textsc{Rudolf, D.} and \textsc{Schweizer, N.} (2015).
\newblock {Perturbation theory for Markov chains via Wasserstein distance}.
\newblock \textit{arXiv preprint arXiv:1503.04123}.

\bibitem[{Rue et~al.(2009)Rue, Martino and Chopin}]{rue:martino:chopin:2009}
\textsc{Rue, H.}, \textsc{Martino, S.} and \textsc{Chopin, N.} (2009).
\newblock Approximate {B}ayesian inference for latent {G}aussian models using
  integrated nested {L}aplace approximations (with discussion).
\newblock \textit{J. Royal Statist. Society Series B}, \textbf{71} 319--392.

\bibitem[{Saksman and Vihola(2010)}]{saksman2008ergodicity}
\textsc{Saksman, E.} and \textsc{Vihola, M.} (2010).
\newblock On the ergodicity of the adaptive {M}etropolis algorithm on unbounded
  domains.
\newblock \textit{The Annals of Applied Probability}, \textbf{20} 2178--2203.

\bibitem[{Salimans and Knowles(2013)}]{salimans:knowles:2013}
\textsc{Salimans, T.} and \textsc{Knowles, D.~A.} (2013).
\newblock Fixed-form variational posterior approximation through stochastic
  linear regression.
\newblock \textit{Bayesian Anal.}, \textbf{8} 837--882.

\bibitem[{Saloff-Coste(1997)}]{MR1490046}
\textsc{Saloff-Coste, L.} (1997).
\newblock Lectures on finite {M}arkov chains.
\newblock In \textit{Lectures on probability theory and statistics
  ({S}aint-{F}lour, 1996)}, vol. 1665 of \textit{Lecture Notes in Math.}
  Springer, Berlin, 301--413.

\bibitem[{Schreck et~al.(2013)Schreck, Fort, Corff and
  Moulines}]{schreck2013shrinkage}
\textsc{Schreck, A.}, \textsc{Fort, G.}, \textsc{Corff, S.~L.} and
  \textsc{Moulines, E.} (2013).
\newblock A shrinkage-thresholding {Metropolis adjusted Langevin algorithm for
  Bayesian} variable selection.
\newblock \textit{arXiv preprint arXiv:1312.5658}.

\bibitem[{Schr{\"o}dle and Held(2011)}]{schrodle:held:2011}
\textsc{Schr{\"o}dle, B.} and \textsc{Held, L.} (2011).
\newblock A primer on disease mapping and ecological regression using
  {{\texttt{INLA}}}.
\newblock \textit{Computational Statistics}, \textbf{26} 241--258.

\bibitem[{Scott et~al.(2013)Scott, Blocker, Bonassi, Chipman, George and
  McCulloch}]{scott:etal:2013}
\textsc{Scott, S.}, \textsc{Blocker, A.}, \textsc{Bonassi, F.},
  \textsc{Chipman, H.}, \textsc{George, E.} and \textsc{McCulloch, R.} (2013).
\newblock Bayes and big data: The consensus {M}onte {C}arlo algorithm.
\newblock \textit{EFaBBayes 250 conference}, \textbf{16}.

\bibitem[{Searle et~al.(1992)Searle, Casella and
  McCulloch}]{searle:casella:mcculloch:1992}
\textsc{Searle, S.}, \textsc{Casella, G.} and \textsc{McCulloch, C.} (1992).
\newblock \textit{Variance Components}.
\newblock John Wiley, New York.

\bibitem[{Sherlock(2014)}]{sherlock2014optimal}
\textsc{Sherlock, C.} (2014).
\newblock Optimal scaling for the pseudo-marginal random walk {M}etropolis:
  insensitivity to the noise generating mechanism.
\newblock \textit{arXiv preprint arXiv:1408.4344}.

\bibitem[{Sherlock et~al.(2014)Sherlock, Thiery, Roberts and
  Rosenthal}]{sherlock2014efficiency}
\textsc{Sherlock, C.}, \textsc{Thiery, A.~H.}, \textsc{Roberts, G.~O.} and
  \textsc{Rosenthal, J.~S.} (2014).
\newblock On the efficiency of pseudo-marginal random walk {M}etropolis
  algorithms.
\newblock \textit{The Annals of Statistics}, \textbf{43} 238--275.

\bibitem[{Smith et~al.(1985)Smith, Skene, Shaw, Naylor and
  Dransfield}]{smith:sken:1985}
\textsc{Smith, A.}, \textsc{Skene, A.}, \textsc{Shaw, J.}, \textsc{Naylor, J.}
  and \textsc{Dransfield, M.} (1985).
\newblock The implementation of the {B}ayesian paradigm.
\newblock \textit{Comm. Statist.-Theory Methods}, \textbf{14} 1079--1102.

\bibitem[{Smith et~al.(1987)Smith, Skene, Shaw and Naylor}]{smithssn}
\textsc{Smith, A.}, \textsc{Skene, A.~M.}, \textsc{Shaw, J. E.~H.} and
  \textsc{Naylor, J.~C.} (1987).
\newblock Progress with numerical and graphical methods for practical
  {B}ayesian statistics.
\newblock \textit{J. Roy. Statist. Soc. Series D}, \textbf{36} 75--82.

\bibitem[{Solonen et~al.(2012)Solonen, Ollinaho, Laine, Haario, Tamminen and
  J{\"a}rvinen}]{solonen2012efficient}
\textsc{Solonen, A.}, \textsc{Ollinaho, P.}, \textsc{Laine, M.},
  \textsc{Haario, H.}, \textsc{Tamminen, J.} and \textsc{J{\"a}rvinen, H.}
  (2012).
\newblock Efficient {MCMC} for climate model parameter estimation: Parallel
  adaptive chains and early rejection.
\newblock \textit{Bayesian Analysis}, \textbf{7} 715--736.

\bibitem[{{Stan Development Team}(2014)}]{stan-software:2014}
\textsc{{Stan Development Team}} (2014).
\newblock {STAN}: A {C++} library for probability and sampling, version 2.5.0.
\newblock \urlprefix\url{http://mc-stan.org/}.

\bibitem[{Stephens(2000)}]{stephens:2000}
\textsc{Stephens, M.} (2000).
\newblock {B}ayesian analysis of mixture models with an unknown number of
  components---an alternative to reversible jump methods.
\newblock \textit{Ann. Statist.}, \textbf{28} 40--74.

\bibitem[{Stramer and Tweedie(1999{\natexlab{a}})}]{stramer1999langevinI}
\textsc{Stramer, O.} and \textsc{Tweedie, R.} (1999{\natexlab{a}}).
\newblock Langevin-type models {I}: Diffusions with given stationary
  distributions and their discretizations.
\newblock \textit{Methodology and Computing in Applied Probability}, \textbf{1}
  283--306.

\bibitem[{Stramer and Tweedie(1999{\natexlab{b}})}]{stramer1999langevinII}
\textsc{Stramer, O.} and \textsc{Tweedie, R.} (1999{\natexlab{b}}).
\newblock Langevin-type models {II}: Self-targeting candidates for {MCMC}
  algorithms.
\newblock \textit{Methodology and Computing in Applied Probability}, \textbf{1}
  307--328.

\bibitem[{Strathmann et~al.(2015)Strathmann, Sejdinovic and
  Girolami}]{strathmann2015unbiased}
\textsc{Strathmann, H.}, \textsc{Sejdinovic, D.} and \textsc{Girolami, M.}
  (2015).
\newblock Unbiased {B}ayes for big data: Paths of partial posteriors.
\newblock \textit{arXiv preprint arXiv:1501.03326}.

\bibitem[{Strid(2010)}]{strid:2010}
\textsc{Strid, I.} (2010).
\newblock Efficient parallelisation of {M}etropolis--{H}astings algorithms
  using a prefetching approach.
\newblock \textit{Computational Statistics \& Data Analysis}, \textbf{54}
  2814--2835.

\bibitem[{Suchard et~al.(2010)Suchard, Wang, Chan, Frelinger, Cron and
  West}]{suchard2010}
\textsc{Suchard, M.}, \textsc{Wang, Q.}, \textsc{Chan, C.}, \textsc{Frelinger,
  J.}, \textsc{Cron, A.} and \textsc{West, M.} (2010).
\newblock Understanding {GPU} programming for statistical computation: studies
  in massively parallel massive mixtures.
\newblock \textit{Journal of Computational and Graphical Statistics},
  \textbf{19} 418--438.

\bibitem[{Tanner and Wong(1987)}]{tanner:wong:1987}
\textsc{Tanner, M.} and \textsc{Wong, W.} (1987).
\newblock The calculation of posterior distributions by data augmentation.
\newblock \textit{J. American Statist. Assoc.}, \textbf{82} 528--550.

\bibitem[{Tavar{\'e} et~al.(1997)Tavar{\'e}, Balding, Griffith and
  Donnelly}]{tavare:balding:griffith:donnelly:1997}
\textsc{Tavar{\'e}, S.}, \textsc{Balding, D.}, \textsc{Griffith, R.} and
  \textsc{Donnelly, P.} (1997).
\newblock Inferring coalescence times from {DNA} sequence data.
\newblock \textit{Genetics}, \textbf{145} 505--518.

\bibitem[{Taylor(2014)}]{WolnyPHD}
\textsc{Taylor, K.} (2014).
\newblock Ph{D} thesis.
\newblock \textit{University of Warwick}.

\bibitem[{Teh et~al.(2014)Teh, Thi{\'e}ry and Vollmer}]{teh2014consistency}
\textsc{Teh, Y.~W.}, \textsc{Thi{\'e}ry, A.} and \textsc{Vollmer, S.} (2014).
\newblock Consistency and fluctuations for stochastic gradient {L}angevin
  dynamics.
\newblock \textit{arXiv preprint arXiv:1409.0578}.

\bibitem[{Templeton(2008)}]{templeton:2008}
\textsc{Templeton, A.} (2008).
\newblock Statistical hypothesis testing in intraspecific phylogeography:
  nested clade phylogeographical analysis vs. {A}pproximate {B}ayesian
  computation.
\newblock \textit{Molecular Ecology}, \textbf{18(2)} 319--331.

\bibitem[{Templeton(2010)}]{templeton:2010}
\textsc{Templeton, A.} (2010).
\newblock Coherent and incoherent inference in phylogeography and human
  evolution.
\newblock \textit{Proc.~National Academy of Sciences}, \textbf{107(14)}
  6376--6381.

\bibitem[{Tierney(1998)}]{tierney1998note}
\textsc{Tierney, L.} (1998).
\newblock {A note on Metropolis--Hastings kernels for general state spaces}.
\newblock \textit{Annals of Applied Probability}, \textbf{8} 1--9.

\bibitem[{Toni et~al.(2009)Toni, Welch, Strelkowa, Ipsen and
  Stumpf}]{toni:etal:2009}
\textsc{Toni, T.}, \textsc{Welch, D.}, \textsc{Strelkowa, N.}, \textsc{Ipsen,
  A.} and \textsc{Stumpf, M.} (2009).
\newblock {Approximate {B}ayesian computation scheme for parameter inference
  and model selection in dynamical systems}.
\newblock \textit{Journal of the Royal Society Interface}, \textbf{6} 187--202.

\bibitem[{{VanDerwerken} and {Schmidler}(2013)}]{vanderwerken:schmidler:2013}
\textsc{{VanDerwerken}, D.~N.} and \textsc{{Schmidler}, S.~C.} (2013).
\newblock Parallel {M}arkov chain {M}onte {C}arlo.
\newblock \textit{ArXiv e-prints}.
\newblock \eprint{1312.7479}.

\bibitem[{Verdinelli and Wasserman(1991)}]{verdinelli:wasserman:1991}
\textsc{Verdinelli, I.} and \textsc{Wasserman, L.} (1991).
\newblock {B}ayesian analysis of outlier problems using the {G}ibbs sampler.
\newblock \textit{{S}tatist. Comput.}, \textbf{1} 105--117.

\bibitem[{Verdu et~al.(2009)Verdu, Austerlitz, Estoup, Vitalis, Georges,
  Th\'ery, Froment, Le~Bomin, Gessain, Hombert, Van~der Veen, Quintana-Murci,
  Bahuchet and Heyer}]{verdu:etal:2009}
\textsc{Verdu, P.}, \textsc{Austerlitz, F.}, \textsc{Estoup, A.},
  \textsc{Vitalis, R.}, \textsc{Georges, M.}, \textsc{Th\'ery, S.},
  \textsc{Froment, A.}, \textsc{Le~Bomin, S.}, \textsc{Gessain, A.},
  \textsc{Hombert, J.-M.}, \textsc{Van~der Veen, L.}, \textsc{Quintana-Murci,
  L.}, \textsc{Bahuchet, S.} and \textsc{Heyer, E.} (2009).
\newblock Origins and genetic diversity of pygmy hunter-gatherers from {Western
  Central Africa}.
\newblock \textit{Current Biology}, \textbf{19} 312--318.

\bibitem[{Vihola(2012)}]{vihola2012robust}
\textsc{Vihola, M.} (2012).
\newblock Robust adaptive {M}etropolis algorithm with coerced acceptance rate.
\newblock \textit{Statistics and Computing}, \textbf{22} 997--1008.

\bibitem[{Wakefield et~al.(1991)Wakefield, Gelfand and
  Smith}]{wakefield:gelfand:smith:1991}
\textsc{Wakefield, J.}, \textsc{Gelfand, A.} and \textsc{Smith, A.} (1991).
\newblock Efficient generation of random variates via the ratio-of-uniforms
  method.
\newblock \textit{{S}tatistics and Computing}, \textbf{1} 129--133.

\bibitem[{Wang and Landau(2001)}]{wangetlandau01}
\textsc{Wang, F.} and \textsc{Landau, D.~P.} (2001).
\newblock Efficient, multiple-range random walk algorithm to calculate the
  density of states.
\newblock \textit{Physical Review Letters}, \textbf{86} 2050--2053.

\bibitem[{Wang and Dunson(2013)}]{wang:dunson:2013}
\textsc{Wang, X.} and \textsc{Dunson, D.} (2013).
\newblock Parallizing {MCMC} via {W}eierstrass sampler.
\newblock \textit{arXiv preprint arXiv:1312.4605}.

\bibitem[{Wang et~al.(2013)Wang, Mohamed and de~Freitas}]{wang2013adaptive}
\textsc{Wang, Z.}, \textsc{Mohamed, S.} and \textsc{de~Freitas, N.} (2013).
\newblock Adaptive {Hamiltonian and Riemann manifold Monte Carlo}.
\newblock In \textit{Proceedings of The 30th International Conference on
  Machine Learning}. 1462--1470.

\bibitem[{Wegmann and Excoffier(2010)}]{wegmann:excoffier:2010}
\textsc{Wegmann, D.} and \textsc{Excoffier, L.} (2010).
\newblock {Bayesian inference of the demographic history of chimpanzees}.
\newblock \textit{Molecular Biology and Evolution}, \textbf{27} 1425--1435.

\bibitem[{Welling and Teh(2011)}]{welling2011bayesian}
\textsc{Welling, M.} and \textsc{Teh, Y.~W.} (2011).
\newblock {Bayesian learning via stochastic gradient Langevin dynamics}.
\newblock In \textit{Proceedings of the 28th International Conference on
  Machine Learning (ICML-11)}. 681--688.

\bibitem[{White(2012)}]{white2012}
\textsc{White, T.} (2012).
\newblock \textit{Hadoop: the definitive guide}.
\newblock O'Reilly Media.

\bibitem[{{Whiteley} et~al.(2010){Whiteley}, {Andrieu} and
  {Doucet}}]{whiteley:andrieu:doucet:2010}
\textsc{{Whiteley}, N.}, \textsc{{Andrieu}, C.} and \textsc{{Doucet}, A.}
  (2010).
\newblock Efficient {B}ayesian inference for switching state-space models using
  discrete particle {Markov chain Monte Carlo} methods.
\newblock \textit{ArXiv e-prints}.
\newblock \eprint{1011.2437}.

\bibitem[{Wilkinson(2005)}]{wilkinson:2005}
\textsc{Wilkinson, D.} (2005).
\newblock Parallel {B}ayesian computation.
\newblock In \textit{Handbook of Parallel Computing and Statistics} (E.~J.
  Kontoghiorghes, ed.). Marcel Dekker/CRC Press, New York, 481--512.

\bibitem[{Wilkinson(2011{\natexlab{a}})}]{wilkinson:2011}
\textsc{Wilkinson, D.} (2011{\natexlab{a}}).
\newblock The particle marginal {M}etropolis--{H}astings {(PMMH)} particle
  {MCMC} algorithm.
\newblock
  https://darrenjw.wordpress.com/2011/05/17/the-particle-marginal-metropolis-hastings-pmmh-particle-mcmc-algorithm/.
\newblock Darren {W}ilkinson's research blog.

\bibitem[{Wilkinson(2011{\natexlab{b}})}]{wilkinson:2011b}
\textsc{Wilkinson, D.~J.} (2011{\natexlab{b}}).
\newblock \textit{Stochastic modelling for systems biology}.
\newblock CRC press, New York.
\newblock (Second edition).

\bibitem[{Wilkinson(2013)}]{wilkinson:2013}
\textsc{Wilkinson, R.} (2013).
\newblock Approximate {B}ayesian computation {(ABC)} gives exact results under
  the assumption of model error.
\newblock \textit{Statistical Applications in Genetics and Molecular Biology},
  \textbf{12} 129--141.

\bibitem[{Xifara et~al.(2014)Xifara, Sherlock, Livingstone, Byrne and
  Girolami}]{xifara2014langevin}
\textsc{Xifara, T.}, \textsc{Sherlock, C.}, \textsc{Livingstone, S.},
  \textsc{Byrne, S.} and \textsc{Girolami, M.} (2014).
\newblock Langevin diffusions and the {M}etropolis-adjusted {L}angevin
  algorithm.
\newblock \textit{Statistics \& Probability Letters}, \textbf{91} 14--19.

\end{thebibliography}
